\documentclass[a4paper,11pt]{article}
\pdfoutput=1 % if your are submitting a pdflatex (i.e. if you have
\usepackage{jheppub} % for details on the use of the package, please
\usepackage{amsmath,amssymb,amsthm,amscd,graphicx,xcolor}
\usepackage{slashed}
\usepackage{braket}
\input epsf.sty

\graphicspath{{figures/}}

\usepackage{texdraw}
\newtheorem{theorem}{Theorem}[section]

\theoremstyle{definition}

\newtheorem{example}[theorem]{Example}

\newcommand{\CC}{{\cal C}}

\newcommand{\CF}{{\cal F}}
\newcommand{\CG}{{\cal G}}

\newcommand{\CI}{{\cal I}}

\newcommand{\CL}{{\cal L}}

\newcommand{\CO}{{\cal O}}

\newcommand{\CT}{{\cal T}}

\def\IZ{{\mathbb Z}}
\def\IR{{\mathbb R}}

\def\IN{{\mathbb N}}

\newcommand{\mR}{\mathsf{R}}
\newcommand{\mQ}{\mathsf{Q}}

\newcommand{\mS}{\mathsf{S}}

\newcommand{\mK}{\mathsf{K}}

\newcommand{\mH}{\mathsf{H}}

\newcommand{\dd}{{\rm d}}

\newcommand{\re}{{\rm e}}
\newcommand{\ri}{{\rm i}}
\newcommand{\rd}{{\rm d}}

\DeclareMathOperator{\Res}{Res}
\DeclareMathOperator{\disc}{disc}

\newcommand{\be}{\begin{equation}}
\newcommand{\ee}{\end{equation}}
\newcommand{\ba}{\begin{aligned}}
\newcommand{\ea}{\end{aligned}}
\newcommand{\ben}{\begin{eqnarray}\displaystyle}
\newcommand{\een}{\end{eqnarray}}

\newcommand{\sectiono}[1]{\section{#1}\setcounter{equation}{0}}

\newdimen\tableauside\tableauside=1.0ex
\newdimen\tableaurule\tableaurule=0.4pt
\newdimen\tableaustep
\def\phantomhrule#1{\hbox{\vbox to0pt{\hrule height\tableaurule width#1\vss}}}
\def\phantomvrule#1{\vbox{\hbox to0pt{\vrule width\tableaurule height#1\hss}}}
\def\sqr{\vbox{%
  \phantomhrule\tableaustep
  \hbox{\phantomvrule\tableaustep\kern\tableaustep\phantomvrule\tableaustep}%
  \hbox{\vbox{\phantomhrule\tableauside}\kern-\tableaurule}}}
\def\squares#1{\hbox{\count0=#1\noindent\loop\sqr
  \advance\count0 by-1 \ifnum\count0>0\repeat}}
\def\tableau#1{\vcenter{\offinterlineskip
  \tableaustep=\tableauside\advance\tableaustep by-\tableaurule
  \kern\normallineskip\hbox
    {\kern\normallineskip\vbox
      {\gettableau#1 0 }%
     \kern\normallineskip\kern\tableaurule}%
  \kern\normallineskip\kern\tableaurule}}
\def\gettableau#1{\ifnum#1=0\let\next=\null\else
\squares{#1}\let\next=\gettableau\fi\next}

\newcommand{\figref}[1]{Fig.~\protect\ref{#1}}
\usepackage{mathtools}

\title{\Huge{\boldmath New renormalons from analytic trans-series}}

\author{Marcos Mari\~no,}
\author{Ramon Miravitllas}
\author{and Tom\'as Reis}

\affiliation{D\'epartement de Physique Th\'eorique et Section de Math\'ematiques\\
Universit\'e de Gen\`eve, Gen\`eve, CH-1211 Switzerland}

\emailAdd{Marcos.Marino@unige.ch}
\emailAdd{Ramon.MiravitllasMas@unige.ch} 
\emailAdd{Tomas.Reis@unige.ch} 

\abstract{
We study the free energy of integrable, asymptotically free field theories in two dimensions coupled to a conserved charge. 
We develop methods to obtain analytic expressions for its trans-series expansion, directly from the Bethe ansatz equations, 
and we use this result to determine the structure of its Borel singularities. We find a new class of infrared renormalons 
which does not fit the traditional expectations of renormalon physics 
proposed long ago by 't~Hooft and Parisi. We check the existence of these new singularities with detailed calculations based 
on the resurgent analysis of the perturbative expansion. Our results show that the structure of renormalons in asymptotically 
free theories is more subtle than previously thought, and that large $N$ estimates of their location might be misleading.}

\begin{document}
\maketitle
\flushbottom
 
%\sectiono{Introduction}

\sectiono{Introduction} 

Understanding the formal properties of perturbative expansions in quantum field theory is an old and venerable problem. Since 
perturbative series are in general factorially divergent, a useful 
approach is to consider their Borel transforms, which are analytic at the origin but have a rich singularity structure in the complex plane. 
These singularities are expected to give important information about non-perturbative physics. 
Some of them are instanton singularities, corresponding to non-trivial saddle-points of the path integral. There are in 
addition renormalon singularities which do not have an obvious semiclassical interpretation. In the case of instanton singularities, 
their location in the complex plane correspond to the 
values of their actions. In the case of renormalons in asympotically free theories, it was argued in \cite{parisi2, parisi1,thooft} that the corresponding singularities occur at points of the form 
\be
\label{slore}
\frac{\ell }{ 2 |\beta_0|} , \qquad  \ell \in \IZ_{\not=0}, 
\ee
where $\beta_0$ is the first coefficient of the beta function. When $\ell>0$, these singularities are called infra-red (IR) renormalons, and they obstruct Borel summability of the perturbative series. The perturbative series is expected to be upgraded to a so-called trans-series, incorporating 
exponentially small corrections associated to the IR renormalon singularities. These corrections are roughly of the form 
\be
\label{lam-power}
\left( \frac{\Lambda }{ \kappa} \right)^{\ell}, \qquad \ell \in \IZ_{>0}, 
\ee
where $\Lambda$ is the dynamically generated scale of the theory and $\kappa\gg \Lambda$ is an external momentum scale. 
For observables with an OPE, these corrections have been related to condensates of 
operators of dimension $\ell$ in the true vacuum \cite{parisi2,itep}. When $\ell<0$, the singularities (\ref{slore}) are called ultra-violet (UV) renormalons. They do not obstruct Borel summability, but they contribute to the large order behavior of perturbation theory. 

Let us note that (\ref{slore}) is supposed to give the position of {\it possible} singularities, and not all of them 
occur in a given observable. For example, in the Adler current of QCD, which is a popular example in renormalon physics, the first IR renormalon singularity occurs at $\ell=4$ (see e.g. \cite{beneke} for a review of renormalons).  

Due to the complexity of realistic quantum field theories, it is not easy 
to test these ideas in detail, and the available evidence relies either on  
large $N$ approximations or on numerical calculations. For example, in QCD, large $N_f$ techniques seem to confirm (\ref{slore}) for certain observables, and numerical calculations of long perturbative series \cite{pineda,pineda2} have established 
the existence of renormalon singularities at $\ell=1,4$. The same techniques can be applied to simpler models 
in lower dimensions. In the two-dimensional, $O(N)$ non-linear sigma model, 
correlation functions can be studied analytically in the $1/N$ expansion, and one finds an infinite sequence of IR 
renormalons of the form (\ref{slore}) with even positive values of $\ell$ \cite{david1,david2,david3, itep2d, beneke-braun, shifman}. 
Numerical evidence for renormalons in the two-dimensional principal chiral field (PCF) has been given in \cite{puhr}. 

The study of the non-linear sigma model suggests that theories in lower dimension and 
with special properties might provide a powerful 
testing ground for renormalon physics, and much interest has been devoted to integrable, asymptotically free theories in two dimensions. 
It has been known for a long time that, 
once one includes an external chemical potential $h$ coupled to a conserved charge, the free energy 
of these models, which we will denote by $\CF(h)$, can be computed exactly by using the Bethe ansatz \cite{pw,wiegmann2,hmn,hn,fnw1,fnw2,pcf,eh-ssm,eh-scpn,eh-review}. At the same time, for large values of $h$ one can use asymptotic freedom to calculate $\CF(h)$ in conventional perturbation theory. Therefore, this observable seems to be rich enough to display all the subtleties of renormalon physics, and at the same time one expects to be able to study it in detail thanks to integrability. 

In spite of these simplifying features, the analysis of the renormalon structure of $\CF(h)$ is not straightforward, and 
the available results are again based on numerical calculations or large $N$ approximations. The numerical analysis 
of the renormalons of $\CF(h)$ was boosted by a new method introduced by Volin in \cite{volin,volin-thesis}, which produces long 
perturbative series for this observable directly from the Bethe ansatz. This method confirmed the presence of a renormalon singularity at $\ell=2$ in the non-linear sigma model \cite{volin} and in many other integrable models, like the Gross--Neveu (GN) model and the PCF \cite{mr-ren}. A comprehensive study of the $O(4)$ non-linear sigma model with these numerical techniques was presented in \cite{abbh1,abbh2}. The free energy of integrable models has been also 
studied in the $1/N$ expansion, both with the Bethe ansatz equations \cite{fkw1,fkw2,ksz,dpmss} and with diagrammatic techniques \cite{mmr,dpmss}, and in this framework one can obtain analytic results for the exponentially small corrections associated to the renormalons. 

% and the  results can be obtained only the singularity at $\ell=2$ of the $O(N)$ non-linear sigma model survives.

The study of renormalons with numerical methods or with the large $N$ approximation has obvious limitations, and it would be desirable to find analytic results at finite $N$. In the case of the free energy of integrable models, one could expect 
that the Bethe ansatz equations encode the full renormalon structure. 
This turns out to be the case, as we explain in this paper. In fact, a closely related analysis of exponentially small corrections 
in the sine--Gordon model, directly from the Bethe ansatz, was already performed by Al. Zamolodchikov 
in \cite{zamo-mass}. By using the Wiener--Hopf techniques of \cite{hmn,hn,fnw1,zamo-mass}, 
we provide for the first time exact analytic results for the trans-series 
in this class of models, at finite $N$, and we compute the very first terms of the 
leading exponentially small corrections, including their Stokes constants. These corrections turn out to be manifestly ambiguous, 
in agreement with the prescient ideas of F. David \cite{david2,david3}, and by requiring the cancellation of ambiguities 
we can determine the position of the renormalon singularities in the Borel plane. 

 \begin{figure}[!ht]
\leavevmode
\begin{center}
\includegraphics[height=8.5cm]{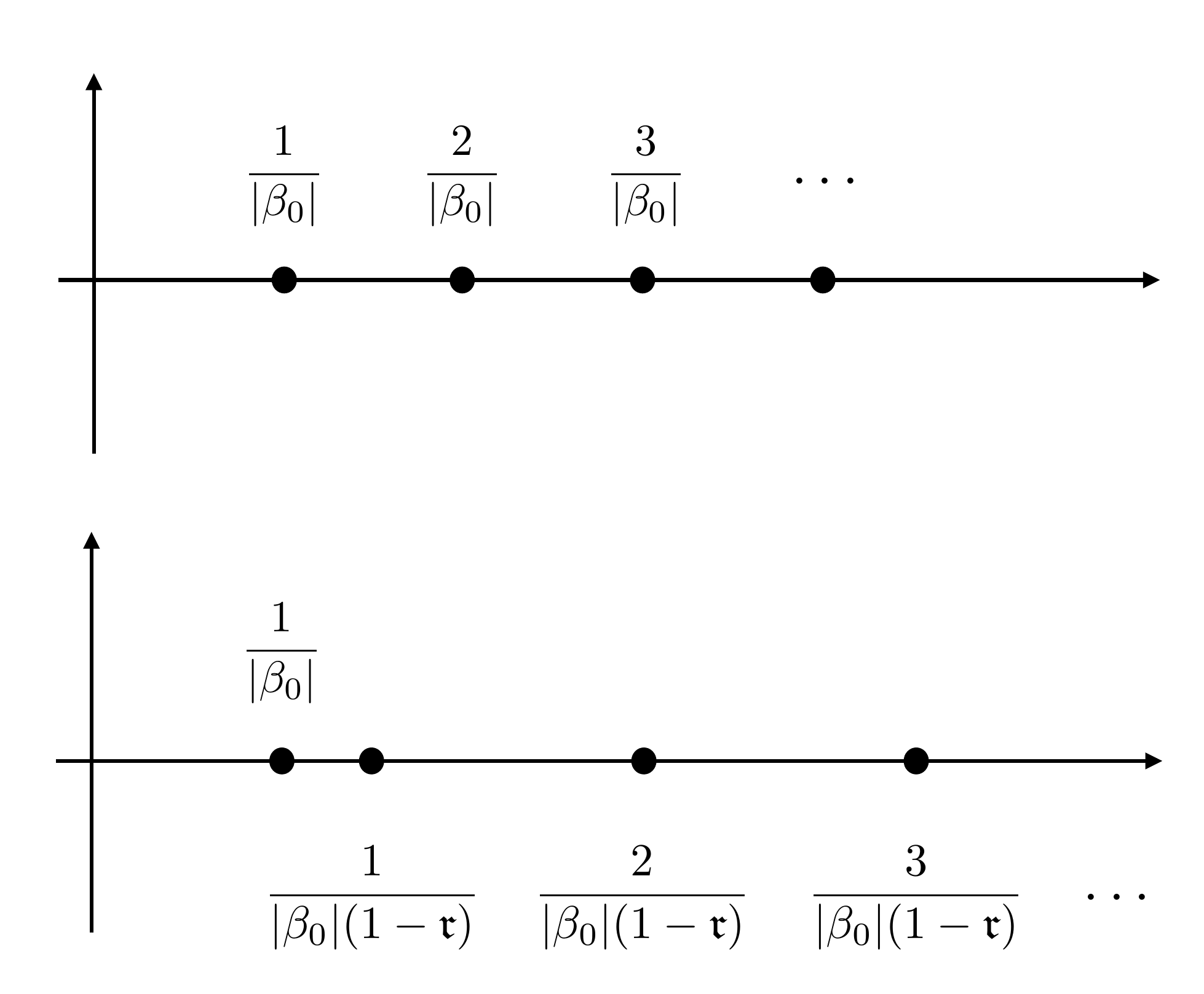}
\end{center}
\caption{The figure at the top shows the traditional picture of IR renormalon singularities in asymptotically free theories, 
corresponding to (\ref{slore}) with an even value of $\ell$. The figure at the bottom 
shows the actual singularities that are found for the free energy of the Gross--Neveu model and the principal chiral field, where the correction $\mathfrak{r}$ is given in (\ref{cor-gn}), (\ref{cor-pcf}), respectively.}
\label{rens-fig}
\end{figure} 

Our most surprising result is that, for the free energy $\CF(h)$, the 
standard expectation (\ref{slore}) about the location of IR singularities turns out to be generically {\it incorrect}. 
For example, in the GN model and the PCF, we find the expected first IR singularity at $1/|\beta_0|$, followed by an infinite sequence of IR singularities at the positions 
\be
\label{corrected}
{\ell \over |\beta_0|} {1\over 1-\mathfrak{r}}, \qquad \ell \in \IZ_{> 0},
\ee
where 
\be
\label{cor-gn}
\mathfrak{r}_{\rm GN}= {2\over N-2}
\ee
for the  $O(N)$ GN model, and
\be
\label{cor-pcf}
\mathfrak{r}_{\rm PCF}= {1\over N} 
\ee
for the $SU(N)$ PCF. These results are illustrated in \figref{rens-fig}. Note that, in the large $N$ limit, the 
singularities (\ref{corrected}) agree with the standard expectation (\ref{slore}) (for even $\ell$), so the discrepancy we find is invisible at large $N$. Similar results hold as well for the supersymmetric $O(N)$ non-linear sigma model, which also presents unconventional Borel singularities of the form \eqref{corrected}.

In the $O(N)$ sigma model, we find an expected first singularity at $1/|\beta_0|$, and then a sequence of singuarities at 
\be
\label{on-insts}
{ \ell (N-2) \over  |\beta_0|}, \qquad \ell \in \IZ_{> 0}. 
\ee
A similar sequence of the form 
\be
\label{pcf-insts}
{ \ell N \over  |\beta_0|}, \qquad \ell \in \IZ_{> 0},
\ee
can also appear in the PCF model. 
Note that the singularities (\ref{on-insts}) and (\ref{pcf-insts}), 
although compatible 
with the standard expectations, are very different from the previous ones, since they go away in the large $N$ limit, and 
thus they might be due to instantons\footnote{As in \cite{mmbook}, we call instanton any solution to the Euclidean equations of motion with 
finite action. Instantons can be unstable, and unstable instantons are sometimes called ``bounces" in the literature. The $O(N)$ sigma model with $N>2$ and the $SU(N)$ PCF with $N \ge 2$ admit unstable instanton configurations.}. %In the PCF we expect to have singularities combining the sequences (\ref{corrected}) and (\ref{pcf-insts}).  
It turns out that the location of UV renormalon singularities can be also determined with this method, and it is compatible with standard expectations. 
We use similar analytic methods to study a well-known non-relativistic model: the 
Gaudin--Yang model. The first Borel singularity of its ground state energy was determined numerically in \cite{mr-ll, mr-long}, and 
we provide here an analytic derivation of its location and its Stokes constant. 

The result (\ref{corrected}) is quite unexpected and goes against the standard lore of renormalon physics. 
Since extraordinary claims call for extraordinary evidence, we provide many tests of our formulae. It follows from the 
theory of resurgence that the analytic results on renormalons obtained in this paper give falsifiable predictions on the behavior of the perturbative series. We then generate long perturbative series with the techniques of \cite{volin} and we verify 
these predictions numerically. In our view, there is very little doubt that renormalon singularities do occur at these unexpected positions.

This paper is organized as follows. In section \ref{rba-sec} we review some background on the theory of resurgence and on the Bethe ansatz solution for integrable quantum field theories. In section \ref{GN-sec} we analyze in detail the free energy of the GN model, 
calculate its trans-series representation at the very first orders, and extract the information about the structure of IR renormalons. These results are then tested in detail, both analytically and numerically. In particular, we give what we find is convincing evidence that unconventional renormalon singularities do really appear. In addition, we show that our methods can also handle UV 
renormalons. In section 
\ref{bos-sec} we develop our analytic formalism for bosonic models, and we present general results for 
their trans-series structure. We give all the details for the non-linear $O(N)$ sigma model, its supersymmetric 
extension, and the PCF, and we present tests of our results. In section \ref{sec-nr} we extend our 
methods to the Gaudin--Yang model, which is a non-relativistic version of the 
GN model. We study analytically the resurgent structure of its ground state energy, 
and we find agreement with the numerical results obtained previously in \cite{mr-ll, mr-ren}. 
Section \ref{sec-conclusions} contains a discussion of the technical and conceptual issues raised by our results, as well as 
some prospects for future developments. The paper 
contains two Appendices.
In Appendix~\ref{pert-calculation}, we explain in detail the perturbative calculation of the free energy for bosonic models, which was first sketched and numerically computed in \cite{hmn,hn, pcf}. In contrast, our computation is performed analytically, by using the mathematical tools  introduced in Appendix \ref{airy-kernel}.
%In Appendix \ref{pert-calculation}, we explain in detail the perturbative calculation 
%of the free energy sketched in \cite{hmn,hn, pcf}. By using some mathematical results explained in Appendix \ref{airy-kernel}, 
%we are able to obtain a fully analytic derivation of the results presented in \cite{hmn,hn,pcf}. 

\sectiono{Resurgence and the Bethe ansatz}
\label{rba-sec}

\subsection{Resurgent structures in QFT}
\label{sec-resQFT}
We will first discuss some basic aspects of the theory of resurgence which will be needed in this paper. An excellent introduction to the mathematical formalism of resurgence can be found in \cite{ss}, see also \cite{abs} for a more comprehensive exposition. 
A presentation in the context of instanton and renormalon physics can be found in \cite{mmbook,mmlargen}. 
A phenomenologically-oriented review of renormalons can be found in \cite{beneke}. 

Let us consider a formal perturbative series $\varphi(\alpha)$, obtained as an asymptotic expansion 
of an observable $\CG(\alpha)$. Here, 
$\alpha$ will denote a convenient coupling constant, and we will assume that $\varphi(\alpha)$ has the form
\be
\label{p-series}
\varphi(\alpha)= \sum_{k \ge 0} e_k \alpha^k. 
\ee
Generically, the coefficients $e_k$ grow as $e_k \sim k!$, so the above series is purely formal and has a zero radius of convergence. 
The Borel transform of $\varphi(\alpha)$, defined as 
\be
\widehat \varphi(\zeta)= \sum_{k \ge 0} {e_k \over k!} \zeta^k,
\ee
is analytic at the origin. We will assume that it can be analytically continued to the full complex plane, and we would like to find the 
structure of its singularities. This is in general a difficult problem. One way of detecting these singularities is by looking at the discontinuities of the Borel resummation of $\varphi(\alpha)$. Let us define the Borel resummation of $\varphi(\alpha)$ as
\begin{equation}
  s(\varphi)(\alpha)=
  \int_0^\infty \widehat \varphi (x\alpha) \re^{-\zeta} \rd \zeta=
  \frac{1}{\alpha} \int_{\CC^\theta}
 \widehat \varphi (\zeta)\re^{-\zeta/\alpha} \rd \zeta
  \label{eq:brlsum}
\end{equation}
where $\CC^\theta=\re^{\ri\theta}\IR_+$ and $\theta = \arg \alpha$. This function is ill-defined if $\CC^\theta$ passes through a Stokes ray, joining the origin to a singularity of the Borel transform. One can make sense of the Borel resummation in this situation by deforming  $\CC^\theta$ slightly above (respectively, below) the Stokes ray, leading to contours $\CC_\pm^\theta$. We then define the {\it lateral Borel resummations} as
\be
\label{lbr}
 s_{\pm \theta} (\varphi)(\alpha)=
  \frac{1}{\alpha} \int_{\CC^\theta_\pm}
 \widehat \varphi (\zeta)\re^{-\zeta/\alpha} \rd \zeta. 
 \ee
The difference between the two lateral resummations gives precise information about the singularities of the Borel transform. 
Let us assume that along the Stokes ray forming an angle $\theta$ with the real axis there are singularities of the Borel transform at the locations $A_\ell$, $\ell=1, 2, \cdots$. Then, for a large class of 
perturbative series, the Stokes discontinuity, defined by
\be
\label{disc-def}
{\rm disc}_\theta s(\varphi)(\alpha)= s_{+ \theta} (\varphi)(\alpha)-s_{-\theta} (\varphi)(\alpha),
\ee
is given by 
\be
\label{st-def}
{\rm disc}_\theta s(\varphi)(\alpha)= s_{-\theta}\left(\Sigma\right) (\alpha), 
\ee
where $\Sigma (\alpha)$ is a trans-series, i.e. a formal linear combination of factorially divergent power series 
involving exponentially small terms (see \cite{ss,abs,mmlargen} for further details). It has the form, 
\be
\label{sal}
\Sigma (\alpha)=\ri \sum_{\ell=1}^\infty \mathsf{S}_{\ell} \, 
\alpha ^{-b_\ell} \re^{-A_\ell /\alpha} \psi_\ell(\alpha).   
\ee
In this equation, $\psi_\ell (\alpha)$ is a formal power series in $\alpha$, and $\mathsf{S}_\ell$ is called the Stokes constant associated to the 
singularity at $A_\ell$. Note that the value of this constant depends on a normalization of $\psi_\ell (\alpha)$. In this paper we will choose 
the normalization $\psi_\ell (\alpha)=1+ \CO(\alpha)$. We will also focus on singularities located on the positive real axis, so that $\theta=0$, and we will remove the subscript indicating the angle in (\ref{lbr}) and (\ref{st-def}). We finally note that the (lateral) Borel resummation of a trans-series, which we used in (\ref{st-def}), is simply obtained by replacing the formal series $\psi_\ell(\alpha)$ in the trans-series by their lateral Borel resummations. 

Let us note that the series $\psi_\ell(\alpha)$ and the Stokes constants $\mathsf{S}_{\ell}$ are in principle 
completely determined by the perturbative series $\varphi(\alpha)$, although it is not easy to obtain them explicitly. 
One way to obtain numerical information about them is to exploit 
their connection to the large order behavior of $\varphi(\alpha)$, since the discontinuity equation (\ref{st-def}) determines the behavior of the coefficients $e_k$ in (\ref{p-series}) at large $k$. Let $A_1$ be the closest singularity to the origin. Then, one has the following 
asymptotic formula, 
\be
\label{loek}
e_k \sim{\mathsf{S}_1\over 2 \pi}  A_1^{-k-b_1}\Gamma(k+ b_1) \left( \psi_{1,0} + \CO(k^{-1}) \right), \qquad k \gg 1, 
\ee
where we have written 
\be
\psi_{\ell}(\alpha)= \sum_{k \ge 0} \psi_{\ell,k} \alpha^k. 
\ee
The subleading singularities $A_\ell$, with $\ell >1$, give exponential 
corrections to this asymptotics which can be incorporated systematically (see e.g. \cite{abs}). 

The existence of a trans-series (\ref{sal}) giving the discontinuity is closely related to the existence of a trans-series expansion for the observable $\CG(\alpha)$. In this paper we will consider very general trans-series of the form 
\be
\label{phi-psi}
\Phi^\pm (\alpha)= \varphi(\alpha)+ \sum_{\ell=1}^\infty \mathsf{C}^\pm_\ell  \alpha^{-b_\ell} \re^{-A_\ell/\alpha}  \varphi^\pm_\ell(\alpha). 
\ee
Here, $\varphi^\pm_\ell(\alpha)$ are formal power series (normalised to $\varphi^\pm_\ell(\alpha) = 1+ \CO(\alpha)\,$), and 
$\mathsf{C}^\pm_\ell$ are complex constants (sometimes called trans-series parameters). Let us now 
assume that $\CG(\alpha)$ can be obtained in two different ways, by performing a lateral Borel resummation of $\Phi^\pm(\alpha)$ from above (respectively, below), i.e. 
\be
\CG(\alpha)= s_\pm \left( \Phi^\pm \right)(\alpha). 
\ee
The fact that the two lateral resummations of the trans-series $\Phi^\pm (\alpha)$ are equal gives an equation for the 
discontinuity (\ref{st-def}), and relates the trans-series (\ref{sal}) to $\Phi^\pm (\alpha)$. The results for the discontinuity obtained in this way can then be tested from the large order behavior formula (\ref{loek}). This is the strategy we will follow in this paper to obtain 
information about the Borel singularities. 

In \cite{dpmss} two different versions of the resurgence program were distinguished. 
According to the weak version, 
observables in QFT with an asymptotic expansion can be written as generalized Borel--\'Ecalle resummations of 
trans-series. According to the strong version, all ingredients of the trans-series can be extracted from the Borel 
singularities of the perturbative series\footnote{More precisely, this includes all the formal power series obtained by acting with all possible alien derivatives on the perturbative series.} (except the trans-series parameters, 
which have to be fixed by other means). In this paper we will also make some comments on which version 
of the program might apply to the cases at hand. A more precise diagnosis of this issue requires however a deeper analysis.
  
\subsection{Integrable field theories and the Bethe ansatz}

In this paper we will consider integrable, asymptotically free field theories in two dimensions. We will focus on three examples: the $O(N)$ 
GN model \cite{gross-neveu}, the $O(N)$ non-linear sigma model \cite{polyakov} and its supersymmetric version \cite{witten}, 
and the $SU(N)$ PCF.  Starting with the work of \cite{zz,zamo-zamo}, exact expressions for the 
$S$-matrix of these theories have been conjectured and passed many checks. These $S$-matrix expressions 
make possible the following exact computation \cite{pw,wiegmann2}. Let $\mH$ be the Hamiltonian of the 
model, and let $\mQ$ be a conserved charge, associated to a global conserved 
current. Let $h$ be an external field coupled to $\mQ$, which can be regarded as a chemical potential. 
As usual in statistical mechanics we can consider the ensemble defined by the operator
\be
\label{HQ}
\mH- h \mQ, 
\ee
as well as the corresponding free energy per unit volume 
\be
\label{free-en}
F(h) =-\lim_{V, \beta \rightarrow \infty} {1\over V\beta } \log {\rm Tr}\,  \re^{-\beta (\mH-h \mQ)}, 
\ee
where $V$ is the volume of space and $\beta$ is the total length of Euclidean time. As pointed out in \cite{pw}, we can compute 
\be
\label{cfh}
 \CF (h)= F(h)- F(0) 
\ee
by using the exact $S$ matrix and the Bethe ansatz.  One considers the following integral equation for a Fermi density $\epsilon(\theta)$ %
\be
\label{eps-ie}
\epsilon(\theta)-\int_{-B}^B  \rd \theta' \, K(\theta-\theta') \epsilon (\theta')=h-m \cosh(\theta),\quad \theta \in[-B,B]. 
\ee
In this equation, $m$ is the mass of the charged particles, and with a clever choice of $\mQ$, it is directly related to the mass gap of the theory. 
The kernel of the integral equation is given by 
\be
K(\theta)={1\over 2 \pi \ri} {\rd \over \rd\theta} \log S(\theta),
\ee
where $S(\theta)$ is the $S$-matrix appropriate for the scattering of the charged particles. 
The endpoints $\pm B$ are fixed by the condition 
\be 
\label{boundary}
\epsilon(\pm B)=0. 
\ee
The free energy is then given by
\be
\label{fh-eps}
\CF(h)=-{m \over 2 \pi} \int_{-B}^B \epsilon(\theta) \cosh(\theta) \rd \theta. 
\ee
It will also be convenient to use a ``canonical" formalism and introduce the density of particles $\rho$ and energy density $e$ through 
a Legendre 
transform of $\CF(h)$, 
\be
\ba
\rho&=-\CF'(h), \\
e(\rho)-\rho h&=\mathcal{F}(h).
\ea
\ee
The canonical observables can also be calculated directly from a Bethe ansatz integral equation for a rapidity density $\xi(\theta)$
\be
\label{eps-xi}
\chi(\theta)-\int_{-B}^B  \rd \theta' \, K(\theta-\theta') \chi (\theta')=m \cosh(\theta),\quad \theta \in[-B,B]. 
\ee
Then $\rho$ and $e$ relate to $B$ through
\begin{equation}
\rho = \frac{1}{2\pi}\int_{-B}^B \chi(\theta) \rd\theta,\quad e = \frac{m}{2\pi}\int_{-B}^B \chi(\theta)\cosh(\theta) \rd\theta.
\end{equation}
This formulation is sometimes more convenient. For example, the integral equation is easier to solve numerically.

Unfortunately, the solution of the integral equation (\ref{eps-ie}) is not known in closed form. 
One can solve it either numerically, or in a perturbative 
expansion for $B$ large. It turns out that $B$ large means $h$ large, which is the regime in which one can use 
conventional perturbation theory, due to asymptotic freedom. The evaluation of $\CF(h)$ at the very first orders in a 
large $B$ expansion was done for the non-linear sigma model in \cite{hmn,hn}, for the Gross--Neveu model in \cite{fnw1,fnw2}, and for the principal chiral field in \cite{pcf}. By comparing this result to a conventional perturbative calculation 
in the $\overline{\rm MS}$ scheme, it is possible to obtain an exact expression for the mass gap in terms of the 
dynamically generated scale $\Lambda_{\overline{\rm MS}}$ (which we will henceforth call simply $\Lambda$). 

More recently, Volin found an efficient method \cite{volin,volin-thesis} to obtain long perturbative series for $\CF(h)$ 
 at large $B$, starting from the canonical formalism. In \cite{mr-long,mr-ll, mr-ren,abbh1,abbh2} Volin's method was used to find trans-series representations for $\CF(h)$, including exponentially small corrections due to IR renormalons, in many integrable models. However, most of the results of this type have been numerical. Analytic results are available only in exceptional cases 
 (like the one-dimensional Hubbard model at half-filling \cite{mr-hubbard}) or by working in the $1/N$ expansion \cite{fkw1,fkw2, ksz,dpmss}. 

It turns out that, to understand analytically the trans-series structure of $\CF(h)$, it is convenient to use the 
Wiener--Hopf approach of \cite{jap-w,hmn,fnw1}, which relies on writing \eqref{eps-ie} in Fourier space.
The standard procedure is to first extend \eqref{eps-ie} to $\theta\in\mathbb{R}$. To do so, we extend $\epsilon(\theta)$ to the zero function outside $[-B,B]$ and we introduce
\begin{equation}
g(\theta) = 
\begin{cases}
      - \displaystyle\frac{m}{2}\re^\theta & \text{if } \theta < -B, \\[1.5mm]
     h - m \cosh \theta & \text{if } \theta\in[-B,B], \\[1.5mm]
     - \displaystyle\frac{m}{2}\re^{-\theta} & \text{if } \theta > B.  
\end{cases}
\label{gt_def}
\end{equation}
We also introduce an unknown function $Y(\theta)$, defined as the 0 function for $\theta<0$, and defined for $\theta>0$ such that
\begin{equation}
\epsilon(\theta)-\int_{-B}^B  \rd \theta' \, K(\theta-\theta') \epsilon (\theta')=g(\theta)+Y(\theta-B)+Y(-\theta-B)
\label{eps-line}
\end{equation}
is satisfied for all $\theta\in\IR$, where $\epsilon$ is the solution to the original problem \eqref{eps-ie}.\footnote{There is some freedom in the extension of $g(\theta)$ to the real line which amounts to redefinitions of the unknown function $Y(\theta)$. The choice \eqref{gt_def}, which is inspired by \cite{zamo-mass,sb-book}, minimizes irrelevant terms in intermediate calculations.}

We consider the Fourier transform of the kernel, 
\be
\tilde K(\omega)= \int_\IR \rd \theta \, \re^{\ri \omega \theta} K(\theta), 
\ee
and its Wiener--Hopf factorization
\be
\label{wh-fact}
1- \tilde K (\omega)= {1\over G_+(\omega) G_-(\omega)},
\ee
where $G_\pm (\omega)$ is analytic in the upper (respectively, lower) complex half plane. 
We will 
only consider the case in which $K(\theta)$ is an even function, therefore $G_-(\omega)= G_+(-\omega)$. 
We then introduce the 
Fourier transform of the function $g$:
\be
\tilde g(\omega) = \frac{2 h \sin (B \omega)}{\omega} + \frac{\ri m\re^B}{2}\left(\frac{\re^{ \ri B \omega}}{\omega-\ri}-\frac{\re^{- \ri B \omega}}{\omega+\ri}\right)
\ee
and define
\be
\label{gdef}
g_\pm (\omega)= \re^{\pm \ri B \omega} \tilde g(\omega).
\ee
Similarly, we define the function 
\be
\epsilon_\pm (\omega)= \re^{\pm \ri B \omega} \tilde \epsilon(\omega),
\ee
where $\tilde\epsilon(\omega)$ is the Fourier transform of $\epsilon(\theta)$.
Lastly we introduce the convenient definitions
\begin{align}
\label{sig-ome}
\sigma(\omega) &= {G_-(\omega) \over G_+(\omega)},\\
Q(\omega) &= G_+(\omega) \tilde Y(\omega),
\end{align}
where $\tilde Y(\omega)$ is the Fourier transform of $Y(\theta)$.
The %original integral equation can 
Fourier transform of \eqref{eps-line} can
then be written as 
\be
\label{ft-ie}
{1\over G_+(\omega) G_-(\omega)} \tilde \epsilon (\omega) = \tilde g(\omega)+ \re^{\ri B \omega} G_+^{-1}(\omega) Q (\omega) + 
 \re^{-\ri B \omega} G_-^{-1}(\omega) Q (-\omega). 
 \ee
It is shown in \cite{hmn,fnw1} that $Q(\omega)$ satisfies the integral equation
\be
\label{Q-eq}
Q(\omega)-{1\over 2 \pi \ri} \int_\IR {\re^{2 \ri B\omega'} \sigma (\omega') Q(\omega') \over \omega+ \omega'+ \ri 0} \rd \omega'= {1\over 2 \pi \ri} \int_\IR {G_-(\omega') g_+(\omega') \over \omega+ \omega'+ \ri 0} \rd \omega'. 
\ee
The solution $Q(\omega)$ determines $\epsilon_+(\omega)$ through the following equation 
\be
\label{epsom}
{\epsilon_+(\omega) \over G_+(\omega)} = {1\over 2 \pi \ri} \int_\IR {G_-(\omega') g_+(\omega') \over  \omega'-\omega- \ri 0}\rd \omega'+ 
{1\over 2 \pi \ri} \int_\IR {\re^{2 \ri B\omega'} \sigma (\omega')Q(\omega') \over\omega'- \omega- \ri 0} \rd \omega'. 
\ee
Equations \eqref{Q-eq} and \eqref{epsom} are, respectively, the 
%positive and, respectively, negative Wiener--Hopf projections of equation \eqref{ft-ie}. Schematically, they correspond to the part analytic in the upper and, respectively, lower complex half plane. 
projections of \eqref{ft-ie} into analytic functions in the upper and lower half plane.
They are obtained with the Wiener--Hopf formalism, see appendix A of \cite{fnw1} for a detailed derivation.

The relationship between $h,m$ and $B$ is determined by the boundary condition \eqref{boundary}, which in Fourier space takes the form
\be
\label{bc-eps}
\lim_{\kappa \to +\infty} \kappa \epsilon_+(\ri \kappa)=0. 
\ee
The free energy is then given by 
\be
\label{fh-eps+}
\CF(h)= -{1\over 2 \pi} m \re^B \epsilon_+(\ri). 
\ee

The above formalism is general and can be applied to all integral equations appearing in the different integrable models. 
We will revisit it in some detail in the section on the bosonic models. However, as pointed out in \cite{fnw1,zamo-mass}, when $G_+(0)$ is finite 
and non-vanishing, there is an alternative, simpler formulation (see \cite{sb-book} for a nice presentation). This 
happens in the sine--Gordon model analyzed in \cite{zamo-mass}, and in the Gross--Neveu model \cite{fnw1}. In this case, 
one obtains an integral equation for an auxiliary function $u(\omega)$ defined in Fourier space. This equation has the form
\be
\label{uom-int}
u(\omega) = \frac{\ri}{\omega} + \frac{1}{2\pi\ri} \int_{\IR}  \frac{\re^{2\ri B\omega'} \rho(\omega')u(\omega')}{\omega+\omega'+ \ri 0}\rd\omega', 
\ee
where
\be
\rho(\omega) = - \frac{\omega+\ri}{\omega-\ri} \frac{G_-(\omega)}{G_+(\omega)}.
\label{eq_rho_def}
\end{equation}
%
%and $\CC$ is a contour, as shown in the figure, which encircles 
The boundary condition (\ref{bc-eps}) fixes the value of $u(\ri)$ as
\begin{equation}
\label{uh}
u(\ri) = \frac{m \re^B }{2h} \frac{G_+(\ri)}{G_+(0)}, 
\end{equation}
and this can be used to determine the relationship between $h,m$ and $B$. Finally, once $u(\omega)$ is known, one can find 
the free energy from the equation 
\be
\label{fh-rho}
\CF(h)= -{h^2 \over 2 \pi} u(\ri) G_+(0)^2 \left\{1-  \frac{1}{2 \pi \ri} \int_{\IR} {\re^{2 \ri B \omega'} \rho(\omega') u(\omega')  \over \omega'-\ri} 
\rd \omega' \right\}.
\ee

In this paper we will use these Wiener--Hopf integral equations to obtain information about the trans-series structure of $\CF(h)$ and the corresponding Borel singularities.

\sectiono{Trans-series and renormalons in the Gross--Neveu model}
\label{GN-sec}
\subsection{Analytic solution}
In the $O(N)$ Gross--Neveu model, the basic field is an $N$-uple of Majorana fermions $\boldsymbol{\chi}$. 
The Lagrangian density describing the theory is 
\be
\CL= {\ri \over 2} \overline{\boldsymbol{\chi}} \cdot \slashed{\partial} \boldsymbol{\chi}+ {g^2\over 8} \left(\overline{\boldsymbol{\chi}} \cdot \boldsymbol{\chi}  \right)^2. 
\ee
Our convention for the beta function is 
\be
\label{betaf}
\beta(g)= \mu {\rd g \over \rd \mu} =-\beta_0 g^3 - \beta_1 g^5- \cdots, 
\ee
This model is asymptotically free, and the first two coefficients of its beta function are (see e.g. \cite{gracey})
\be
\beta_0= {1 \over 4 \pi \Delta}, \qquad \beta_1=- {1\over 8 \pi^2 \Delta}, 
\ee
where 
\be
\Delta={1\over N-2}. 
\ee
We consider the setting of \cite{fnw1,fnw2}, where the charge in (\ref{HQ}) is the quantum 
version of $Q^{12}$, associated to the global $O(N)$ symmetry. We restrict ourselves to $N>4$, 
since the cases $N\le 4$ are somewhat special, see \cite{fnw1,dpmss}. The relevant kernel can be 
found in \cite{fnw1}, and its Wiener--Hopf decomposition is determined by
\begin{equation}
G_+(\omega)=\frac{\re^{-\frac{1}{2}\ri\Upsilon\omega[1-\log(-\frac{1}{2}\ri\Upsilon\omega)]}}{\re^{-\frac{1}{2}\ri\omega[1-\log(-\frac{1}{2}\ri\omega)]}} \frac{\Gamma\big(\frac{1}{2} - \frac{1}{2}\ri\Upsilon\omega\big)}{\Gamma\big(\frac{1}{2} - \frac{1}{2}\ri\omega\big)}, 
\end{equation}
where
\be
\Upsilon=1-2 \Delta. 
\ee
Since $G_+(0)=1$ is finite and nonvanishing, we can obtain the free energy from the integral equation (\ref{uom-int}). In this equation, 
the key object is the function $\rho(\omega)$ introduced in (\ref{eq_rho_def}). In this case it is given by
\begin{equation}
\label{rhoGN}
\rho(\omega) = \frac{\re^{\frac{1}{2}\ri\Upsilon\omega\left[ 2 - \log\left( - \frac{1}{2}\ri\Upsilon\omega \right) - \log\left(\frac{1}{2}\ri\Upsilon\omega \right)\right]}}{\re^{\frac{1}{2}\ri\omega\left[ 2 - \log\left( - \frac{1}{2}\ri\omega \right) - \log\left(\frac{1}{2}\ri\omega \right)\right]}}  \frac{\Gamma\big(\frac{3}{2}-\frac{1}{2}\ri\omega\big)\Gamma\big(\frac{1}{2}+\frac{1}{2}\ri\Upsilon\omega\big)}{\Gamma\big(\frac{3}{2}+\frac{1}{2}\ri\omega\big)\Gamma\big(\frac{1}{2}-\frac{1}{2}\ri\Upsilon\omega\big)}. 
\end{equation}
The analytic structure of $\rho(\omega)$ is the same as $\sigma(\omega)$, defined in (\ref{sig-ome}). 
Due to the Gamma functions, it has simple poles along the imaginary axis. 
In the complex upper half plane the poles occur at $\omega_n= \ri \xi_n$, with
\be
\xi_n= {2n+1 \over \Upsilon}, \qquad n \in \IZ_{\ge 0}. 
\ee
As we will see, these poles will eventually lead to renormalon singularities. 
At the same time, the logarithms in (\ref{rhoGN}) lead to two branch cuts starting at $\omega = 0$, going respectively upwards and downwards along the imaginary axis. In the upper half plane, the discontinuity is given by
\begin{equation}
\delta \rho(\ri \xi) = - 2\ri \re^{[2\Delta (1+\log 2) +\Upsilon \log\Upsilon]\xi-2\Delta \xi \log \xi} \sin(\pi \Delta \xi)  \frac{\Gamma\big(\frac{1}{2}-\frac{1}{2}\Upsilon\xi\big)\Gamma\big(\frac{3}{2}+\frac{1}{2}\xi\big)}{\Gamma\big(\frac{1}{2}+\frac{1}{2}\Upsilon\xi\big)\Gamma\big(\frac{3}{2}-\frac{1}{2}\xi\big)}.
\label{rhodisc}
\end{equation}
This expression is obtained with the convention $\delta \rho(\omega) = \rho(\omega(1-\ri 0) ) - \rho(\omega(1+\ri 0) )$, which we will use consistently in this paper.
 \begin{figure}[!ht]
\leavevmode
\begin{center}
\includegraphics[height=5.8cm]{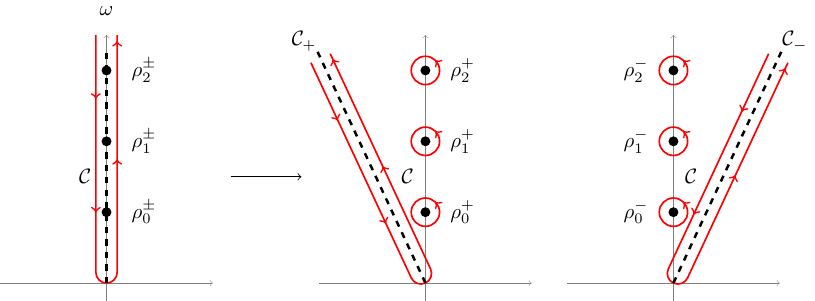}
\end{center}
\caption{The Hankel contour $\CC$ can be deformed into an integral along the discontinuity of $\rho(\omega)$, denoted by the dashed line, plus a sum over residues. However, due to 
the branch cut along the imaginary axis, this can be done in two different ways, which leads to two different integrations along the discontinuity, corresponding to the contours $\CC_\pm$. The residues of the poles $\rho_n^\pm$ will also depend on this choice, as shown explicitly in (\ref{polepm}).}
\label{contourdef-fig}
\end{figure}

We can now deform the integration contour appearing in (\ref{uom-int}) into a Hankel contour $\CC$ around the positive imaginary axis. This contour is made of two rays, one of them to the left of the imaginary axis, 
and the other one to the right. If $\rho(\omega)$ had only poles, the contour integral 
could simply be evaluated by residues. This is exactly what happens when one does this calculation in 
the sine--Gordon model \cite{zamo-mass}. 
However, since there is also a branch cut along the imaginary axis we have to be careful. A convenient way to proceed is to move the branch cut away from the imaginary axis by a small angle $\delta$. Then, as seen in \figref{contourdef-fig}, the discontinuity and the poles become disentangled, and the integral along the path $\CC$ can be separated into an integral along the discontinuity with angle $\delta$,  and a sum over the residues. The resulting tilted paths corresponding to $\delta> 0$ (respectively, $\delta<0$) will be denoted by 
$\CC_\pm$. In the variable $\xi=-\ri\omega$, $\CC_\pm$ correspond simply to the integrals over $\re^{\ri\delta}\IR_+$, with the respective sign of $\delta$, in harmony with the notation introduced in section \ref{sec-resQFT}. The crucial point is that the value of the residues is sensitive to the sign of $\delta$, that is, to the branch choice of $\rho(\omega)$. Explicitly, the residues are given by
\begin{equation}
\begin{aligned}
\rho^\pm_n &= \textrm{Res}_{\xi=\xi_n\mp\ri 0} \, \rho(\ri\xi)\\
&= \re^{\mp \ri \pi \Delta\frac{2n+1}{\Upsilon}}  \frac{2}{\Upsilon} \frac{(-1)^{n+1}}{(n!)^2} \left(\frac{2n+1}{2\re}\right)^{2n+1} \left( \frac{2n+1}{2\Upsilon\re} \right)^{-\frac{2n+1}{\Upsilon}} \frac{\Gamma\big(\frac{3}{2} + \frac{2n+1}{2\Upsilon}\big)}{\Gamma\big(\frac{3}{2} - \frac{2n+1}{2\Upsilon}\big)},
\end{aligned}
\label{polepm}
\end{equation}
where the plus (minus) sign in $\rho^\pm_n$ has to be paired with the branch choice $\delta >0$ ($\delta < 0$).
As we will see, this ambiguity in the residues will lead to
the renormalon ambiguity discovered by F.~David in \cite{david2}.

From the construction above we obtain the following expression for the function $u(\ri \xi)$:
\be
u(\ri\xi) = \frac{1}{\xi} + \frac{1}{2\pi\ri} \int_{\CC_\pm} \frac{\re^{-2B\xi'}\delta\rho(\ri\xi')u(\ri\xi') }{\xi+\xi'} \dd\xi'  + \sum_{n\ge 0} \frac{\re^{-2B\xi_n} \rho_n^\pm u_n}{\xi+\xi_n}.
\label{eq_u_im}
\ee
A similar argument can be applied in the calculation of the free energy (\ref{fh-rho}), and we obtain 
\be
\mathcal{F}(h) = \begin{multlined}[t][0.8\textwidth]
-\frac{h^2}{2\pi} u(\ri) G_+(0)^2 \biggl\{  1 - \frac{1}{2\pi \ri} \int_{\CC_\pm}  \frac{\re^{-2B\xi'}\delta \rho(\ri\xi')u(\ri\xi')}{\xi'-1} \dd \xi'\\
- \re^{-2B}\rho(\ri\pm 0)u(\ri) - \sum_{n\ge 0} \frac{\re^{-2B\xi_n}\rho_n^{\pm} u_n}{\xi_n-1}\biggr\},
\label{eq_freeF}
\end{multlined}
\ee
where $u_n = u(\ri\xi_n)$. In the second line, the ambiguity due to the branch cut also applies to the value of $\rho(\omega)$ at $\omega=\ri$. After using the boundary condition (\ref{uh}), we can write this term as
\be
\re^{-2B}\rho(\ri\pm 0)u(\ri)
% = \frac{m\re^{-B}}{2h} \sqrt{2}\, \re^{\mp\ri\pi\Delta+\Delta+\frac{1}{2}\Upsilon\log\big(\frac{\Upsilon}{2}\big)}\Gamma\bigg( \frac{1}{2} - \frac{\Upsilon}{2} \bigg)
=  \frac{m\re^{-B}}{2h}\tilde{\rho}^\pm,
\ee
where
\be
\label{tilderho}
\tilde{\rho}^\pm = \re^{\mp \ri \pi  \Delta  }(2\re)^{\Delta } (1-2 \Delta )^{\frac{1}{2}-\Delta } \Gamma (\Delta ) .
 \ee

In the following, we explicitly check the ambiguity cancellation between the integral in \eqref{eq_freeF} and the exponential terms. 
It is convenient to first compute the difference between the two directions of integration, which can be written as a contour encircling the poles of $\delta \rho(\ri\xi')$ and the explicit pole at $\xi' = 1$: 
\be
 \label{eq_direction_ambiguity}
{1\over 2 \pi \ri} \left( \int_{\CC_-}- \int_{\CC_+} \right)\frac{\re^{-2B\xi'}\delta \rho(\ri\xi')u(\ri\xi')}{\xi'-1} \dd\xi' = \re^{-2 B} \delta \rho(\ri) u (\ri) + \sum_{n \ge 0}  \frac{\re^{-2B\xi_n}}{\xi_n-1} u_n {\rm Res}_{\xi=\xi_n} \delta \rho(\ri\xi) .
\ee
In particular, we note that $\delta\rho(\ri\xi)$ has no branch along the positive real line and thus, the integral only picks the residues of the function, which are related to the residues of $\rho(\ri\xi)$ by
\begin{equation}
\Res_{\xi=\xi_n} \delta\rho(\ri\xi) = \rho_n^+ \re^{\ri\pi\Delta\frac{2n+1}{\Upsilon}} (-2\ri) \sin\left(\pi\Delta\frac{2n+1}{\Upsilon} \right).
\label{eq_disc_rho_residues}
\end{equation}
Accordingly, the difference between the two choices of residues in \eqref{eq_freeF} yields
\begin{equation}
\re^{-2B} [\rho(\ri-0) + \rho(\ri+0)] u(\ri) + \sum_{n\ge 0} \frac{\re^{-2B\xi_n}}{\xi_n-1}(\rho_n^- - \rho_n^+) u_n.
\end{equation}
We have $\rho(\ri-0) - \rho(\ri+0) = -\delta\rho(\ri)$, which cancels the first term in the r.h.s of \eqref{eq_direction_ambiguity}. In addition, from the expression for $\rho_n^\pm$ in \eqref{polepm}, we find
\begin{equation}
\rho_n^- - \rho_n^+ =  \rho_n^+ \re^{\ri\pi\Delta\frac{2n+1}{\Upsilon}} (2\ri)  \sin\left(\pi \Delta\frac{2n+1}{\Upsilon} \right).
\end{equation}
It is now clear that the contribution from $\rho_n^- - \rho_n^+$ cancels the sum in \eqref{eq_direction_ambiguity}. This completes the check that \eqref{eq_freeF} is unambiguous and, in particular, imaginary ambiguities in the exponential corrections arising from the residues cancel exactly with the imaginary ambiguity arising from the integral. A similar argument can also be applied to $u(\ri\xi)$ and its integral equation \eqref{eq_u_im}.

The cancellation mechanism that we have just analyzed is reminiscent of the ambiguity cancellation between perturbative and non-perturbative sectors typical of the theory of the resurgence. At the same time, there are obvious differences between the two. 
For example, the integral in \eqref{eq_freeF} looks like a Borel resummation in the variable $B$, but one has to be reminded that the factor $u(\ri\xi')$ inside the integral depends also on $B$ and, in particular, comes with its own exponential corrections. However, we conjecture that both mechanisms are closely related. Indeed, the asymptotic expansion of the integrals above will lead to formal power series which can be resummed in two different ways, by lateral Borel resummation. We will assume that these two choices are correlated to the two choices of branch cuts in the formulae above, and in particular to the two choices for the residues  $\rho_n^\pm$, $\tilde \rho^\pm$. 

In \cite{hmn,hn,fnw1}, the integrals appearing in the Wiener--Hopf method \eqref{eq_u_im}--\eqref{eq_freeF} were also calculated
by deforming the contour and picking the discontinuity of the integrand. This is enough to obtain perturbative 
expansions, and in those papers the contribution of the poles was neglected. We will now keep these contributions, which are exponentially 
small for large $B$, but at the same time we will expand the remaining quantities in power series in $1/B$. The result will 
have the structure of a trans-series, with small parameters $\re^{-B}$, $1/B$ and $\log B/B$.

As noted in \cite{fnw1}, it is useful to change variables from $\xi, B$ to $\eta, v$, as follows:
\begin{equation}
\label{veta}
\frac{1}{v} - 2 \Delta \log v = 2 B,\qquad\xi = v \eta.
\end{equation}
This change of variables combines $1/B$ and $\log B/B$ terms into $v$ terms with no logarithms.
We will write $u(\eta)$ for the function obtained from $u(\ri \xi)$ after the change of variables, and we introduce the function $P(\eta)$ through
\begin{equation}
\label{P-def}
\re^{-2 B \xi}\delta \rho(\ri \xi)  = - 2 \ri \,v \re^{-\eta} P(\eta).
\end{equation}
 The integral equation reads now 
\be
\label{int-eq-2}
u(\eta)= {1\over v \eta}-{v \over \pi}  \int_{\CC_\pm} \frac{\re^{- \eta' } P(\eta') u(\eta')}{\eta+\eta'}  \rd \eta' + \Upsilon \sum_{n \ge 0} \frac{q^{2n+1}  \rho^\pm _n u_n}{\Upsilon v\eta + 2n+1},
\ee
where $q$ is defined by  
\be
q= \exp\left(-\frac{1}{\Upsilon v}\right) v^{\frac{2 \Delta }{\Upsilon}}. 
\ee
This is the exponentially small variable in the trans-series expansion. 
The equation (\ref{int-eq-2}) is solved by iteration, as follows. The ``seed" of the integral equation is 
\be
\mathfrak{u}(\eta) =  {1\over v \eta}+ \Upsilon \sum_{n \ge 0} \frac{ q^{2n+1} \rho^\pm _n u_n}{ \Upsilon v\eta + 2n+1}. 
\ee
Let us introduce the integral operator 
\be
\label{D-op}
\left({\cal D} f \right) (\eta)= -{v \over \pi}  \int_{\CC_\pm}\frac{\re^{- \eta' } P(\eta') f(\eta')}{\eta+\eta'} \rd \eta' , 
\ee
as well as  
\be
{\cal T}= \sum_{\ell=0}^\infty {\cal D}^\ell. 
\ee
Then, 
\be
u(\eta)= \left({\cal T} \mathfrak{u} \right)(\eta). 
\ee
We will now perform a systematic expansion in powers of $q$. First, we note that the unknowns $u_k$ will have $q$-series expansions of the form 
\be
u_k =\sum_{s\ge 0}  u_k ^{(s)} q^s. 
\ee
They satisfy the equation 
\be
{1\over \Upsilon} u_k= {1 \over 2k+1} + v {\cal D}_k u  +  {1 \over 2}  \sum_{n \ge 0} \frac{ q^{2n+1} \rho^\pm _n u_n}{ 1+ n+k },
\ee
where
\be
{\cal D}_k u = -{v  \over \pi} \int_{\CC_\pm} {  \re^{-\eta} P(\eta) u (\eta) \over 1+ 2k+ \Upsilon v \eta} \rd \eta, 
\ee
and does not depend on $\eta$. We will also write the ``seed" of the integral equation as a $q$-series:
\be
\mathfrak{u} (\eta)= \sum_{s\ge 0}  \mathfrak{u}^{(s)}(\eta) q^s , 
\ee
where
\be
\mathfrak{u}^{(0)}(\eta)={1\over v \eta}, \qquad \mathfrak{u}^{(s)}(\eta) = \Upsilon \sum_{\ell=0}^{s-1} {\rho^\pm_{s-1-\ell \over 2} u^{(\ell)}_{s-1-\ell \over 2} \over 
s-\ell+ \Upsilon v \eta}.
\ee
In the sum over $\ell$, it is understood that only values such that $s-1-\ell$ is even occur. For example, we have
\be
\mathfrak{u}^{(1)} (\eta)= \Upsilon {\rho^\pm_0 u_0^{(0)} \over 1+ \Upsilon v \eta}.
\ee
This leads to a decomposition of the 
full solution, 
\be
u(\eta)= \sum_{s\ge 0} \mathfrak{u}^{(s)}(\eta) q^s , 
\ee
where
\be
u^{(s)}(\eta)= \left( {\cal T} \mathfrak{u}^{(s)}\right)(\eta). 
\ee
We can now plug in this decomposition into the equation for the residues, and we find
\be
{1\over \Upsilon} u_k^{(0)}={1\over 2k+1} + v {\cal D}_k  \CT {1\over v\eta}, 
\ee
while for $r \ge 1$ we have
\be
{1\over \Upsilon} u_k^{(r)}= v {\cal D}_k  \CT \mathfrak{u}^{(r)}(\eta) + \sum_{\ell=0}^{r-1} {\rho^\pm_{r-1-\ell \over 2} u^{(\ell)}_{r-1-\ell \over 2} \over 1+ 2k+ r-\ell}. 
\ee
This gives a recursive equation to solve for the $u_k^{(r)}$. For example, we obtain 
\be
{1\over \Upsilon} u_0^{(1)}= {1\over 2} \rho^\pm_0 u_0^{(0)} \left( 1+ 2 v \Upsilon {\cal D}_0 \CT {1 \over 1+ \Upsilon v \eta}  \right). 
\ee

So far we have taken into account the expansion in $q$, leading to exponentially small corrections, but we also 
want to perform a conventional weak coupling expansion. 
To do this, we expand the discontinuity function $P(\eta)$ appearing in the integral operators in power series, as
\be
\label{P-exp}
P(\eta)\sim \sum_{n=1}^\infty  v^{n-1}  \sum_{m=0}^{n-1} d_{n,m} 
%\re^{-\eta} 
 (\log\eta)^m \eta^n. 
\ee
The coefficients $d_{n,m}$ are explicitly computable. In this way, we obtain a systematic 
expansion in both $v$ and $q$ with the 
structure of a trans-series. Note that the iteration of the operator ${\cal D}$ defined in (\ref{D-op}) 
will involve multiple integrals with the kernel $1/(\eta + \eta')$. These integrals are easy to calculate up to two iterations, 
but beyond that they are not straightforward. This already happens in the purely perturbative sector, and that's one of the 
reasons why the method of \cite{volin} is more powerful. For this reason, 
in this paper we will not obtain long perturbative series attached to the exponentially 
small corrections, but only the very leading terms. Once the double expansion 
of $u(\ri \xi)$ in $v$ and $q$ has been worked out, we can plug it in (\ref{eq_freeF}) 
to obtain the corresponding equation for the free energy. Finally, the boundary condition 
(\ref{uh}) gives a trans-series expression for $B$ as a function of $\log(m/h)$ and $m/h$. 

Let us present some explicit results that are obtained with this procedure. For the very first values of $k=0,1$ one finds, 
\be
\ba
u_0&= \Upsilon -\frac{ d_{1,0} \Upsilon }{\pi }v +\CO\left(v^2\right)
+q \left( \frac{\Upsilon ^2 \rho^\pm _0}{2}-\frac{ d_{1,0}\Upsilon ^2 \rho^\pm_0}{2 \pi }v+\CO(v^2) \right)+ \CO(q^2), \\
u_1&=\frac{\Upsilon }{3}-\frac{ d_{1,0} \Upsilon }{3 \pi }v +\CO\left(v^2\right)
+q \left(\frac{\Upsilon ^2 \rho^\pm_0}{4}-\frac{ d_{1,0} \Upsilon ^2 \rho^\pm_0}{4 \pi } v +\CO\left(v^2\right)\right)+ \CO(q^2), 
\ea
\ee
where the coefficients $d_{n,m}$ are defined in (\ref{P-exp}). The free energy reads 
\begin{multline}
\CF(h)= {h^2 \over 2 \pi} u(\ri) G_+(0)^2
\biggl\{-1+\frac{m\re^{-B}}{2h}\tilde{\rho}^\pm +\frac{d_{1,0}}{\pi }v +\CO\left(v^2\right)
\\
+q \left(\frac{\Upsilon^2  \rho^\pm _0}{1-\Upsilon}-\frac{d_{1,0}\Upsilon ^2 \rho^\pm_0 }{\pi  (1-\Upsilon )}v +\CO\left(v^2\right)\right)+ \CO(q^2)\biggr\}.
\end{multline}
Finally, one needs to calculate $u(\ri)$ to implement the boundary condition. One can also calculate this 
as a trans-series expansion, and at the very first orders we obtain 
\be
u(\ri)=1-\frac{d_{1,0}}{\pi }v +\CO\left(v^2\right)
+q \left(\frac{ \Upsilon ^2 \rho^\pm_0}{\Upsilon +1}-\frac{ d_{1,0} \Upsilon ^2 \rho^\pm_0 }{\pi  (\Upsilon +1) } v +\CO\left(v^2\right)\right) + \CO(q^2). 
\ee

In order to make contact with the perturbative expansion obtained in \cite{mr-ren}, we have to use 
the appropriate coupling constant, i.e. appropriate schemes. There are two useful schemes that have 
been proposed in this context \cite{bbbkp,volin,mr-ren}. In the first scheme, one introduces a coupling constant 
$\tilde \alpha$ satisfying 
\be
\label{alphatilde}
\frac{1}{\tilde{\alpha}}-\Delta \log\tilde \alpha = \log \left( \frac{h}{\Lambda} \right),
\ee
where $\Lambda$ is the dynamically generated scale in the ${\overline{\text{MS}}}$ scheme, and it is related to the mass gap by \cite{fnw1}
\be
 \frac{m}{\Lambda} = \frac{(2\re)^\Delta}{\Gamma(1-\Delta)}.
 \ee
Comparing (\ref{alphatilde}) to the renormalization group equation it can be easily seen that 
\be
\label{talg2}
\tilde \alpha = 2 |\beta_0| \overline g^2 (h)+ \CO\big(\overline g^4(h)\big), 
\ee
where $\overline g^2 (h)$ is the running coupling constant at the scale $h$ in the $\overline{\text{MS}}$ scheme. 
We now need a dictionary relating $v$ to $\tilde \alpha$. This follows from the boundary condition (\ref{uh}), and it will involve non-perturbative corrections. 
%Let us first introduce the counterpart of $q$ when written in terms of $\tilde \alpha$:  
%%
%\begin{equation}
%\kappa = \re^{-\frac{2}{\Upsilon \tilde{\alpha}}}\left(\frac{\tilde{\alpha}}{2}\right)^{\frac{2\Delta}{\Upsilon}}.
%\end{equation}
%%
One finds 
\begin{multline}
v = \frac{\tilde\alpha }{2}+\frac{1}{4}  \big[ (\Delta-1) \log (4)-\Upsilon  \log (\Upsilon )\big]\tilde\alpha ^2 + \CO\left(\tilde\alpha ^3\right)\\
+\re^{-\frac{2}{\Upsilon \tilde{\alpha}}}\left(\frac{\tilde{\alpha}}{2}\right)^{\frac{2\Delta}{\Upsilon}}  \left(-\frac{2^{-\frac{1}{\Upsilon }-2} \Upsilon \rho^\pm_0 }{\Upsilon +1}\tilde\alpha ^2 + \CO\left(\tilde\alpha ^3\right)\right)+ \CO\left(\re^{-\frac{4}{\Upsilon \tilde{\alpha}}}\right).
\end{multline}
Putting all these ingredients together, one finally obtains the trans-series expansion of the free energy in terms of $\tilde \alpha$:
\begin{equation}
\ba
\CF(h)\sim  
- \frac{h^2}{2\pi}\Biggl\{&1-\Delta  \tilde{\alpha }+\frac{1}{2} \Delta  \big[\Delta -2+2 \log (2)\big]\tilde{\alpha }^2 + \CO\left(\tilde{\alpha }^3\right)
%- \re^{-\frac{2}{\tilde{\alpha}}}\tilde{\alpha}^{2\Delta} \frac{(2\re)^\Delta (1-2\Delta) ^{-\frac{1 }{2}+\Delta} \tilde{\rho}^\pm}{4\Gamma(1-\Delta)}
\\
&+\re^{-\frac{2}{\Upsilon \tilde{\alpha}}}\left(\frac{\tilde{\alpha}}{2}\right)^{\frac{2\Delta}{\Upsilon}}\left(
%\frac{\tilde{\alpha}}{2}\right)^{\frac{2\Delta}{(1-2\Delta)}}  \left(\frac{2^{\frac{1}{2 \Delta -1}-2} (2 \Delta -1) (3 \Delta -1) \rho_0^\pm}{(\Delta -1) \Delta }-\frac{2^{\frac{1}{2 \Delta -1}-1} \Delta  (3 \Delta -1) \rho_0^\pm \tilde{\alpha }}{\Delta -1}
\frac{2^{\frac{1}{2 \Delta -1}-2} (1-2 \Delta )^2 \rho _0^\pm}{\Delta (\Delta -1)}-\frac{ 2^{\frac{1}{2 \Delta -1}-1} \Delta  (2 \Delta -1) \rho _0^\pm}{\Delta -1}\tilde{\alpha}
+\CO\left(\tilde{\alpha }^2\right)\right)
\\
&+\re^{-\frac{4}{\Upsilon \tilde{\alpha}}}\left(\frac{\tilde{\alpha}}{2}\right)^{\frac{4\Delta}{\Upsilon}}\left(
%\frac{\tilde{\alpha}}{2}\right)^{\frac{4\Delta}{(1-2\Delta)}}  \left(\frac{2^{\frac{2}{2 \Delta -1}-4} (2 \Delta -1) \left(7 \Delta ^2+2 \Delta -1\right) \rho _0^2}{(\Delta -1)^2 \Delta }
\frac{2^{\frac{2}{2 \Delta -1}-3} (1-2 \Delta )^2 \left( \rho^\pm_0\right)^2}{(\Delta -1)^2}
+\CO\left(\tilde{\alpha}\right)\right)
+\CO\left(\re^{-\frac{6}{\Upsilon \tilde{\alpha}}}\right)\Biggr\}\\
& \mp \frac{\ri m^2}{8}+\frac{m^2}{8} \cot (\pi  \Delta )
.
\ea
\label{fhWHZ}
\end{equation}
The last line is an $h$-independent term in the free energy. Its imaginary part leads to an IR renormalon pole. Following e.g. \cite{zamo-mass}, 
its real part can be identified with $-F(0)$, i.e. 
\begin{equation}
\label{f0GN}
F(0)=-\frac{m^2}{8}  \cot (\pi  \Delta ).
\end{equation}
It will also be useful to give the result for the normalized energy density $e/\rho^2$, since this is the observable 
studied in \cite{bbbkp,volin,mr-ren}. This requires using yet another scheme, and we introduce the coupling constant $\alpha$ as
\be
\label{alpha-GN}
\frac{1}{\alpha}-\Delta  \log\alpha=\log\left(\frac{2\pi\rho}{\Lambda}\right). 
\end{equation}
We note that 
\be
\alpha = \tilde \alpha+ \CO(\tilde \alpha^2).  
\ee
In terms of this coupling constant, we find
\be
\ba
\frac{e}{2\pi\rho^2} \sim \frac{1}{4} & +\frac{\Delta}{4}\alpha +\frac{1}{8} \Delta  (\Delta +2) \alpha ^2+\CO\left(\alpha ^3\right)
 +\re^{-\frac{2}{\alpha}}\alpha^{2\Delta} 
 %-
 \frac{2^{\Delta -2} \re^{\Delta } (1-2 \Delta )^{\Delta -\frac{1}{2}} \tilde{\rho}^\pm}{\Gamma (1-\Delta )}\\
&+\re^{-\frac{2}{\Upsilon\alpha}}\alpha^\frac{2\Delta}{\Upsilon}  \left(-\frac{(1-2 \Delta )^2 \rho^\pm_0}{8 \Delta (\Delta -1)}
+\frac{\Delta (1-2 \Delta ) \rho^\pm_0  }{4 (\Delta -1)}\alpha+\CO\left(\alpha ^2\right)\right)
\\&+ \re^{-\frac{4}{\Upsilon\alpha}}\alpha^\frac{4\Delta}{\Upsilon} \left(\frac{(1-2 \Delta )^2  (\rho^\pm_0)^2}{8 (\Delta -1)^2}+\CO(\alpha)\right)
+\CO\left(\re^{-\frac{6}{\Upsilon\alpha}}\right).
\label{efromWHZ}
\ea
\ee
This is our final result for the normalized energy density, which is given in terms of a trans-series in $\alpha$. There are 
various observations that we would like to make on this result. 

First of all, note that the first few terms in the r.h.s. of the first line give the perturbative expansion of this observable,  
which was computed in \cite{mr-ren} to much higher order. Then, 
we have two types of exponentially small corrections. The last term in the r.h.s. of the first line, which is proportional to 
\be
\re^{-{1 \over |\beta_0| \overline g^2(h)}}
\ee
corresponds to an IR renormalon singularity at the expected location (\ref{slore}) with $\ell=2$. However, there is an infinite series of corrections with exponentials of the form
\be
\re^{-{\ell \over \Upsilon |\beta_0|  \overline g^2(h)}}, \qquad \ell \in \IZ_{>0}. 
\ee
The first two corrections of this type are displayed in (\ref{efromWHZ}). 
They lead to the new IR renormalon singularities located at (\ref{corrected}). 

Our second observation is that 
the exponentially small contributions are inherently ambiguous, as indicated by the $\pm$ signs in the residues $\tilde \rho^\pm$, 
$\rho^\pm_n$. This ambiguity is ultimately due to the logarithmic %and square root 
branch cut in the 
discontinuous function (\ref{rhoGN}). 
We know since the work of David \cite{david2,david3} that this is a standard feature of renormalon contributions in QFT. 
As noted above, it is then natural to expect that 
these two choices in the exponentially small contributions are correlated with the two possible choices of lateral Borel resummation 
of the perturbative series (which, as we know from \cite{mr-ren}, is not Borel summable along the positive real axis). 
More precisely, let us write the 
r.h.s. of (\ref{efromWHZ}) as a formal trans-series 
\be
\label{transerho}
\Phi^\pm (\alpha)= \varphi(\alpha) + \CC_0^\pm  \re^{-2/\alpha} \alpha^{2 \Delta} + \sum_{\ell=1}^\infty \CC_\ell ^\pm \re^{-\frac{2 \ell}{\Upsilon\alpha}}\alpha^\frac{2 \ell \Delta}{\Upsilon} \varphi_\ell^{\pm}(\alpha), 
\ee
where 
\be
\label{var-p}
\varphi(\alpha)\equiv \sum_{k \ge 0} e_k \alpha^k=\frac{1}{4}+\frac{\Delta}{4}\alpha +\frac{1}{8} \Delta  (\Delta +2) \alpha ^2+\CO\left(\alpha ^3\right)
\ee
is the perturbative series, 
\be
\label{coefC0}
\CC_0^\pm = 
\frac{2^{\Delta -2} \re^{\Delta } (1-2 \Delta )^{\Delta -\frac{1}{2}}  }{\Gamma (1-\Delta )}\tilde{\rho}^\pm, 
\ee
is the coefficient of the first non-perturbative correction, and $\varphi_\ell^\pm (\alpha)$ are formal power series associated to the $\ell$-th exponentially small correction of the form (\ref{corrected}). Their first overall coefficient is given by
\be
 \label{coef_C10}
\CC_1^\pm =-\frac{(1-2 \Delta )^2 }{8  \Delta (\Delta -1) }\rho _0^\pm.
\ee
We expect the following exact result 
\be 
\label{exacterho}
{e \over 2\pi\rho^2} = s_\pm (\Phi^\pm)(\alpha), 
\ee
where $s_\pm$ are lateral Borel resummations along the positive real axis. We will test some aspects of this proposal in the next subsection.

We can improve upon the general form \eqref{transerho} by noticing that we can factor out the ambiguous part of the residues in \eqref{polepm}. In all of our equations, we can replace
\begin{equation}
\re^{-2B\xi_n}\rho^\pm_n  =  \left(\re^{-2B} \re^{\mp \ri \pi\Delta}\right)^{\xi_n}r_n,
\end{equation}
where $r_n$ are real factors. %determined in \eqref{polepm}
Therefore, all exponential terms of the same order have the same ambiguous factor, and we can write
%\be
%\label{transerho_strong}
%\Phi^\pm (\alpha)= \varphi(\alpha) + \CC_0^\pm  \re^{-2/\alpha} \alpha^{2 \Delta} + \sum_{\ell=1}^\infty 
%\left(
%{\mathsf r}_\ell
%\re^{\mp \ri\ell\pi\frac{ \Delta}{\Upsilon}}
%\right) 
%\re^{-\frac{2 \ell}{\Upsilon\alpha}}\alpha^\frac{2 \ell \Delta}{\Upsilon} \varphi_\ell(\alpha), 
%\ee
\be
\label{transerho_strong}
\CC_\ell^\pm =
%{\mathsf r}_\ell \re^{\mp \ri\ell\pi\frac{ \Delta}{\Upsilon}}
{\mathsf r}_\ell \re^{\mp\ri\ell\frac{\pi}{N-4}}%\left\{\cos\left(\frac{\ell \pi \Delta}{\Upsilon}\right)\mp\ri \sin\left(\frac{\ell \pi \Delta}{\Upsilon}\right)\right\}
%\left\{\cos\left(\frac{\ell \pi }{N-4}\right)\mp\ri \sin\left(\frac{\ell \pi }{N-4}\right)\right\}
, \qquad
\varphi^\pm_\ell(\alpha) = \varphi_\ell(\alpha),
\ee
where ${\mathsf r}_\ell$ are real constants and $\varphi_\ell (\alpha)$ are real formal power series. The first one is given by 
\be
\varphi_1(\alpha)= 1+ c_1^{(1)} \alpha+ \CO(\alpha^2),
\ee
with 
\be 
c^{(1)}_1 = -\frac{2  \Delta ^2}{1-2 \Delta }.
\ee
 Due to the factorization \eqref{transerho_strong}, the real part of the trans-series 
 is identical to the ambiguous imaginary part, up to overall constants. Since there is only one independent 
 formal power series associated to each exponentially small correction, it is likely that they can be all detected through the Borel singularities of the perturbative series, and therefore that the strong resurgence program defined in \cite{dpmss} holds in this case.

\subsection{Testing the analytic results}
\label{sec-tar-gn}
Although the result (\ref{efromWHZ}) has been found analytically, we have not provided a rigorous derivation that it leads to the correct 
trans-series representation. The reason is that in the calculation above we have replaced some quantities by their conventional 
asymptotic expansions (like for example in the expressions involving the operator (\ref{P-exp})). It might happen that this replacement is not 
valid when we upgrade the expansion to an exact statement involving Borel resummations, and therefore that we are missing exponentially small corrections in the trans-series. 

We will give now extensive evidence that the trans-series (\ref{transerho}) that we have obtained is indeed correct, and in particular that 
it leads to the right results for the singularities of the Borel transform of $\varphi(\alpha)$. 

A first test is to consider the $1/N$ expansion of the free energy $\CF(h)$. As 
shown in \cite{fnw1,fnw2}, $\CF(h)$ can be expanded as 
\be
\CF(h)= \sum_{k \ge 0} \Delta^k \CF_k(h), 
\ee
and the functions $\CF_k(h)$ for $k=0,1,$ can be computed in closed form either from the Bethe ansatz \cite{fnw1} or 
directly in field theory \cite{fnw2} (higher order terms were computed numerically in \cite{dpmss}). Each of these functions is given by a trans-series which was written down explicitly in \cite{dpmss}. To compare with the results in \cite{dpmss}, it is useful to introduce yet another coupling $\bar\alpha$ through the following equation\footnote{We apologize for the proliferation of couplings.}
\begin{equation}
\frac{1}{\bar{\alpha}}-\Delta\log\bar{\alpha}=\log \left(\frac{2h}{m} \right),
\end{equation}
which is related to $\tilde \alpha$ through
\begin{equation}
\tilde{\alpha}=\bar{\alpha}+ \bigl[-\Delta  (1+\log (2))+\log \Gamma (1-\Delta )+\log (2)\bigr]\bar{\alpha}^2 +\CO\left(\bar{\alpha}^3\right).
\end{equation}
After changing couplings we can expand (\ref{fhWHZ}) in a series in $\Delta$ with $\bar \alpha$ fixed. Note that, since $\bar \alpha$ is related to the running coupling constant by (\ref{talg2}), this is indeed a conventional large $N$ 't Hooft limit in which $N \overline{g}^2(h)$ is fixed. In this limit the location of the singularities (\ref{corrected}) becomes the conventional one, and we obtain an infinite tower of IR renormalons at the expected locations. One finds, 
%
%\begin{equation}
%\CF(h) =
%\begin{multlined}[t]
%- \frac{h^2}{2\pi}\Biggl\{
%1+\CO\left(\bar{\alpha}^3\right)
%+\re^{-\frac{2}{\bar{\alpha}}} 
%\left(-\frac{4}{\bar{\alpha}}-2+\CO\left(\bar{\alpha} ^2\right)\right)
%+\re^{-\frac{4}{\bar{\alpha}}} 
%\left(2+\CO\left(\bar{\alpha}\right)\right)
%+\CO\left(\re^{-\frac{6}{\bar{\alpha}}}\right)
%\\
%+\Delta
%\left(
%-\bar{\alpha} -\bar{\alpha}^2+O\left(\bar{\alpha}^3\right)
%+\re^{-\frac{2}{\bar{\alpha}}} 
%\left(\frac{8}{\bar{\alpha}^2}+\frac{-8 \log (\bar{\alpha} )+4 \pi C_\pm}{\bar{\alpha} }-4+\CO\left(\bar{\alpha}\right)\right)
%\right.
%\\
%\left.
%+\re^{-\frac{4}{\bar{\alpha}}} 
%\left(-\frac{16}{\bar{\alpha}}+\CO\left(\bar{\alpha} ^0\right)\right)
%+\CO\left(\re^{-\frac{6}{\bar{\alpha}}}\right)
%\right)
%+\CO\left(\Delta^2\right)
%\Biggr\},
%\end{multlined}
%\label{largeNFGN}
%\end{equation}
\begin{equation}
\ba
\CF_0(h) &=
\begin{multlined}[t]
- \frac{h^2}{2\pi}\Biggl\{
1+\CO\left(\bar{\alpha}^3\right)
+\re^{-\frac{2}{\bar{\alpha}}} 
\left(-\frac{4}{\bar{\alpha}}-2+\CO\left(\bar{\alpha} ^2\right)\right)
+\re^{-\frac{4}{\bar{\alpha}}} 
\big(2+\CO(\bar{\alpha})\big)
+\CO\left(\re^{-\frac{6}{\bar{\alpha}}}\right)\Biggr\},
\end{multlined}
\\
\CF_1(h) &=
\begin{multlined}[t]
- \frac{h^2}{2\pi}\Biggl\{
-\bar{\alpha} -\bar{\alpha}^2+O\left(\bar{\alpha}^3\right)
+\re^{-\frac{2}{\bar{\alpha}}} 
\left(\frac{8}{\bar{\alpha}^2}+\frac{-8 \log (\bar{\alpha} )+4 \pi C_\pm}{\bar{\alpha} }-4+\CO\left(\bar{\alpha}\right)\right)
\\
+\re^{-\frac{4}{\bar{\alpha}}} 
\left(-\frac{16}{\bar{\alpha}}+\CO\left(\bar{\alpha} ^0\right)\right)
+\CO\left(\re^{-\frac{6}{\bar{\alpha}}}\right)
\Biggr\},
\end{multlined}
\ea
\label{largeNFGN}
\end{equation}
where $C_\pm=\pm \ri$. This matches precisely the trans-series obtained in \cite{dpmss}, which was extracted from the exact results in \cite{fnw1,fnw2}. One interesting 
aspect of this calculation is that the first two exponential corrections in (\ref{efromWHZ}), proportional to
\be
\re^{-{2 \over \alpha}}, \qquad \re^{-{2 \over \Upsilon \alpha}},
\ee
combine in the large $N$ limit. In particular, the ambiguous term in (\ref{largeNFGN})
\be
\Delta {4 \pi C_\pm \over \bar{\alpha}}\re^{-\frac{2}{\bar{\alpha}}}
\ee
comes from the non-conventional exponentially small correction $\re^{-{2 \over \Upsilon \alpha}}$. This ambiguous term is what controls the large order behavior of the non-trivial perturbative series at order $\Delta$, which is due to ring diagrams (similarly to what happens to the non-linear sigma model analyzed in \cite{mmr}). This means that the singularity at $1/\Upsilon |\beta_0|$ encodes the information about the large order behavior or renormalon diagrams, and it is indeed a renormalon singularity. 

The result above has implications for the large $N$ determination of renormalon singularities. In many examples in field theory, 
one establishes the existence of renormalon singularities by studying renormalon diagrams in an appropriate large $N$ limit. 
This typically leads 
to a singularity at $1/|\beta_0^{\rm N}|$, where $\beta_0^{\rm N}$ is the large $N$ limit of $\beta_0$. One then hopes that 
subleading $1/N$ corrections change this into $1/|\beta_0|$. However, in the case at hand, the $1/N$ corrections {\it split} the large $N$ singularity in the Borel plane of the coupling at $\zeta=2$, into two different singularities at 
\be
\label{twosings}
\zeta=2, \qquad \zeta=2 {N-2 \over N-4},  
\ee
which correspond to the conventional singularity at $1/|\beta_0|$ and the non-conventional one at $1/\Upsilon |\beta_0|$, respectively. 

\begin{figure}
\center
\includegraphics[width=0.45\textwidth]{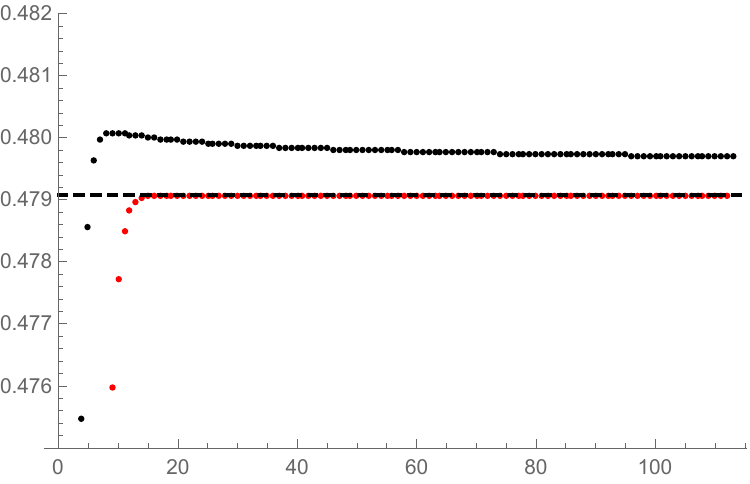} \qquad 
\includegraphics[width=0.45\textwidth]{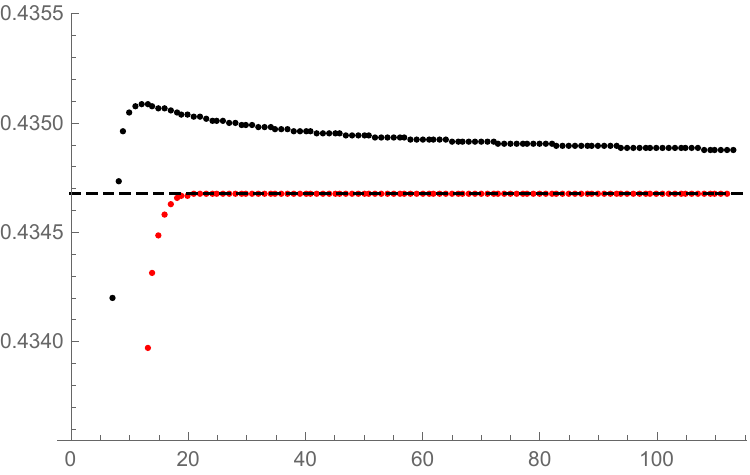}
\caption{Plot of the sequence $s_k$ in \eqref{seq_s_k} for the Gross--Neveu model with 
$N=7$ (left, black) and $N=8$ (right, black) as well as their respective second Richardson 
transforms (red). The dashed line is the predicted value $2^{2\Delta}\mS_0/(2\pi)$.
}
\label{fig-asym-beh-GN}
\end{figure}

Let us now consider the normalized energy density and its trans-series, (\ref{efromWHZ}). The perturbative series (\ref{var-p}) is known analytically 
up to order $45$ from \cite{mr-ren}, and we have generated many more terms numerically for low values of $N$. 
As we explained in section \ref{sec-resQFT}, one way of accessing its Borel 
singularities on the positive real axis is to compute the discontinuity of its lateral Borel resummation. At 
the same time, since the discontinuity is imaginary, our working hypothesis (\ref{exacterho}) indicates that this 
ambiguous imaginary piece has to cancel against the imaginary part of the trans-series. 
Therefore, we should have
\be
\label{discformula}
{\rm disc} \, s(\varphi)(\alpha) \sim \ri \, \mathsf{S}_0 \re^{-\frac{2}{\alpha}}\alpha^{2\Delta} 
+ \ri \, \mathsf{S}_{1} \re^{-\frac{2}{\Upsilon\alpha}}\alpha^\frac{2\Delta}{\Upsilon} \Big(1+ c_1^{(1)} \alpha+\CO\big(\alpha^2\big)\Big) + \CO\left(\re^{-\frac{4}{\Upsilon\alpha}}\right),
\end{equation}
where we recall that ${\rm disc} \, s(\varphi)(\alpha)$ is the discontinuity (\ref{disc-def}) at $\theta=0$. The Stokes constants 
$\mS_{0,1}$ can be read from (\ref{coefC0}), (\ref{coef_C10}) and are given by
\begin{align}
\mathsf{S}_0&= -\ri \big( \CC_0^- - \CC_0^+ \big) =
%\frac{2^{2 \Delta -1} \re^{2 \Delta } \sin (\pi  \Delta ) \Gamma (\Delta )}{\Gamma (1-\Delta )} 
\frac{\pi (2\re)^{2 \Delta } }{2\Gamma (1-\Delta )^2}
,\\
\mathsf{S}_1 &=  -\ri \big( \CC_1^- - \CC_1^+ \big) = -\frac{(2 \re (1-2 \Delta ))^{\frac{2 \Delta }{1-2 \Delta }}}{2 \pi } \left[\sin \left(\frac{\pi  \Delta }{1-2 \Delta }\right) \Gamma \left(\frac{\Delta }{1-2 \Delta }\right)\right]^2
%-\Delta^{-1}2^{\frac{1}{1-2 \Delta }-2} \re^{\frac{2 \Delta }{1-2 \Delta }} (1-2 \Delta )^{\frac{1}{1-2 \Delta }}  \frac{\cos \left(\frac{\pi }{2-4 \Delta }\right)\Gamma \left(\frac{\Delta }{1-2 \Delta }\right)}{  \Gamma \left(\frac{\Delta }{2 \Delta -1}\right)}
.
\end{align}
The discontinuity formula (\ref{discformula}) implies that indeed the first two singularities of the Borel transform of $\varphi (\alpha)$ are at (\ref{twosings}). The singularity at $\zeta=2$ controls the leading large order asymptotics, together with an UV renormalon singularity at $\zeta=-2$, as noted in \cite{mr-ren}. We can now refine the large order analysis of \cite{mr-ren}. To get rid of the effect of the UV 
renormalon at leading order, we define the auxiliary sequence 
\begin{equation}
s_k=
\frac{2^{2 m-1} e_{2m}}{\Gamma \left(2 m-2\Delta\right)} + \frac{2^{2 m} e_{2m+1}}{\Gamma \left(2 m-2\Delta+1\right)},
\label{seq_s_k}
\end{equation}
where $e_{m}$ are the coefficients in (\ref{var-p}). By using the relationship between the large order behaviour of this series and the discontinuity of the Borel sum, it is easy to see that \eqref{discformula} implies the asymptotic behavior
\begin{equation}
s_k = 2^{2 \Delta} \frac{\mathsf{S}_0}{2\pi}+\CO\left(\frac{1}{k^{1-4\Delta}}\right),\quad k\gg 1.
\label{lim_s_k}
\end{equation}
This makes it possible to test our calculation of $\mS_0$ for various values of $N$. In \figref{fig-asym-beh-GN} we show the sequence $s_k$ for $N=7,8$ and its second Richardson acceleration, by using 230 coefficients of the perturbative series. The straight line is the predicted value of $\mS_0$, which we can match with 20 digits of precision. 

\begin{figure}
\center
\includegraphics[width=0.45\textwidth]{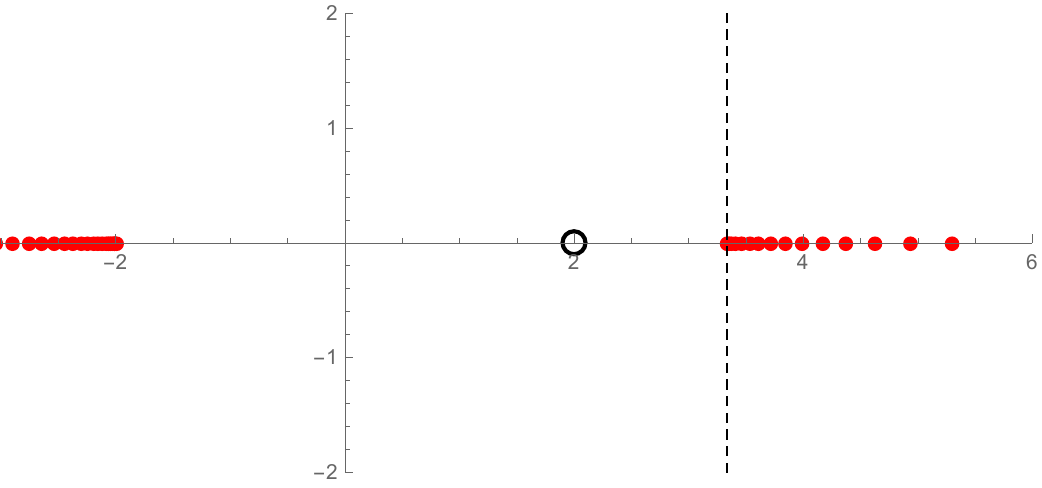} \qquad 
\includegraphics[width=0.45\textwidth]{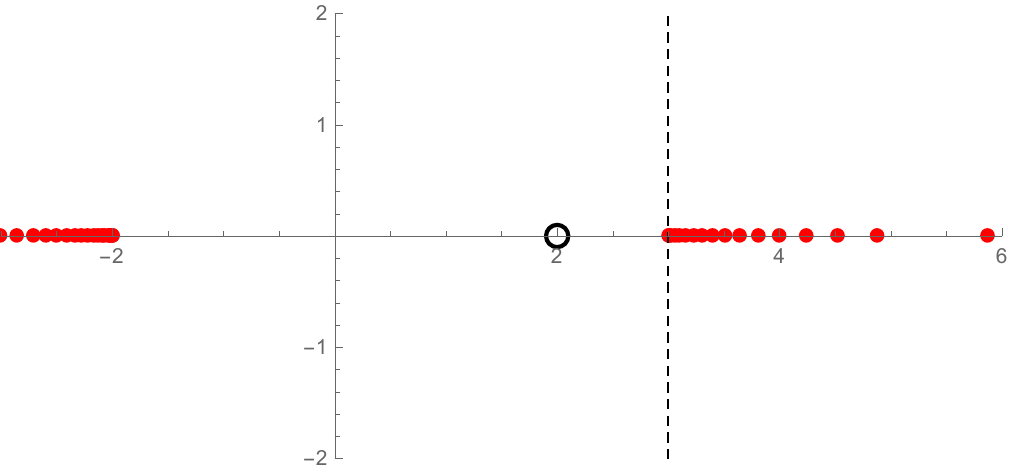}
\caption{The poles of the Borel-Padé approximant of the series $\bar{e}_m$ in 
\eqref{seq_no_IR} truncated at 120 terms. The plots correspond to the Gross--Neveu model with
$N=7$ (left) and  
$N=8$ (right). The dashed vertical line indicates the predicted position of first 
unconventional renormalon singularity $\zeta = 2/\Upsilon$. The black circle indicates the 
position of the removed IR singularity at $\zeta=2$.
}
\label{fig-borel-poles-GN}
\end{figure}

The analysis above tests the first singularity and its Stokes constant, but we would like to test as well 
the presence of the second singularity at the non-conventional location $\zeta=2/\Upsilon$. One possibility is to remove from the perturbative series the effect of the first singularity at $\zeta=2$, and locate the remaining singularities by 
using Pad\'e approximants of the resulting Borel transform. We consider then the auxiliary series $\bar{e}_m$, which is 
obtained by subtracting the effect of the first IR renormalon:
\begin{equation}
\bar{e}_m =  e_m - 2^{-m+2\Delta}\frac{\mS_0}{2\pi}\Gamma\left(m-2\Delta\right).
\label{seq_no_IR}
\end{equation}
We can then inspect the poles of Borel-Pad\'e approximants to this series to see where they accumulate. As shown in \figref{fig-borel-poles-GN}, which considers the cases $N=7,8$, the singularities occur at $2/\Upsilon$, as expected from (\ref{discformula}). 

\begin{figure}
\begin{center}
\includegraphics[width=0.65\textwidth]{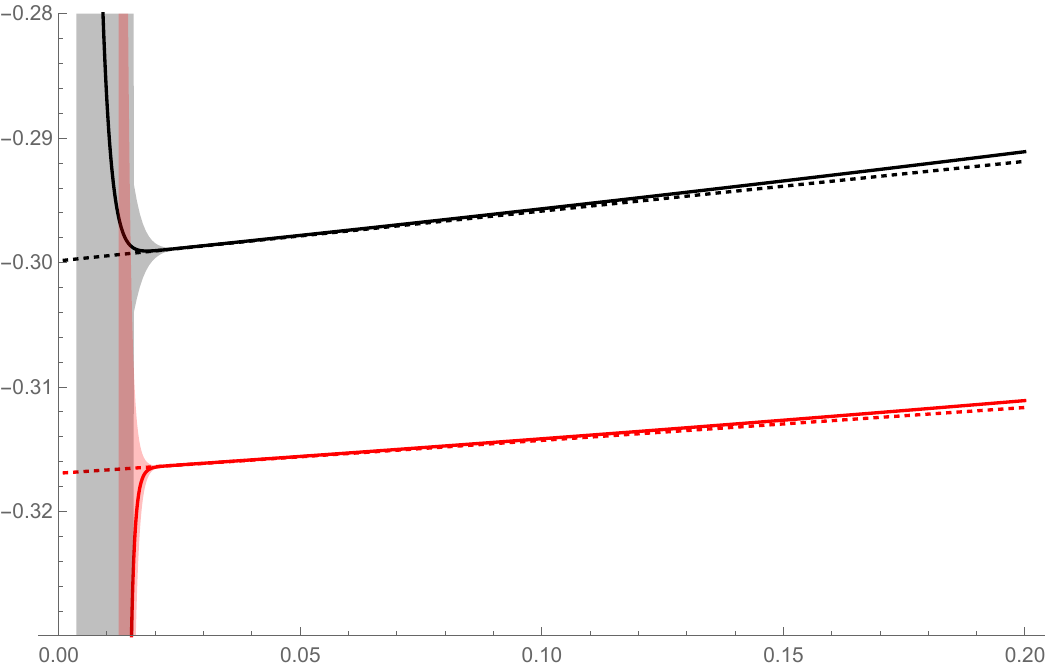}
\end{center}
\caption{We plot an approximation to $f(\alpha)$, defined in \eqref{f1def_p}, 
for the Gross--Neveu model with $N=7$ (black) and $N=8$ (red). The shaded areas 
represent the error of the corresponding color and the dashed lines represent the asymptotic behaviour 
for $\alpha\ll 1$ in \eqref{f1def_p}. For this plot we truncated the energy series at 71 coefficients, calculated numerically with at least 400 digits of precision, and used a $[35/35]$ Padé approximant.}
\label{fig-disc-f1-GN}
\end{figure}

It is possible to do a more quantitative test of the unconventional renormalon singularity 
appearing in (\ref{discformula}): one can 
calculate the discontinuity of the Borel-Padé resummation, remove the contribution of the 
first singularity, and inspect its asymptotic behavior as $\alpha$ becomes small. We then 
consider the quantity
\begin{equation}
f(\alpha) = \re^{\frac{2}{\Upsilon\alpha}}\alpha^{-\frac{2\Delta}{\Upsilon}}\left(\frac{\text{disc} \, s(\varphi)(\alpha)}{2\pi \ri}- \re^{-\frac{2}{\alpha}}\alpha^{2\Delta}\frac{\mS_0}{2\pi}\right) \sim \frac{\mS_1}{2\pi}\left(1+c_1^{(1)}\alpha + \CO\left(\alpha^2\right)\right)+\CO\left(\re^{-\frac{2}{\Upsilon\alpha}}\right).
\label{f1def_p}
\end{equation}
The computational strategy is the following: we take a sufficiently high truncation of the 
Borel transform of $\varphi(\alpha)$ 
and calculate its highest diagonal Padé approximant. To calculate the discontinuity numerically, 
one could integrate along the rays $\theta=\pm \epsilon$ and calculate the difference between these 
integrals (or just their imaginary part). It turns out to be better to obtain the discontinuity by 
calculating the numerical residues of the Padé approximant. The main source of numerical error is the 
convergence of the Padé approximation. To estimate it, we follow \cite{dpmss} and 
calculate the difference between using the highest diagonal Padé approximant or the one of one degree lower. 
In \figref{fig-disc-f1-GN} we plot $f(\alpha)$ in the cases $N=7,8$, and we compare it to the expected asymptotic behaviour\footnote{Note that, in practice, instead of calculating $s(\varphi)(\alpha)$ and then subtracting the contribution of the IR renormalon, it is numerically more stable 
to calculate instead the discontinuity of the Borel resummation of the series \eqref{seq_no_IR}, particularly for small $\alpha$.}.

\begin{figure}
\center
\includegraphics[width=0.45\textwidth]{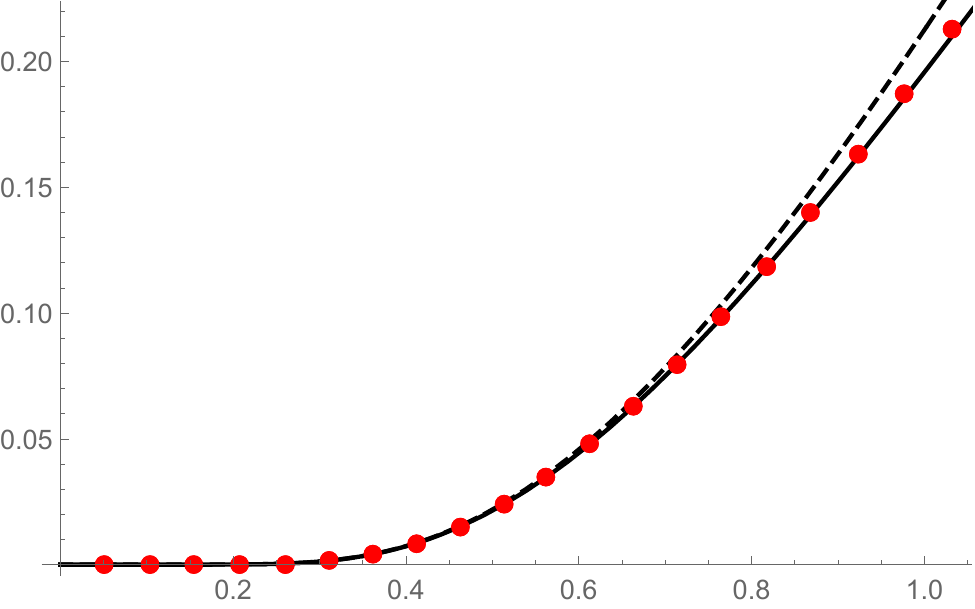} \qquad 
\includegraphics[width=0.45\textwidth]{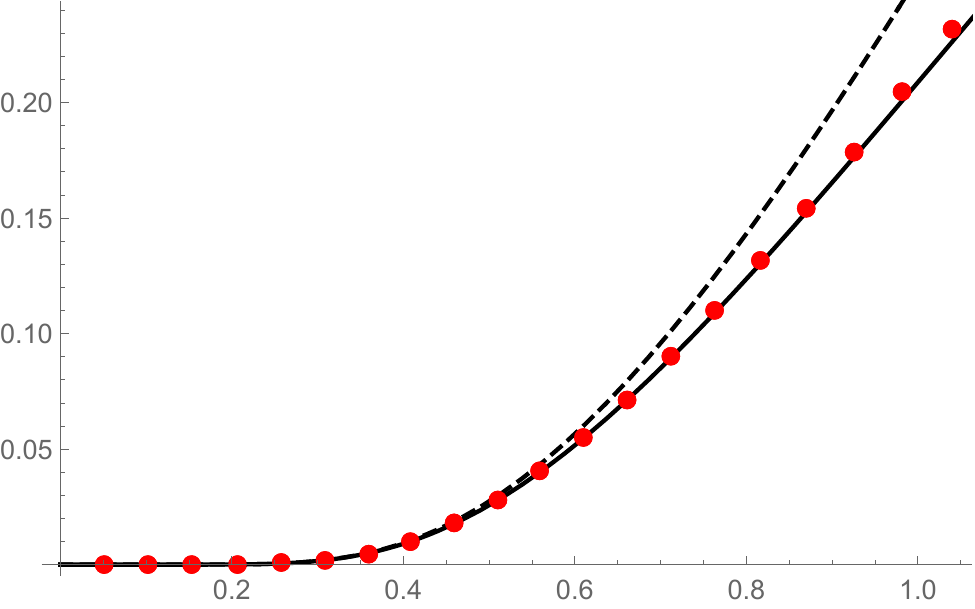}
\caption{In these figures we compare the difference between the normalized energy density and the real part of the Borel resummation of the perturbative series, against the theoretical predictions \eqref{realGN} for the Gross--Neveu model with $N=7$ (left) and $N=8$ (right). The $x$-axis is the value of $\alpha$. The dots (red) are the numerical calculations of the l.h.s. of \eqref{realGN}, using a discretisation of 50 points in the integral equation and 120 coefficients in the perturbative series, evaluated at $B=20/k$ for $k=1,\dots,20$. The dashed line (black) is the contribution in \eqref{realGN} coming from the leading IR singularity, while the full line (black) includes also the first two terms of the new, unconventional renormalon sector. %Error estimates are too small to be plotted.
}
\label{fig-real-GN}
\end{figure}

In our view, these tests give very convincing evidence that (\ref{discformula}) is correct and that the 
perturbative series $\varphi(\alpha)$ has an IR singularity at the unconventional location $\zeta=2/\Upsilon$. 
We now provide evidence for the stronger statement (\ref{exacterho}), which tests also the real part of the coefficients $\CC_{0,1}^\pm$ in (\ref{coefC0}) and (\ref{coef_C10}). The conjectural equation (\ref{exacterho}) leads to the asymptotic behavior for small $\alpha$, 
\be
\label{realGN}
{e\over 2\pi\rho^2}- {\rm Re}\left(s_\pm (\varphi)(\alpha) \right) \sim \mR_0 \, \re^{-\frac{2}{\alpha}}\alpha^{2\Delta} 
+ \mR_{1} \, \re^{-\frac{2}{\Upsilon\alpha}}\alpha^\frac{2\Delta}{\Upsilon} \left(1+ c_1^{(1)} \alpha + \CO\left(\alpha^2\right)\right) + \CO\left(\re^{-\frac{4}{\Upsilon\alpha}}\right), 
\ee
where 
\be
\mR_0=\frac{\mathcal{C}_0^++\mathcal{C}_0^-}{2},\qquad 
\mR_1=\frac{\mathcal{C}_1^++\mathcal{C}_1^-}{2}.
\ee
In order to test (\ref{realGN}), we first calculate $e/\rho^2$ from a numerical solution of the Bethe ansatz integral equations. To obtain 
the Borel resummation of the perturbative series, we use $\sim 100$ coefficients and we 
improve the numerical result with a conformal mapping, a strategy similar to  the ``Padé-Conformal-Borel'' method in \cite{costin-dunne-conformal}. This makes it possible to compute the l.h.s. of (\ref{realGN}), 
which can then be compared to the r.h.s. We show such a comparison in 
\figref{fig-real-GN} for $N=7$ (left) and $N=8$ (right). Here the $x$-axis represents the value of $\alpha$, the red 
dots are the values of the l.h.s. of (\ref{realGN}), the dashed line is the contribution of the first IR renormalon in the r.h.s., while the continuous line is the full r.h.s., including the unconventional renormalon contribution. 

\subsection{UV renormalons}
\label{uv-sec}

So far we have focused on renormalon singularities in the positive real axis. These are 
eventually due to poles of $\sigma(\omega)$ in (\ref{sig-ome}) in the complex upper half-plane. It is tempting to 
believe that the poles of $\sigma(\omega)$ in the complex lower half-plane will lead to UV renormalons. 
In order to pick up these poles in the observables, we need to deform the problem such that it has the same perturbative series but with negative $B$. 
%We will indeed see that this is the case, and we will obtain analytically the first few terms of the trans-series corresponding to the first UV renormalon. 
Let us then assume that $B<0$ and change\footnote{This change is analogous to the Gaudin--Yang model with repulsive coupling, see \cite{mr-long}.} $\rho\rightarrow-\rho$. We are able to check the perturbative expansion is the same and obtain analytically the first few terms of the trans-series corresponding to the first UV renormalon. 

In this setting, we must deform the contour in \eqref{uom-int} downwards in the complex 
plane, leading to
\begin{equation}
u(-\ri \xi) = -\frac{1}{\xi}+\frac{1}{2\pi\ri } \int_{\CC_\pm} \frac{\re^{ 2 B\xi'}\delta^{\text{UV}}\rho(-\ri \xi')u(-\ri\xi')}{\xi+\xi'} \rd \xi' + \sum_{n\ge 1} \frac{\re^{ 2 B \xi^{\text{UV}}_n }\rho^{\text{UV},\pm}_n u^{\text{UV}}_n}{\xi+\xi_n},
\label{UVeq}
\end{equation}
where $\delta^{\text{UV}}\rho (\omega) $ denotes the discontinuity of $\rho (\omega) $ along the negative imaginary axis, 
which is due to $G^{-1}_+(\omega)$ rather than $G_-(\omega)$. Explicitly, it is given by 
\begin{equation}
\label{disc-uv}
\ba
\delta^{\text{UV}}\rho(-\ri\xi) &= 2\ri\re^{-[2\Delta(1+\log 2) + \Upsilon\log\Upsilon]\xi + 2\Delta\xi\ln\xi} \sin(\pi\Delta\xi)\frac{\Gamma\big(\frac{1}{2}+\frac{1}{2}\Upsilon\xi\big)\Gamma\big(\frac{3}{2}-\frac{1}{2}\xi\big)}{\Gamma\big(\frac{1}{2}-\frac{1}{2}\Upsilon\xi\big)\Gamma\big(\frac{3}{2}+\frac{1}{2}\xi\big)}.
%\\&= \left(\delta \rho\right)(-\ri\xi).
\ea
\end{equation}
Similarly, the poles $\xi_n^{\text{UV}}$ are given by the zeroes of 
$G_+(\omega)$, which are located at:
\begin{equation}
\xi_n^{\text{UV}} = 2n+1,\quad n\geq 1.
\end{equation}
These poles will lead to a different trans-series structure than in the IR case. The residues $\rho^{\text{UV},\pm}_n$ and unknowns $u^{\text{UV}}_n$ are defined in strict analogy with the IR case. We now introduce the analogue of (\ref{veta}), 
\begin{equation}
-\frac{1}{w}-2\Delta\log(w) = 2B,\qquad \xi = w\eta. 
\end{equation}
The variable $w$, the analogue of the previously introduced $v$, is positive. With this convention we have
\begin{equation}
\re^{2B\xi}\delta^{\rm UV}\rho(-\ri\xi) = -2\ri (-w) \re^{-\eta} P(\eta,-w), 
\end{equation}
where the last argument means that we replace $v$ with $-w$ in the previous definition of $P(\eta)$. 
Thus our integral equation becomes
\begin{equation}
\ba
u\big(\ri (-w) \eta\big) &= \frac{1}{(-w)\eta} -
 \frac{(-w)}{\pi}\int_{\CC_\pm} \frac{\re^{-\eta'}P(\eta',-w)}{\eta+\eta'}u\big(\ri (-w) \eta'\big) \rd\eta'
\\
&+\sum_{n\geq 1}\frac{q_{\text{UV}}^{2n+1}\rho_n^{\text{UV},\pm}u_n^{\text{UV}}}{w\eta+2n+1},
\ea
\label{uUV}
\end{equation}
where we introduced
\begin{equation}
q_{\text{UV}}= \re^{2B} = \re^{-\frac{1}{w}}w^{-2\Delta}.
\end{equation}
It follows from \eqref{uUV} that, under $w\rightarrow -v$, we find the same perturbative solution as in the previous analysis.

Let us now consider the boundary condition (\ref{uh}), which can be obtained 
again from (\ref{uom-int}) by setting $\omega=\ri$. In deforming the contour downwards we pick up 
an additional residue at $\omega'=-\ri$, and we find
\be
\ba
u(\ri) &= 1 - \frac{(-w)}{\pi}\int_{\CC_\pm} \frac{\re^{-\eta'}P(\eta',-w)}{\eta'-1/w}u\big(\ri (-w) \eta'\big)\rd\eta' +\sum_{n\geq 1}\frac{q_{\rm UV}^{2n+1}\rho_n^{\rm{UV},\pm}u_n^{\rm{UV}}}{2n} \\
& +\re^{2B}\rho^{\rm{UV},\pm}_0  u(-\ri),
\ea
\ee
where
\begin{equation}
\rho^{\rm{UV},\pm}_0 = \rho(-\ri\mp 0) = \re^{\pm \ri \pi  \Delta}(2 \re)^{-2 \Delta } (1-2 \Delta )^{2 \Delta -1}\frac{\Gamma (1-\Delta )}{ \Gamma (\Delta )}.
\end{equation}
%
%We will denote $u_0^{\text{UV}}=u(-\ri)$. It 
The value $u(-\ri)$ can be calculated by using equation \eqref{uUV}, and one finds 
at leading order
\begin{equation}
%u_0^{\rm{UV}}
u(-\ri) = -1-\frac{d_{1,0}}{\pi }w +\CO\big(w^2\big)+\CO\big(q_{\rm{UV}}^3\big).
\end{equation}
We now extend the definition of $\tilde{\alpha}$ in \eqref{alphatilde} to account for negative values, in such a way that the perturbative coefficients of the free energy $\CF(h)$ 
remain the same. %(up to a change of sign). 
The appropriate choice is
\begin{equation}
\frac{1}{\tilde{\alpha}}-\Delta\log|\tilde{\alpha}|=\log \left(\frac{h}{\Lambda}\right).
\end{equation}
%
%and we understand that $\tilde{\alpha}$ is negative. 
%Similarly, we need an appropriate non-perturbative scale:
%%
%\begin{equation}
%\tilde\kappa = \re^{\frac{2}{\tilde{\alpha}}}\left(-\tilde{\alpha}\right)^{-2\Delta}.
%\end{equation}
%%
In terms of these variables we obtain the following expression for the free energy:
\begin{multline}
\CF(h)=- \frac{h^2}{2\pi}\bigg\{ 1- \Delta \tilde{\alpha} +\frac{1}{2} \Delta  \big[ \Delta -2+2 \log (2)\big]\tilde{\alpha}^2 + \CO\left(\tilde{\alpha}^3\right)\\
- \re^{\frac{2}{\tilde{\alpha}}}\left|\tilde{\alpha}\right|^{-2\Delta} \left(4 (1-2 \Delta )^{1-2 \Delta }\rho_0^{\rm{UV},\pm} -8 \Delta (1-2 \Delta )^{1-2 \Delta }\rho_0^{\rm{UV},\pm} \tilde{\alpha}+\CO\left(\tilde{\alpha}^2\right)\right)+\CO\left(\re^{\frac{4}{\tilde{\alpha}}}\right)\bigg\}.
\end{multline}
Notice that the perturbative part is the same as in (\ref{fhWHZ}), but now $\tilde{\alpha}$ is negative. We can now make a Legendre transform to obtain the normalized energy density. We need again to extend the definition of the coupling $\alpha$ in (\ref{alpha-GN}) to
\begin{equation}
\frac{1}{\alpha}-\Delta \log|\alpha| = \log\left(\frac{2\pi\rho}{\Lambda}\right).
\end{equation}
We find
\begin{multline}
\frac{e}{2\pi\rho^2}=\frac{1}{4}+\frac{\Delta }{4}\alpha +\frac{1}{8}\Delta  (\Delta +2)  \alpha^2 +\CO\left(\alpha^3\right)
\\
+\re^{\frac{2}{\alpha}}|\alpha|^{-2\Delta}
\left(\frac{1}{4} (1-2 \Delta )^{1-2 \Delta }\rho_0^{\rm{UV},\pm} +\frac{1}{2}\Delta  (1-2 \Delta )^{1-2 \Delta }\rho_0^{\rm{UV},\pm} \alpha +\CO\big(\alpha^2\big)\right)+\CO\left(\re^{\frac{4}{\alpha}}\right),
\label{eUV}
\end{multline}
where $\alpha$ is again understood to be negative. The second line gives the trans-series associated to the first UV renormalon. Of course, one can 
push the calculation to obtain higher order corrections in both $\alpha$ and $\re^{2/\alpha}$. 

\begin{figure}
\center
\includegraphics[width=0.45\textwidth]{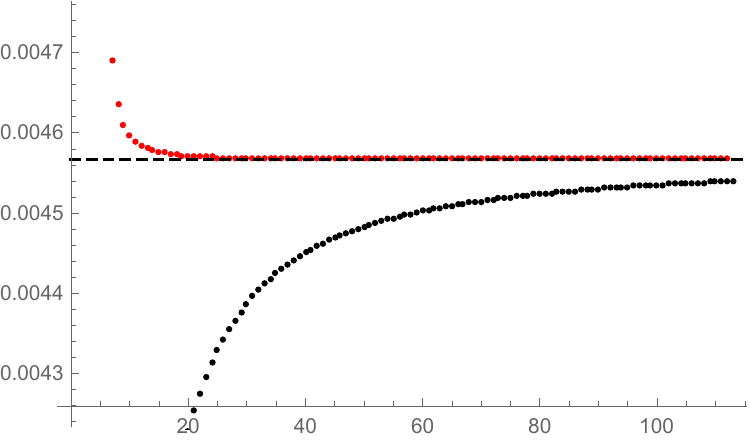} \qquad 
\includegraphics[width=0.45\textwidth]{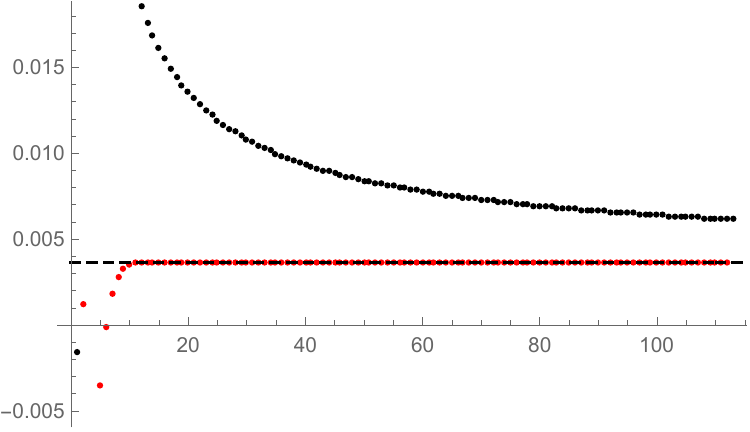}
\caption{Plot of sequences $d_k$ (left, black) and $2k\left(U_0-d_k\right)$ (right, black) for the Gross--Neveu model with $N=7$ as well as their respective second Richardson transforms (red). The dashed lines are the predicted values $U_0$ (left) and $U_1$ (right).
}
\label{fig-UV-GN}
\end{figure}

We can again test our analytic calculation with a 
resurgent study of the perturbative series $\varphi(\alpha)$, since the UV renormalon contributes to its large order 
behavior. We first define an auxiliary sequence which removes the effect of the IR renormalon at leading order, 
\begin{equation}
d_k = \frac{2^{2 m} e_{2m+1}}{\Gamma \left(2 m+2\Delta+1\right)}-\frac{2^{2 m-1} e_{2m}}{\Gamma \left(2 m+2\Delta\right)},
\label{seq_d_k}
\end{equation}
in analogy with \eqref{seq_s_k} and \cite{mr-ren}. From \eqref{eUV} we deduce the large $k$ asymptotics
\begin{equation}
\label{ddk}
d_k \sim  U_0, \quad 2k\left(U_0-d_k\right) \sim  U_1, \qquad k\gg 1,
\end{equation}
where
\begin{equation}
U_0 = %\frac{2^{-2 \Delta } \left(\re^{-2 \Delta  (1+\log (2))}\right)}{4 \Gamma (\Delta )^2},
\frac{(4 \re)^{-2 \Delta }}{4 \Gamma (\Delta )^2}, \qquad
U_1 = %\frac{\Delta  \re^{-2 \Delta  (1+2\log (2))} \sin (\pi  \Delta ) \Gamma (1-\Delta )}{\pi  \Gamma (\Delta )}
4\Delta U_0.
\end{equation}
We match these two coefficients with great precision for all values of $N$ between $5$ and $12$. In \figref{fig-UV-GN} we plot the sequences in (\ref{ddk}), as well as their Richardson transforms and their asymptotic values for $N=7$, for which we can get an agreement of 16 digits of precision for $U_0$ and 12 digits for $U_1$.

\sectiono{Trans-series and renormalons in bosonic models}
\label{bos-sec}

\subsection{Analytic solution}

In this section we will consider the free energy $\CF(h)$ for various ``bosonic" models: 
the non-linear sigma model and its supersymmetric version, and the 
PCF with two different choices of charges \cite{pcf,fkw1,fkw2}. The analysis of the Bethe ansatz equations of 
these models is different from the one we did in the GN model. The reasons is that 
$G_+(\omega) \sim \omega^{-1/2}$ as $\omega \to 0$, and 
we cannot use the equations (\ref{uom-int}) and (\ref{fh-rho}). Instead, we have to go back to the more 
general equations (\ref{Q-eq}) and (\ref{epsom}). The procedure is slightly more involved than in the GN case, since the 
integral equations cannot be simply solved by iteration, but we will eventually obtain similar results for the trans-series. In particular, 
we will be able to establish the existence of unconventional renormalons in the supersymmetric non-linear sigma model and in the PCF. 

In order to incorporate non-perturbative corrections, we deform the integration contour around the positive imaginary axis 
in the second term of (\ref{Q-eq}) 
and pick the poles and the discontinuity of the function, just as we did in (\ref{eq_u_im}). 
We will denote the poles of $\sigma(\ri \xi)$ along the positive imaginary axis by $\xi_n$, 
and we will label them by an integer $n \in \IZ_{>0}$. Their precise location depends on the particular model one is 
considering, but for the moment being we will write general formulae, valid for all bosonic models. One finds,  
\be
\label{Q-eq-np}
Q(\ri \xi)-{1\over2 \pi \ri } \int_{\CC_\pm} { \re^{-2 B \xi'} \delta \sigma (\ri \xi') Q(\ri \xi') \over \xi + \xi'} \rd \xi'+ \sum_{n \ge 1} {\re^{-2B \xi_n} \ri\sigma^\pm_n  Q_n  \over \xi + \xi_n} ={1\over 2 \pi \ri} \int_\IR {G_-(\omega') g_+(\omega' ) \over \ri\xi+ \omega'+\ri 0}\dd\omega'.
\ee
In this equation, $Q_n \equiv Q(\ri \xi_n)$, $\delta \sigma (\ri \xi)$ is the discontinuity of $\sigma(\omega)$ across the positive 
imaginary axis, and $\sigma^\pm_n$ is the residue of $\sigma(\omega)$ 
at $\ri\xi_n\pm 0$. 
A similar expression, including exponentially small corrections, can be obtained from (\ref{epsom}): 
\be
\label{eps-disc}
\frac{\epsilon_+(\ri\kappa)}{G_+(\ri\kappa)} = {1\over 2 \pi \ri} \int_\IR {G_- (\omega') g_+ (\omega')
 \over \omega'-\ri\kappa-\ri 0}  \rd \omega'+ 
{1\over 2 \pi \ri } \int_{\CC_\pm}  {\re^{-2 B \kappa'} \delta \sigma (\ri \kappa') Q (\ri \kappa') \over  \kappa'-\kappa} \rd \kappa' 
- \sum_{n \ge 1}  {\re^{-2B \xi_n} \ri \sigma^\pm_n Q_n \over \xi_n-\kappa}. 
\ee
The free energy can then be obtained from (\ref{fh-eps+}). 

The integral equation (\ref{Q-eq-np}) was analyzed in detail in \cite{hmn,hn,pcf,fnw1} at the perturbative level, 
in order to obtain an exact expression for the mass gap. In the perturbative approximation 
we neglect all exponentially small corrections and introduce a function 
\be
\label{qx-def}
q(x)= Q_{(0)}\left( {\ri x \over 2 B}\right), 
\ee
where the subscript $(0)$ means that we keep only the perturbative part. This function satisfies the integral equation \cite{hmn,hn}
\begin{equation}
q(x) + \frac{1}{\pi} \int_0^\infty \frac{\re^{-y}\gamma(y/2B)}{x+y}q(y)\rd y = r(x),
\label{eq_q}
\end{equation}
where the function $\gamma(\xi)$ is defined by 
\be
\delta \sigma( \ri \xi)= -2 \ri \gamma(\xi), 
\ee
and $r(x)$ is the perturbative part of 
\be
\label{eq_r}
\frac{1}{2\pi\ri} \int_\mathbb{R} \frac{G_-(\omega)g_+(\omega)}{\omega + \ri x/2B} \rd\omega . 
\ee
In the bosonic models we will consider, the functions $G_+(\ri \xi)$ and $\gamma(\xi)$ have the following expansion around the origin:
\be
\label{Gxi-exp}
\ba
G_+(\ri \xi)&=  {k \over {\sqrt{\xi}}} \left( 1- a \xi \log \xi-b \xi+ \CO(\xi^2) \right), \\
\gamma(\xi) &= 1+ 2 b \xi + 2 a \xi \log \xi + \CO(\xi^2). 
\ea
\end{equation}
The coefficients $a,b$ depend on the details of the model. 
As noted in \cite{hmn,hn}, the function $r(x)$ can be expanded in a series in $1/B$ and 
$\log(B)/B$, and this suggests a similar ansatz for $q(x)$. As shown in Appendix \ref{pert-calculation} one has, 
\be
\label{lead-qr}
r(x)= -k h (2B)^{1/2} \Big[ B r_0(x)+ \CO\big(B^{-1/2}\big)\Big], \qquad q(x)= -k h (2B)^{1/2} \Big[ Bq_0(x)+ \CO\big(B^{-1/2}\big)\Big],
\ee
where we have only written down the dominant terms. By plugging this expansion in (\ref{eq_q}), 
one obtains a series of $B$-independent integral equations which can be solved for $q_0(x)$ and the 
subleading functions in the expansion. This solution leads then to a perturbative expression for $\CF(h)$. 
In \cite{hmn, hn, pcf}, the integral equations were solved numerically, and the result for $\CF(h)$ 
involved numerical constants which were fitted to known numbers (like $\gamma_E$ and $\pi$). 

Although in our study of the non-perturbative corrections we will only need the leading contributions 
in $1/B$, we have obtained a fully {\it analytic} derivation of the perturbative expression for $\CF(h)$ quoted in \cite{hmn,hn,pcf}, 
at next-to-leading order in the $1/B$ expansion. For example, in 
(\ref{qx-eq}) we give an explicit solution for the function $q_0(x)$ appearing in (\ref{lead-qr}). 
This derivation, which is of independent interest, is presented in Appendix 
\ref{pert-calculation}, while Appendix \ref{airy-kernel} explains how to solve the integral equations explicitly. 
Some ingredients of this computation will be used in the following. 

In order to derive the exponentially small corrections, we first have to calculate $Q_1=Q(\ri\xi_1)$. Since we are only after the leading contribution in  $1/B$ and $\re^{-2B}$, $\re^{-2B\xi_1}$, we can focus on the leading order term 
of the perturbative part of $Q_1$. This quantity satisfies the equation
\begin{equation}
\label{Q1eq}
Q_1+\frac{1}{\pi}\int _0^\infty \frac{\re^{-2B \xi} \gamma(\xi)Q(\ri\xi)}{\xi_1+\xi} \rd\xi  +\CO\big(\re^{-2B\xi_1}\big) = \frac{1}{2\pi\ri}\int_\IR \frac{G_-(\omega)g_+(\omega)}{\ri\xi_1+\omega} \rd\omega.
\end{equation}
The integral in the l.h.s. of (\ref{Q1eq}) is calculated in the perturbative approximation. We keep the leading 
order term for $q(x)$ in (\ref{lead-qr}). Using the result (\ref{int-result}) we obtain, 
\begin{equation}
\frac{1}{\pi}\int _0^\infty \frac{\re^{-2B \xi} \gamma (\xi)Q(\ri\xi)}{\xi_1+\xi} \rd\xi 
%& \approx -k h {\sqrt{2}} B^{3/2} \frac{1}{\pi}\int_0^\infty \rd y \frac{\re^{-y} \gamma(y/2B)q_0(y)}{2B\xi_1+y} \\
=-k h {\sqrt{B \over 2 \pi}} {4-\pi   \over \xi_1}\Big(1+\CO\big(B^{-1/2}\big)\Big).
\end{equation}
Let us now calculate the r.h.s. of (\ref{Q1eq}). Splitting $g_+(\omega)$ in terms proportional to $h$ and $m$, the contribution from the $h$ part in (\ref{eq_g(h)}) is
\begin{equation}
\ri h \frac{1}{2\pi\ri}\int_\mathbb{R} \frac{G_-(\omega)}{\ri\xi_1+\omega}\frac{1-\re^{2\ri B \omega}}{\omega} \rd \omega = - kh\sqrt{B \over 2 \pi} \frac{4}{\xi_1} \Big(1+\CO\big(B^{-1/2}\big)\Big).
\end{equation}
This result is computed in the same way as the perturbative part $r(x)$, see \figref{fig-crh}:
for the term $1/\omega$ in the integrand, we deform the contour downwards, picking the 
pole at $\omega = -\ri\xi_1$, which gives a contribution subleading in $1/B$. 
For the term $-\re^{2\ri B \omega}/\omega$, we deform the contour upwards, picking the discontinuity of $G_-(\omega)$. This integral 
has to be computed by using the same trick as in \eqref{eq_hankel_contour_h_term}.
Let us now consider the contribution from the $m$ part in (\ref{eq_g(m)}):
\begin{equation}
\frac{\ri m\re^B}{2}\frac{1}{2\pi\ri}\int_\IR \frac{ G_-(\omega)}{\ri \xi_1+\omega} \left( \frac{\re^{2\ri B \omega}}{\omega-\ri} - \frac{1}{\omega+\ri} \right) \rd \omega = -\frac{m\re^B}{2}\frac{G_+(\ri\xi_1)-G_+(\ri)}{\xi_1-1} \Big( 1 + \CO\big(B^{-1/2}\big) \Big).
\end{equation}
To compute this term we proceed as before: we deform the contour downwards for the term $1/(\omega+\ri)$ and 
upwards for the term $\re^{2\ri B \omega}/(\omega-\ri)$. This last term yields a subleading contribution, which we ignore. 
Plugging all the above results in \eqref{Q1eq}, we obtain
\begin{equation}
Q_1 =  -kh(2B)^{1/2}\frac{\sqrt{\pi}}{2} \frac{1}{\xi_1}-\frac{m\re^B}{2}\frac{G_+(\ri\xi_1)-G_+(\ri)}{\xi_1-1}+\CO\big(B^0\big).
\end{equation}

%The next step is identifying the exponentially small contributions to the boundary condition (\ref{bc-eps}). We find, 
%%
%\begin{equation}
%\lim_{\kappa\rightarrow\infty} \kappa \epsilon(\ri \kappa) = \text{perturbative part}
%+ \ri\re^{-2B\xi_1}\sigma_1^\pm\left(- h \frac{G_+(\ri \xi_1) }{\xi_1}
%+\frac{m\re^B}{2} \frac{G_+(\ri \xi_1) }{\xi_1-1} + Q_1 +\cdots\right)
%+\cdots
%\end{equation}
%%
%%\mmm{Should check the notation for $\sigma_1$}
%where the perturbative part is computed in (\ref{eq_int_G-g+}) and (\ref{eq_int_gamma_q}). 
%By using this, we derive the corrected condition
%%
%\begin{equation}
%m\re^B= kh(2B)^{1/2} \frac{\sqrt{\pi }}{G_+(\ri)}\left[ 1+\CO\big(B^{-1}\big) +
%\re^{-2 B \xi _1} \left( \frac{\ri \sigma^\pm_1}{\left(\xi _1-1\right) \xi _1} + \CO\big(B^{-1}\big) \right)
%+ \CO\big( \re^{-2 B \xi _2} \big)\right].
%\label{bc_nonpert}
%\end{equation}
%However,  we will see later that this contribution actually cancels.
In principle, there are exponential correction to the boundary condition that one needs to calculate. However, they do not contribute at leading order to the free energy.
We can then turn our attention to (\ref{eps-disc}). There are three sources of leading 
non-perturbative corrections to this 
quantity. The first one is due to the exponentially small corrections in the last term of (\ref{eps-disc}). The two other sources 
of corrections are in the first term of (\ref{eps-disc})\footnote{There is a fourth potential source of non-perturbative corrections in the second term of the r.h.s. of \eqref{eps-disc} that originate from the non-perturbative corrections to $Q$, however these are subleading in $1/B$.}. In the integrand of
\be
\label{int-poles}
{1\over 2 \pi \ri} \int_\IR {G_- (\omega) g_+ (\omega)
 \over \omega-\ri}  \rd \omega,
 \ee
 there is a simple pole at $\omega=\ri$, coming from (\ref{eq_g(m)}), as well as a pole at $\omega=\xi_1$. The first pole is responsible 
 for the first IR renormalon, located at (\ref{slore}) with $\ell=2$. The second pole leads generically to an IR singularity %renormalon 
 in an unconventional location, as we will see. We find
\be
\ba
{1\over 2 \pi \ri} \int_\IR {G_- (\omega) g_+ (\omega)
 \over \omega-\ri}  \rd \omega& = \text{perturbative part}+ \frac{m\re^B}{4\pi} \re^{-2B} \rho^\pm
\\
&+ \ri h \re^{-2B\xi_1}\sigma_1^\pm \frac{G_+(\ri\xi_1)}{\xi_1(\xi_1-1)}
- \ri \frac{m \re^B}{2} \re^{-2B\xi_1}\sigma_1^\pm \frac{G_+(\ri\xi_1)}{(\xi_1-1)^2}
+\cdots
\ea
\ee
where 
\begin{equation}
\rho^\pm = 2\pi\ri\, \text{Res}_{\omega=\ri\pm 0} \, \frac{G_-(\omega)}{(\omega-\ri)^2}.
\end{equation}
After using the boundary condition, we obtain the simple expression
\begin{equation}
\CF(h)= - \frac{k^2 h^2}{4} B \left\{1- 
\frac{2\ri \sigma^\pm_1 \re^{-2 B \xi _1}}{\left(\xi _1-1\right)^2 \xi _1}+ \cdots
\right\}- \frac{m^2}{8\pi^2} \rho^\pm G_+(\ri). 
\end{equation}
where the $\cdots$ include both perturbative and non-perturbative corrections. The term outside the brackets is due to the contribution of the pole at $\omega=\ri$ to the integral (\ref{int-poles}), 
and it is independent of $B$. 

In order to make contact 
with the standard expansions we need to introduce an appropriate coupling constant. 
Following \cite{bbbkp,volin,mr-ren}, we introduce
\begin{equation}
\frac{1}{\tilde{\alpha}}+\xi \log\tilde\alpha = \log\left( \frac{h}{\Lambda} \right),
\end{equation}
where
\begin{equation}
\xi= \frac{\beta_1}{2\beta_0^2} = a+\frac{1}{2}, 
\end{equation}
and $a$ is the constant appearing in the expansion (\ref{Gxi-exp}). As in the GN model, this coupling is related to 
the running coupling constant in the $\overline{\text{MS}}$ scheme by (\ref{talg2}). In terms of this coupling, we have 
\begin{equation}
\label{fh-bos}
\ba
\CF (h) &= - \frac{k^2 h^2}{4\tilde\alpha}\left\{ 1
-
2\ri \sigma^\pm_1\left(\re^{-\frac{2}{\tilde{\alpha}}}\tilde{\alpha}^{1-2 \xi }\right)^{\xi _1}\! 
\frac{1}{\left(\xi _1-1\right)^2 \xi _1}
\left(\frac{G_+(\ri)^2}{2 \pi  k^2}
\left(\frac{m}{\Lambda}\right)^{2}\right)^{\xi_1}+ \cdots \right\}\\
&\qquad - \frac{m^2}{8\pi^2} \rho^\pm G_+(\ri).
\ea
\end{equation}
The term in the last line is independent of $h$. Its imaginary part leads to an IR renormalon, while its real part can be identified, as in the 
GN model, with $-F(0)$, i.e. the ground state energy. We then obtain the general formula, 
\be
\label{F0-gen}
F(0)=  \frac{m^2}{4\pi} {\rm Re}\left(\ri G_+(\ri) G'_-(\ri\pm 0)\right). 
\ee
%
%where
%%
%\begin{equation}
%q= \re^{-2/\tilde{\alpha}}\tilde{\alpha}^{1-2 \xi }.
%\end{equation}
%
It is convenient to pass to the canonical formalism with $e, \rho$, in which one uses instead the coupling 
\begin{equation}
\frac{1}{\alpha}+(\xi-1)\log\alpha= \log\left(\frac{\rho}{2\mathfrak{c}\beta_0\Lambda}\right). 
\end{equation}
Here, $\mathfrak{c}$ is a convenient constant introduced in \cite{mr-ren}, which varies from model to model. 
One obtains in the end
\begin{equation}
\label{bos-ts}
\frac{e}{\rho^2}= \frac{\alpha}{k^2}\left\{\varphi(\alpha)
+\CC_0^\pm \re^{-\frac{2}{\alpha}}  \alpha ^{1-2 \xi }
+\CC_1^\pm \left(\re^{-\frac{2}{\alpha}}  \alpha ^{1-2 \xi }\right)
^{\xi_1}(1+\cdots)+\cdots \right\}. 
\end{equation}
In this expression we have included the full perturbative series $\varphi(\alpha)$, 
which is of the form (\ref{p-series}). The coefficients of the exponentially small corrections are given by
\be
\label{2cos}
\ba
\CC_0^\pm&=-\rho^\pm {G_+(\ri) k^2 \over 32 \pi^2  \beta _0^2 \mathfrak{c}^2} \left(\frac{m}{\Lambda}\right)^2,\\
\CC_1^\pm&={ 2 \ri \sigma_1^\pm \over \left(\xi _1-1\right)^2 \xi _1}\left({G_+(\ri)^2 k^2 \over 32 \pi  \beta _0^2 \mathfrak{c}^2} \left(\frac{m}{\Lambda}\right)^2 \right)^{\xi_1}. 
\ea
\ee
(\ref{bos-ts}) is our final expression for the trans-series expansion of the normalized energy density, displaying two 
different types of 
exponentially small corrections. They lead to two IR renormalon singularities, at $\zeta=2$ and $\zeta=2 \xi_1$. 
Let us note that, as we will see in the next section, 
the coefficients $\CC_{0,1}^\pm$ are simpler than they look, since 
$G_+(\ri)$, $\beta_0$ and $k$ are the ingredients that compute the mass gap $m/\Lambda$ in most models.

Although we have not worked out the details, the trans-series corresponding to UV renormalons could be obtained exactly as we did in the GN model in section \ref{uv-sec}.

\subsection{Results for the different models}
\label{bos-results}
In the previous section we derived a general formula (\ref{bos-ts}) for the trans-series of all bosonic models. We will now 
write it in some detail for each specific model, and we will compare it with the large $N$ results obtained previously in \cite{mmr,dpmss}. This will provide a first, analytic test of our results. 

\vskip .2cm

(i) {\it Non-linear $O(N)$ sigma model}. We consider the choice of charges made in \cite{hmn,hn}. The different parameters characterizing this model are given by
\begin{equation}
\Delta=\frac{1}{N-2},\quad \beta_0 = \frac{1}{4\pi\Delta},\quad \xi = \Delta,\quad \mathfrak{c} = 1,\quad k = \frac{1}{\sqrt{\pi \Delta}},  
\end{equation}
and the relation between the mass gap and the dynamically generated scale in the $\overline{\rm{MS}}$ scheme is given by \cite{hmn,hn}
\be
\frac{m}{\Lambda}=\left(\frac{8}{\re}\right)^{\Delta }\frac{1}{\Gamma (\Delta +1)}.
\ee
The Wiener--Hopf decomposition of the kernel was determined in \cite{hmn,hn}, and one has 
\be
G_+(\omega)=\frac{\re^{-{1 \over 2} \ri \omega [(1-2\Delta)(\log(-\frac{1}{2}\ri \omega)-1)-2\Delta\log(2\Delta)]}}{\sqrt{-\ri \Delta \omega}}\frac{\Gamma(1-\ri \Delta \omega)}{\Gamma\big(\tfrac{1}{2}-\tfrac{1}{2}\ri\omega\big)}. 
\label{Gplus_ON}
\ee
The structure of IR singularities in the Borel transform is mostly determined by the poles of $\sigma(\ri \xi)$, 
in addition to the $\omega=\ri$ pole identified in \eqref{int-poles}. The latter is a singularity at $\xi=1$, 
corresponding to (\ref{slore}) with $\ell=2$. Then, there is a sequence of singularities at 
\be
\xi_\ell= {\ell \over \Delta}, \qquad \ell \in \IZ_{>0}. 
\ee
These are the singularities (\ref{on-insts}). They are suppressed at large $N$, and they have 
the right weight to correspond to the action of an $\ell$-instanton (see e.g. \cite{volin-thesis}). 
By using the ingredients above, we find that the coefficients $\CC_{0,1}^\pm$ in (\ref{2cos}), when $N>3$, are explicitly given by
\begin{equation}
\begin{aligned}
\CC_0^\pm &= -\frac{ \re^{\pm \ri \pi  \Delta  }}{2}\left(\frac{64}{\re^2}\right)^{\Delta }\frac{ \Gamma (1-\Delta )}{ \Gamma (1+\Delta)}, \\
\CC_1^\pm &=  \re^{\mp\frac{\ri \pi }{2 }\left(1+\frac{1}{\Delta}\right)}
 \, 16 \left(2^{\Delta -1} \Delta \right)^{1/\Delta }   \frac{\Gamma \left(\frac{1}{2\Delta }-\frac{1}{2}\right)}{ \re^2\Delta ^2 \Gamma \left(\frac{3}{2}-\frac{1}{2 \Delta }\right)}.
\end{aligned}
\end{equation}
In the large $N$ limit (which corresponds to $\Delta\rightarrow 0$), only the first singularity at $\zeta=2$ survives, and we obtain
\begin{equation}
\label{er-NLSM}
\ba
\frac{e}{\pi\Delta\rho^2} &= \alpha -\frac{1}{2} \re^{-\frac{2}{\alpha} } \alpha ^2+\cdots +\Delta  \left\{ \re^{-\frac{2}{\alpha} } \alpha ^2 \left(\mp\frac{\ri \pi}{2}+\log (\alpha )-\gamma_E +1-3 \log (2)\right)+ \cdots \right\}\\
&\quad+\CO\big(\Delta^2\big).
\ea
\end{equation}
The discontinuity matches the one found with renormalon diagrams in \cite{mmr}. It can be verified that (\ref{er-NLSM}) agrees with 
the result of \cite{dpmss} for the non-linear sigma model. In particular, the exponentially small correction given 
by the second term in the r.h.s. of (\ref{er-NLSM}) agrees with the large $N$ correction found in \cite{dpmss}. This is 
easily seen in (\ref{fh-bos}), where the $h$-independent term in the r.h.s. agrees with the corresponding term 
found in \cite{dpmss}. In addition, from (\ref{F0-gen}) we obtain the free energy when $h=0$:
\begin{equation}
F(0) = \frac{m^2}{8}  \cot (\pi  \Delta ).
\end{equation}
This result extends the result of \cite{bcr} to all orders in $1/N$. It was quoted in \cite{saleur}, and by comparing it with the result for the GN model (\ref{f0GN}) we verified the duality $N-2 \rightarrow 2-N$ between the GN and the NLSM noted in \cite{saleur}. 

An additional check of (\ref{er-NLSM}) can be made by comparing with the numerical 
results of \cite{abbh1,abbh2} when $N=4$. We obtain
\begin{equation}
\frac{e}{\rho^2}= \left(\frac{\pi\alpha}{2}+\cdots\right)
\mp \ri \frac{4\pi}{\re}\re^{-\frac{2}{\alpha}}\alpha
\pm \ri \frac{4\pi}{\re^2}\re^{-\frac{4}{\alpha}}\alpha+\cdots
\end{equation}
which matches their results. In this case, $\sigma_1^ \pm$ is purely imaginary, and the first exponentially small correction with a real part should be of order $\re^{-8/\alpha}$, since it comes multiplied by $(\sigma_1^\pm)^2$. This is what is found in \cite{abbh1,abbh2}.

As for the $O(3)$ non-linear sigma model, we cannot use directly our generic $N$ results. %The function $G_+(\omega)$ in \eqref{Gplus_ON} has different complex analytic properties when $\Delta=1$, in particular $G_-(\ri)\neq 0$, 
A key difference in \eqref{Gplus_ON} when $\Delta=1$ is that $G_+(-\ri)\neq 0$ which
%, through the singularity of $g_+(\omega)$ at $\omega=\ri$, 
adds another set of exponentially 
small corrections of order $\re^{-2 B}$ to \eqref{Q-eq-np} and \eqref{eps-disc} (and, in principle, additional $\re^{-2\ell B}$ contributions). These are structurally similar to the ones that come from the poles of $\sigma(\omega)$, 
and they have a non-trivial series in powers of $1/B$ and $\log B/B$. However, their overall coefficient is purely real and unambiguous. 
To leading order we have
%
%\begin{multline}
%\CF(h) = -h^2\Bigg\{\left(\frac{B}{4\pi}+\frac{\log B-2+2\log 2}{8 \pi }+\cdots\right)\mp \ri\frac{m^2 }{16 \re^2}\\
%- \re^{-2B}\left(\frac{B^2}{\re\pi}  + \frac{B}{4\re \pi}\left(2 \log B-2 \gamma_E +3+6 \log 2\right)+\cdots\right)+\CO\left( \re^{-4B}\right)\Bigg\},
%\end{multline}
\begin{multline}
\CF(h) = -h^2\Bigg\{
\frac{1}{4 \pi  \tilde{\alpha }}\left(1-\frac{\tilde{\alpha}}{2}+\cdots\right)
\\
-\frac{16 \re^{-\frac{2}{\tilde{\alpha }}}}{\pi  \re^2 \tilde{\alpha }^3}
\left(1+\frac{1}{4} \tilde{\alpha } \left(2 \log \tilde{\alpha }-2 \gamma_E +6-10 \log 2\right)+\cdots\right)
+ \CO\left(\re^{-\frac{4}{\tilde{\alpha}}}\right)
\Bigg\}\mp \ri\frac{m^2 }{16},
\end{multline}
where the leading ambiguous imaginary term still comes  from the pole at $\omega=\ri$ in \eqref{int-poles}. There are higher order exponentially small corrections coming from the ambiguous poles of $\sigma(\omega)$ at $\xi = 2k$. The normalised 
energy density can be written as
\begin{multline}
\frac{e}{\pi \rho^2}=\alpha+\frac{\alpha^2}{2}+\cdots\mp \ri\frac{16 \pi }{\re^2}  \re^{-\frac{2}{\alpha }}  \\
+\frac{\re^{-\frac{2}{\alpha }}}{\alpha}\left(\frac{64}{\re^2}+\alpha\frac{32}{\re^2} \left(\log \alpha -\gamma_E-5\log 2+3\right)+\cdots\right)+\CO\left(\re^{-\frac{4}{\alpha}}\right).
\label{e_o3}
\end{multline}
Note that at leading exponential order we get a non-trivial, real-valued series in $\alpha$ and $\log \alpha$, but only a 
single imaginary ambiguous 
term. The large order behaviour of the 
perturbative series is only sensitive to the latter term, missing completely the structure of the real exponentially small correction. 
This shows that in this example it is unlikely that the strong version of the resurgence program applies. 
Another particularity of $N=3$ is that due to the simple form of $\sigma(\omega)$ there are no UV renormalons.% and hence \eqref{e_o3} fully captures the leading asymptotic behaviour of the perturbative series.

\vskip .2cm
\vskip .2cm

(ii) {\it Non-linear supersymmetric $O(N)$ sigma model}. We consider this model in the setting of \cite{eh-ssm} (see also \cite{mr-ren} for additional details). Its parameters and mass gap are given by 
\begin{equation}
\Delta=\frac{1}{N-2},\quad \beta_0 = \frac{1}{4\pi\Delta},\quad \xi = 0,\quad \mathfrak{c} = 1,\quad k = \frac{1}{\sqrt{\pi \Delta}},\quad \frac{m}{\Lambda}=\frac{2^{2 \Delta } \sin (\pi  \Delta )}{\pi  \Delta }. 
\end{equation}
As in \cite{eh-ssm}, we consider $N>4$. The Wiener--Hopf decomposition of the kernel was obtained in \cite{eh-ssm}, and one has
\begin{equation}
G_+(\omega) = \frac{\re^{-\frac{1}{2}\ri(1-2\Delta)\omega[1-\log(-\frac{1}{2}\ri(1-2\Delta)\omega)]} \re^{-\ri \Delta \omega[1-\log(-\ri\Delta\omega)]}}{ \re^{-\ri\omega[1-\log(-\frac{1}{2}\ri\omega)]} \sqrt{-\ri\Delta\omega}}\frac{\Gamma \big(\frac{1}{2}-\frac{1}{2} \ri (1-2 \Delta ) \omega \big) \Gamma (1-\ri \Delta  \omega )}{\Gamma \left(\frac{1}{2}-\frac{1}{2}\ri \omega\right)^2}.
\end{equation}
%\begin{equation}
%G_+(\omega) = \frac{\re^{-\frac{1}{2}\ri\omega [-1-2 \Delta  \log (-\ri \Delta  \omega )-(1-2 \Delta ) \log(-\frac{1}{2}\ri (1-2 \Delta ) \omega) + 2 \log(-\frac{1}{2}\ri \omega)]}}{\sqrt{-\ri\Delta\omega}}\frac{\Gamma \big(\frac{1}{2}-\frac{1}{2} \ri (1-2 \Delta ) \omega \big) \Gamma (1-\ri \Delta  \omega )}{\Gamma \left(\frac{1}{2}-\frac{1}{2}\ri \omega\right)^2}.
%\end{equation}
%\mmm{Please add the equation for $G_+(\omega)$}

 The position of the IR singularities 
can be deduced from the poles of $\sigma(\ri \xi)$, and one finds two different sequences:
\be
\label{12sson}
\xi_\ell= {\ell \over 1-2\Delta}, \qquad \xi'_\ell= {\ell \over \Delta}, \qquad \ell \in \IZ_{>0}.
\ee
These correspond to the sequences (\ref{corrected}) and (\ref{on-insts}), respectively. The first sequence is similar to the 
non-conventional IR singularities found in the GN model, We expect these two sequences to 
mix as we calculate non-perturbative corrections, so that the generic singularity occurs at 
\be
\label{mix-seq}
{\ell_1 \over 1-2\Delta}+{\ell_2 \over \Delta}, \qquad \ell_1, \ell_2 \in \IZ_{>0}.
\ee
Since the second sequence in (\ref{12sson}) is suppressed at large $N$ we will focus on the first sequence. 
From the residues at the leading singularities, we find, 
\begin{equation}
\label{susy-cs}
\begin{aligned}
\CC_0^{\pm}&=0, \\
\CC_1^\pm &=  \re^{\mp\frac{\ri \pi}{2}\left(1+\frac{1}{1-2\Delta}\right)}\frac{   \pi  2^{\frac{1}{1-2 \Delta }} (1-2 \Delta )^{\frac{2-2 \Delta }{1-2 \Delta }} }{4 \Delta  \Gamma \big(\frac{\Delta }{2 \Delta -1}\big)^2\sin \big(\frac{\pi -\pi  \Delta }{2 \Delta -1}\big)}.
\end{aligned}
\end{equation}
Therefore, the conventional IR singularity at $\zeta=2$ is absent in this model. 
This clarifies the difficulties found with the numerical analysis of this example in \cite{mr-ren}, at finite $N$. The first singularity in the Borel plane is located at $\zeta=2 \xi_1$, i.e. 
\be
\label{zeta-ab}
\zeta=\frac{2}{1-2\Delta}.
\ee
In the large $N$ limit, this singularity moves to $\zeta=2$ and one finds
\begin{equation}
\ba
\frac{e}{\pi\Delta\rho^2}&=\alpha+\re^{-\frac{2}{\alpha} } \left(-\frac{\alpha ^2}{2}+\CO\left(\alpha ^3\right)\right)+\re^{-\frac{2}{\alpha} }
 \Delta  \left(2 \big(\alpha +\CO\big(\alpha^2\big)\big)\pm\frac{\ri\pi}{2}\big(\alpha^2 +\CO\big(\alpha^3\big)\big)\right)\\
 &
\quad +\CO\left(\Delta ^2\right).
\ea
\end{equation}
Once again, the discontinuity matches the one found with renormalon diagrams in \cite{mmr}. 
As in the case of the GN model, this shows that the unconventional IR singularity at (\ref{zeta-ab}) is indeed of the 
renormalon type, since at large $N$ it encodes the factorially divergent sequence of renormalon ring diagrams studied in \cite{mmr}. 

In this case, the general formula (\ref{F0-gen}) gives $F(0)=0$. 

\vskip .2cm
(iii) {\it Principal chiral field}. We consider the PCF in the setting discussed in \cite{pcf}. One has, 
\begin{equation}
\Delta=\frac{1}{N},\quad \beta_0 = \frac{1}{16\pi\Delta},\quad \xi = \frac{1}{2},\quad \mathfrak{c} = 4,\quad k = \frac{1}{\sqrt{2 \pi  (1-\Delta ) \Delta }},\quad \frac{m}{\Lambda}=\sqrt{\frac{8 \pi }{\re}}\frac{ \sin (\pi  \Delta )}{\pi  \Delta }.
\label{PCFpars}
\end{equation}
The Wiener--Hopf decomposition of the kernel was obtained as well in \cite{pcf}, giving
\begin{equation}
G_+(\omega)=\frac{\re^{- \ri \omega  [-(1-\Delta) \log (1-\Delta)-\Delta  \log (\Delta )]}}{\sqrt{-2 \pi\ri (1-\Delta ) \Delta \omega } }\frac{\Gamma (1-\ri (1-\Delta ) \omega ) \Gamma (1-\ri \Delta  \omega ) }{\Gamma (1-\ri \omega )}.
\end{equation}
%\mmm{Please add the equation for $G_+(\omega)$}

As in previous examples, there is a conventional IR singularity 
at $\zeta=2$, which was already detected in \cite{mr-ren}. The position of the other 
IR singularities is determined by the poles of $\sigma(\ri \xi)$, and one finds a situation 
very similar to the one in the %supersymmetric non-linear sigma model, with two different sequences of poles:
Gross-Neveu model, with a sequence of unconventional renormalons:
\be
\label{xiz-pcf}
\xi_\ell= {\ell \over 1-\Delta}, %\qquad \xi'_\ell= {\ell \over \Delta}, 
\qquad \ell \in \IZ_{>0}.
\ee
These correspond to the sequence (\ref{corrected}). In principle, a distinct sequence $\xi'_\ell= {\ell \over \Delta}$, corresponding to (\ref{pcf-insts}), can appear for general real $\Delta$, similar to the supersymmetric non-linear sigma model. However, at integer $N$ these additional contributions disappear.
% We expect these two sequences to 
%mix, as in (\ref{mix-seq}), so that the generic singularity occurs at 
%
%\be
%{\ell_1 \over 1-\Delta}+{\ell_2 \over \Delta}, \qquad \ell_1, \ell_2 \in \IZ_{>0}.
%\ee
%
For $N\ge 2$, the next-to-leading IR singularity occurs at 
\be
\label{nPCF}
\zeta={2 \over 1-\Delta}. 
\ee
The coefficients in (\ref{bos-ts})\footnote{The following expressions are only valid for $N>2$. For $N=2$, \eqref{2cos} holds but does not correspond to \eqref{pcf-cs}, instead we get the same results as the non-linear sigma model with $N=4$, as expected.} are given by,
\begin{equation}
\label{pcf-cs}
\ba
\CC_0^\pm&=\mp \ri \frac{2}{\re (1-\Delta ) \Delta },\\
\CC_1^\pm &= \pm \ri \frac{2 \Gamma \big(\frac{\Delta }{1-\Delta }\big)}{\re^{\frac{1}{1-\Delta }} (1-\Delta ) \Gamma \big(\frac{1}{1-\Delta }\big)}.
\ea
\end{equation}
In this case, both coefficients are purely imaginary. In the large $N$ limit, the unconventional renormalon at (\ref{nPCF}) 
moves to $\zeta=2$, where it combines with the conventional renormalon, and one finds, 
\begin{equation}
\frac{e}{2\pi\rho^2}= \Delta  \left(\alpha +\cdots \mp \frac{4\ri}{\re}  \re^{-\frac{2}{\alpha }} + \cdots \right)+\CO\left(\Delta ^2\right). 
\end{equation}
This matches the large $N$ result of \cite{dpmss}, which was obtained from the 
large $N$ limit of the Bethe ansatz integral 
equations. Note that the infinite sequence of IR renormalons found for this model at 
large $N$ in \cite{dpmss} is in fact due to the unconventional sequence of IR singularities located at $\zeta=2\xi_\ell$, where $\xi_\ell$ is given in (\ref{xiz-pcf}). 

We can also inspect the $SU(N)$ principal chiral field with a different choice of conserved charges coupled to $h$, as is discussed in \cite{fkw1,fkw2,ksz,mr-ren}. In this setting, the kernel changes so we must use
\begin{equation}
G_+(\omega) = 2^{\ri \Delta  \omega }\frac{(1-\ri \omega )}{\sqrt{-\ri \omega }} \frac{  \Gamma (1-\ri \Delta  \omega )}{ \Gamma \left(1-\frac{\Delta}{2}-\frac{\ri   \Delta \omega}{2}\right) \Gamma \left(1+\frac{\Delta}{2}-\frac{\ri\Delta\omega}{2} \right)},\quad k= \frac{2 \sin \left(\frac{\pi  \Delta }{2}\right)}{\pi  \Delta },
\end{equation}
while $\Delta$, $\beta_0$, $\xi$ and $m/\Lambda$ remain the same as in \eqref{PCFpars}. 
The first notable change is that the analytic structure of $\sigma(\omega)$ changes and we only 
have the singularities
\be
\xi_\ell= N\ell,
\ee
corresponding to \eqref{pcf-insts}, while the sequence \eqref{xiz-pcf} is absent. This means that 
at large $N$ only the conventional IR singularity survives. Due to the presence of 
multiple particles, the definition of the free energy in the setting of \cite{fkw1,fkw2} is now
\begin{equation}
\CF(h) = - \frac{m}{8\sin(\pi\Delta)^2}\int_{-B}^B \epsilon(\theta)\cosh\theta\rd\theta = - \frac{m\re^B }{8\sin(\pi\Delta)^2} \epsilon_+(\ri),
\end{equation}
and we find
\begin{equation}
\CF(h) = - \frac{h^2 \log\frac{h}{m}
}{16\pi\Delta^2\cos\left(\frac{\pi\Delta}{2}\right)^2}
\left\{
1
\mp
\ri
\left(\frac{h^2}{m^2}
\log\frac{h}{m}\right)^{-\frac{1}{\Delta }}
%\left(\frac{\pi  \Delta ^2}{2\sin^2 \left(\frac{\pi  \Delta }{2}\right)}\right)^{\frac{1}{\Delta }} 
\frac{\left(\pi\Delta^2/2\right)^{1+\frac{1}{\Delta}}}{\sin^{2/\Delta} \left(\frac{\pi  \Delta }{2}\right)}
+ \cdots\right\} 
\mp  \frac{\ri m^2}{8\sin ^2(\pi  \Delta )} .
\label{FKWfiniteN}
\end{equation}
We can compare the large $N$ limit of \eqref{FKWfiniteN} with the results of \cite{fkw1,fkw2,ksz}. When $\Delta\rightarrow 0$, it becomes
\begin{equation}
\Delta^2\CF(h) = -\frac{h^2}{16\pi}\left\{\log\frac{h}{m}+\cdots\right\}\mp \ri \frac{m^2 }{8\pi^2},
\label{FKWlargeN}
\end{equation}
where $\cdots$ represents the perturbative series. This series was computed exactly 
and explicitly in \cite{fkw1}. It is simple to check that the imaginary part of the Borel 
resummation of this perturbative series is such that it cancels exactly the ambiguous term in 
\eqref{FKWlargeN}, while the real part matches the exact formula for the free energy at large $N$.
This provides further confirmation that the exponentially suppressed terms we have obtained yield 
in fact the correct trans-series. 

We finally note that the general formula (\ref{F0-gen}) gives $F(0)=0$ for the $SU(N)$ PCF. 

\subsection{Testing the analytic results}

The results of the previous section provide many predictions for the resurgent structure of the perturbative series of $\CF(h)$ in three 
different bosonic models. In this section we give numerical evidence for these predictions. In particular, as in the GN model, we want to establish beyond any reasonable doubt the presence of unconventional IR singularities. One tool that we will use in all examples is the following: in order to display the subleading singularity, it is convenient to extract the effect of the leading IR renormalon 
from the perturbative series. Similarly to the auxiliary sequence (\ref{seq_no_IR}) for the GN model, this is done by looking at
\begin{equation}
\bar{e}_m =  e_m - 2^{1-2\xi-m}\left(\frac{\CC_0^--\CC_0^+}{2\pi\ri}\right)\Gamma\left(m+2\xi-1\right),
\label{seq_no_IR_bosonic}
\end{equation}
where $e_m$ are the coefficients of the series $\varphi(\alpha)$ in \eqref{bos-ts}. Let us now test the results for the different models. 

\vskip .2cm
(i) {\it Non-linear $O(N)$ sigma model}. The first analytic prediction we want to test is the value of the Stokes constant for the 
first IR singularity. To do this, we consider the leading behavior of the discontinuity 
\be
\label{g0}
g(\alpha)={1\over  2\pi\ri} \re^{\frac{2}{\alpha}} \left(\frac{\alpha}{2}\right)^{2\Delta-1} \text{disc} \, s(\varphi)(\alpha). 
\end{equation}
According to our analytic results, when $\alpha \rightarrow 0$ this converges to 
\be
\label{anpred-s}
{ 2^{-2\Delta}\over  \pi} \mS_0, 
\ee
where 
\be
\label{stokeson}
\mS_0=-\ri \left(\CC^-_0-\CC^+_0\right)=
\frac{  64^{\Delta } \re^{-2 \Delta } \pi \Delta }{\Gamma (\Delta +1)^2}
\ee
is the corresponding Stokes constant. The function (\ref{g0}) can be calculated numerically from the perturbative series, for 
small values of $\alpha$, and then extrapolated to $\alpha \rightarrow 0$ using higher Richardson transforms, for various values of $N$. In Table \ref{ONasym} 
we compare this numerical extrapolation to the analytic prediction (\ref{anpred-s}), (\ref{stokeson}) for $5 \le N\le 12$, the last stable digit is underlined. As one can see, 
the agreement is excellent\footnote{In fact, we were able to guess the analytic form of $\mS_0$ from the numerical results.}. An equivalent test can be made for \eqref{e_o3}, where it is easier to simply inspect the large order behaviour of $e_m$ since there are no UV effects, and we find an agreement to 45 digits.

\begin{center}
\renewcommand{\arraystretch}{1.25}
\begin{tabular}{l l | l}
$N$ & $g(\alpha\rightarrow 0)$ numeric & $2^{-2\Delta} \mS_0/\pi$\\
\hline
 5 & 0.540803419\underline 5 & 0.54080341956810599941 \\
 6 & 0.369131065\underline 2 & 0.36913106521716834229 \\
 7 & 0.27687974\underline 8 & 0.27687974850323657588 \\
 8 & 0.220260880\underline 6 & 0.22026088067984630698 \\
 9 & 0.18230784282\underline 8 & 0.18230784282845334782 \\
 10 & 0.155234006\underline 2 & 0.15523400619586855091 \\
 11 & 0.135011430\underline 7 & 0.13501143072958418153 \\
 12 & 0.1193635574\underline 9 & 0.11936355749143292096 \\
 \label{ONasym}
\end{tabular}
\end{center}

\begin{figure}
\center
\includegraphics[width=0.45\textwidth]{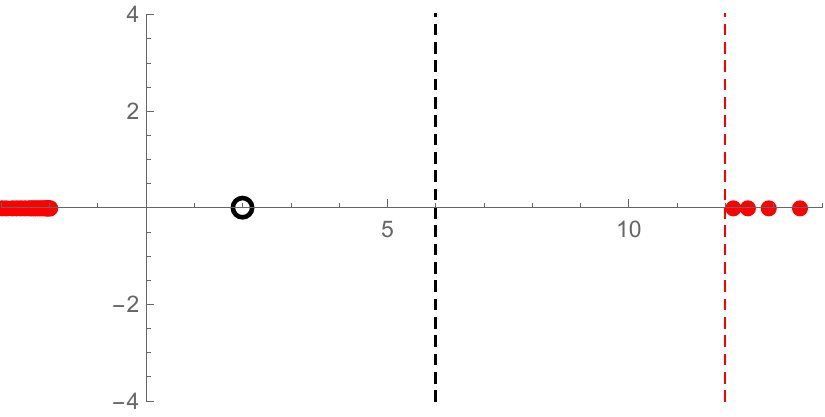} \qquad 
\includegraphics[width=0.45\textwidth]{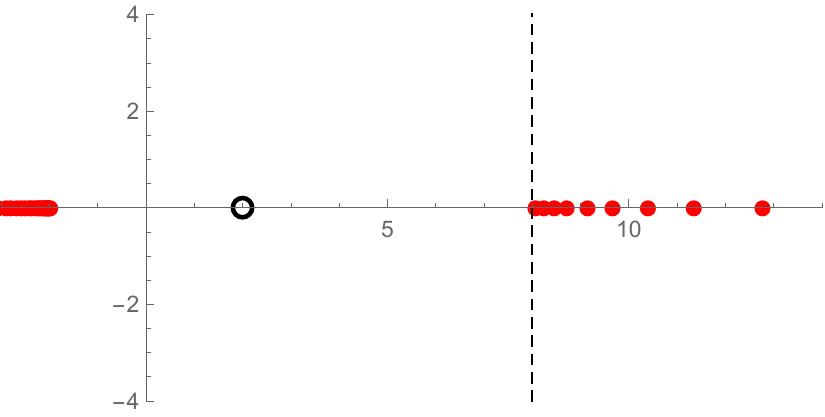}
\caption{The poles of the Borel-Padé approximant of the series $\bar{e}_m$ in \eqref{seq_no_IR_bosonic}, for the $O(N)$ non-linear sigma model, truncated at 120 terms. The plots correspond to $N=5$ (left) and  
$N=6$ (right). The leftmost dashed line (black) indicates the predicted position of the pole $\zeta = 2(N-2)$, the rightmost dashed line (red) is at $\zeta = 4(N-2)$ and the black circle indicates the position of the removed IR singularity at $\zeta=2$. For $N=5$ the first singularity occurs at $\zeta = 4(N-2)$. %Note that since the residues $\rho_n$ are 0 for $N=5$ there are no poles. The imaginary poles are numerical artifacts and are dislodged by taking more coefficients.
}
\label{fig-borel-poles-ON}
\end{figure}

According to our analysis, the next IR singularities occur at the sequence (\ref{on-insts}). Due to the factor of $N-2$ the corresponding exponential corrections are very suppressed for, say, $N\ge 5$, therefore they are relatively difficult to pinpoint. Still, 
they can be seen quite clearly as singularities of the Borel--Pad\'e approximant of the series $\bar{e}_m$. In 
\figref{fig-borel-poles-ON} we show this for $N=5, 6$. When $N=5$ the first singularity at $\zeta = 2(N-2)$ is absent, since the prefactor $\sigma_1^\pm$ vanishes, but we can clearly see the next singularity at $\zeta=4(N-2)=12$. When $N=6$, the singularity at $\zeta=2(N-2)=8$ is also apparent. %However, due to the strong exponential suppression, our tests of the value of $\CC_1^\pm$ have low precision. 

\begin{figure}
\center
\includegraphics[width=0.65\textwidth]{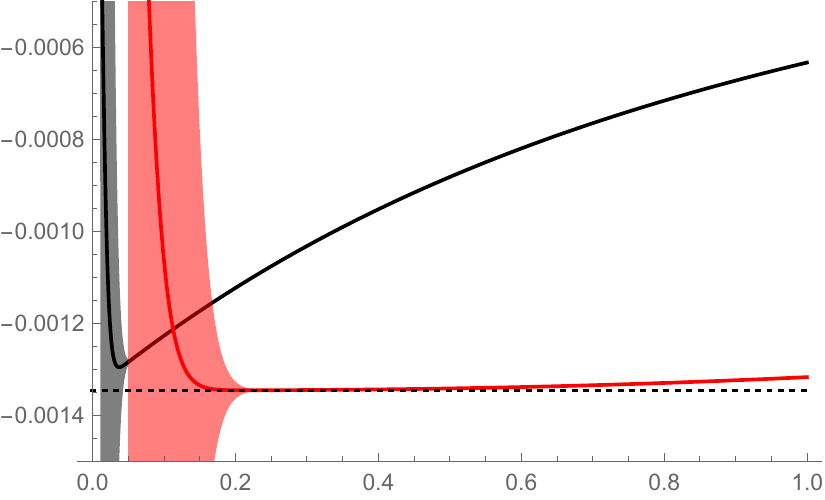} %\qquad 
\caption{Plot of the function $f(\alpha)$ in \eqref{f1_bos} for the $O(N)$ non-linear sigma model with $N=6$ (black), as well as its respective second Richardson transforms (red). The dashed line is the predicted asymptotic value. The shaded areas correspond to error estimates from the convergence of the Padé approximant.}
\label{fig-f1-ON}
\end{figure}

In order to obtain a quantitative test of the coefficient $\CC_1^\pm$ in (\ref{2cos}), we consider the analogue of the 
quantity (\ref{f1def_p}) in the GN model. For a general bosonic model, we define
\begin{equation}
f(\alpha)= \frac{1}{2\pi\ri} \re^{\frac{2\xi_1}{\alpha}}\alpha^{(2\xi-1)\xi_1}\left(\text{disc}\,s(\varphi)(\alpha)- \big(\CC_0^--\CC_0^+\big)\re^{-\frac{2}{\alpha}}\alpha^{1-2\xi}\right). 
\label{f1_bos}
\end{equation}
Its asymptotic behavior at small $\alpha$ is 
\be
f(\alpha) \sim  \frac{1}{2\pi\ri} \big(\CC_1^--\CC_1^+\big), \qquad \alpha \rightarrow 0. 
\ee
We plot $f(\alpha)$ for small values of $\alpha$ with $N=6$ in Figure \ref{fig-f1-ON}. With 120 coefficients in the perturbative series we find agreement to 6 digits.

\begin{figure}
\center
\includegraphics[width=0.55\textwidth]{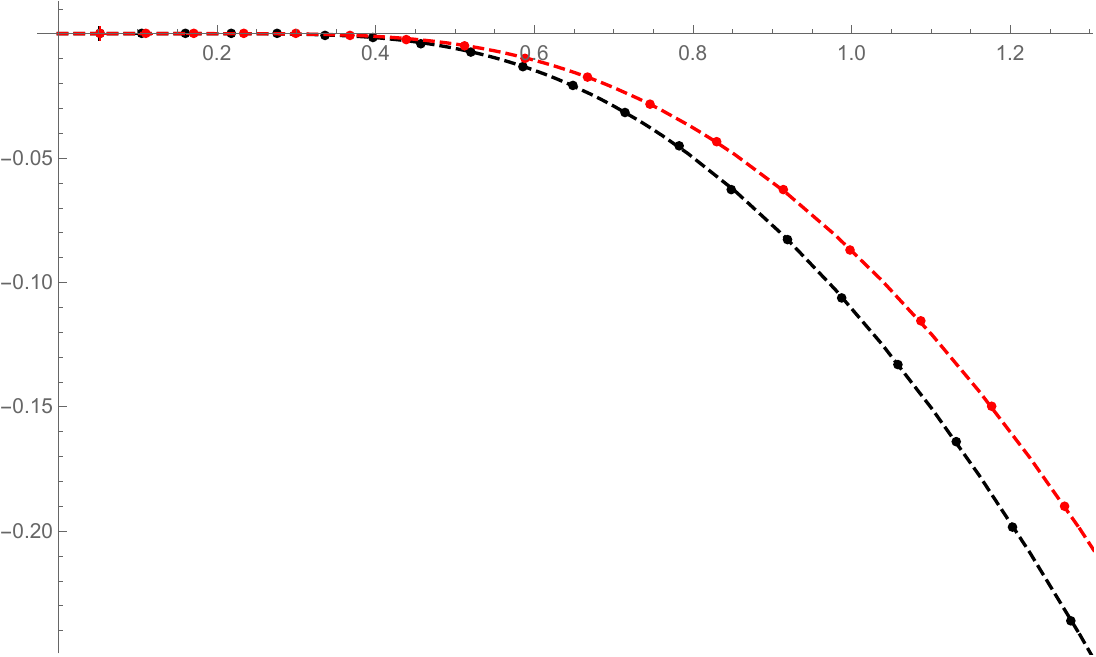}
\caption{In this figure we plot the difference between the normalized energy density and the real part of the Borel resummation of the perturbative series, against the theoretical prediction \eqref{realON} for the $O(N)$ sigma model with $N=5$ (black) and $N=6$ (red). The $x$-axis is the value of $\alpha$. The dots are the numerical calculations of the difference, using a discretisation of 50 points in the integral equation and 90 coefficients in the perturbative series, evaluated at $B=20/k$ for $k=1,\dots,20$. The dashed line is 
the theoretical prediction. 
}
\label{fig-real-ON}
\end{figure}

We can also test the real part of the trans-series parameters, as we did for the Gross--Neveu case. We predict, for the non-linear sigma model, that
\be
\label{realON}
{e\over\rho^2}- \frac{\alpha}{k^2}\rm{Re}\left(s_\pm (\varphi)(\alpha) \right) \sim 
\mR_0 \, \re^{-\frac{2}{\alpha}}\alpha^{2-2\Delta} 
%+ \mR_{1} \, \re^{-\frac{2}{\Delta\alpha}}\alpha^{\frac{2}{\Delta}-2} 
+ \CO\left(\re^{-\frac{4}{\Delta\alpha}}\right), 
\ee
where
\begin{equation}
\ba 
\mR_0 &= \frac{\CC_0^++\CC^-_0}{2k^2} = -\pi ^2\left(\frac{8}{\re}\right)^{2\Delta}\frac{ \cot (\pi  \Delta )}{2\Gamma (\Delta )^2},%\\\mR_1 &= \frac{\CC_1^++\CC^-_1}{2k^2}.
\ea
\end{equation}
There is no contribution from the first pole because
\begin{equation}
\frac{\CC_1^++\CC^-_1}{2k^2}=0 \quad \text{if }\, N\in\IN_{\geq 4}.
\end{equation}
We plot some values of the l.h.s of \eqref{realON} for $N=5,6$ against the prediction of the r.h.s. in Figure \ref{fig-real-ON}. We match $\mR_0$ with 18 digits of agreement. Because the trans-series terms of order $\re^{-\frac{4}{\Delta\alpha}}$ are very exponentially suppressed, we do not have enough precision to discern their effect. We also inspected the real part of \eqref{e_o3} with a similar strategy, finding the correct exponential behaviour to leading and subleading  power of $\alpha$ and matching the coefficients with relative error of $10^{-3}$.% (the presence of $\log\alpha$ terms in the series of this particular case makes the tests slower to converge when compared with the others).

\vskip .2cm

\begin{figure}
\center
\includegraphics[width=0.45\textwidth]{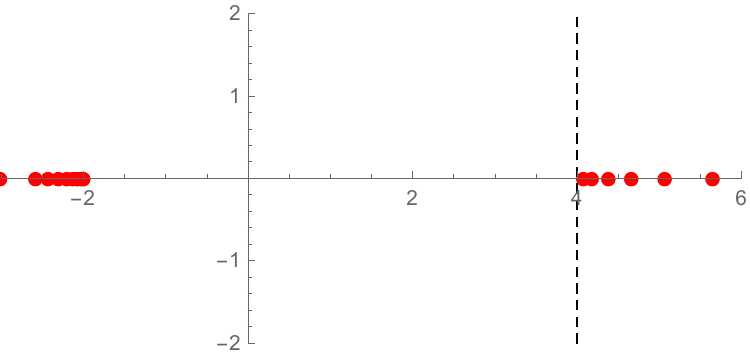} \qquad 
\includegraphics[width=0.45\textwidth]{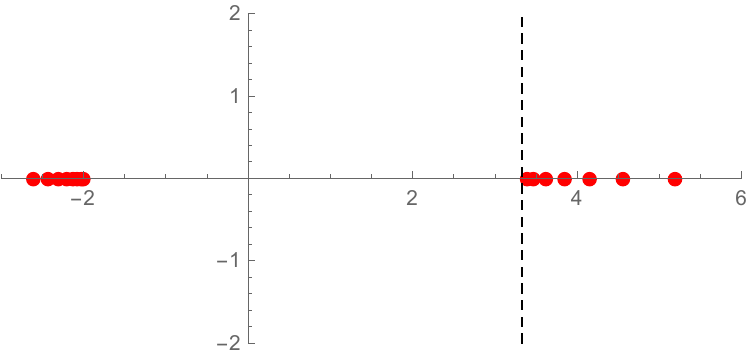}
\caption{The poles of the Borel-Padé approximant of the series $e_m$ truncated at 65 terms. The plots correspond to the $\mathcal{N}=1$ supersymmetric $O(N)$ non-linear sigma model with
$N=6$ (left) and  
$N=7$ (right). The dashed line (black) indicates the predicted position of the pole $\zeta = 2/(1-2\Delta)$. Note that for this model $\CC^\pm_0=0$ so there is no IR renormalon singularity at $\zeta=2$. %Note that since the residues $\rho_n$ are 0 for $N=5$ there are no poles. The imaginary poles are numerical artifacts and are dislodged by taking more coefficients.
}
\label{fig-borel-poles-SUSY}
\end{figure}

\begin{figure}
\center
\includegraphics[width=0.45\textwidth]{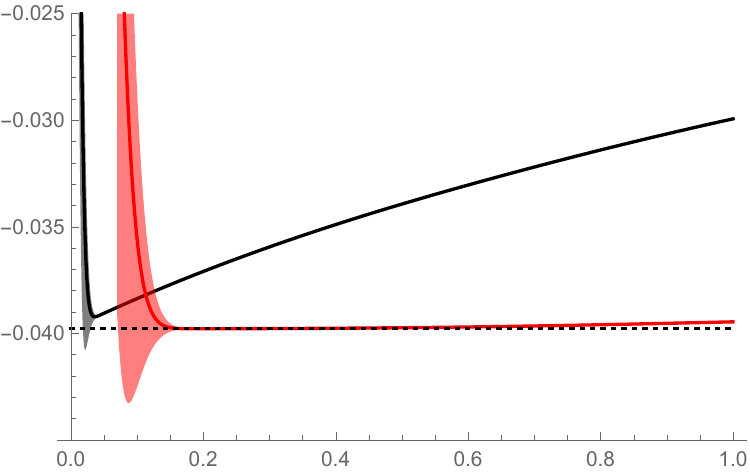} \qquad 
\includegraphics[width=0.45\textwidth]{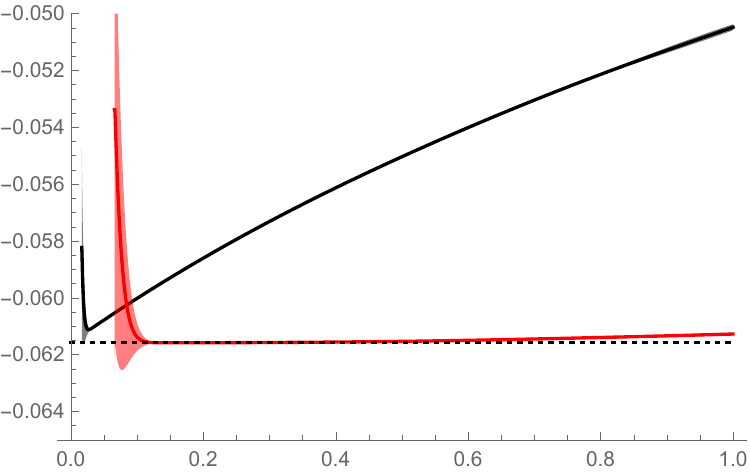}
\caption{Plot of the function $f(\alpha)$ in \eqref{f1_bos} for the $\mathcal{N}=1$ supersymmetric 
$O(N)$ non-linear sigma model with $N=6$ (left, black) and $N=7$ (right, black) as well as their 
respective second Richardson transforms (red). The dashed line is the predicted asymptotic value. 
Note that for this model $C^\pm_0=0$ so this is the leading exponentially small correction. The shaded 
areas correspond to error estimates from the convergence of the Padé approximant.}
\label{fig-f1-SUSY}
\end{figure}

(ii) {\it Supersymmetric non-linear $O(N)$ sigma model}. In this case, the conventional IR singularity at $\zeta=2$ is absent. The unconventional renormalon at (\ref{zeta-ab}) can be clearly seen in the poles of the Borel--Pad\'e approximant of the series $e_m$, as shown in \figref{fig-borel-poles-SUSY} in the cases $N=6,7$. %
Like before, given enough terms in the perturbative series $\varphi (\alpha)$, we can study $\CC_1^\pm$ in (\ref{susy-cs}) 
numerically for very small values of $\alpha$ of \eqref{f1_bos}, and compare the predicted value with the numerical extrapolation. This is shown graphically in \figref{fig-f1-SUSY}, for $N=6,7$. The agreement is again excellent, achieving 6 digits of precision with only 68 coefficients in the perturbative series. 
%\mmm{How many digits of agreement?} 
This provides a clear test of the result for the trans-series (\ref{bos-ts}) in the case of the supersymmetric 
non-linear sigma model. We can also test the real part of the coefficient $\CC_1^\pm$ by comparing the 
Borel resummation with the normalized energy density. We find an agreement of 4 digits  with the predicted value for $N=7,8$, 
 using again only 68 coefficients in the perturbative series.%, which requires 8 digits of precision in the difference between the integral equation and the Borel resummation.

\begin{figure}
\center
\includegraphics[width=0.45\textwidth]{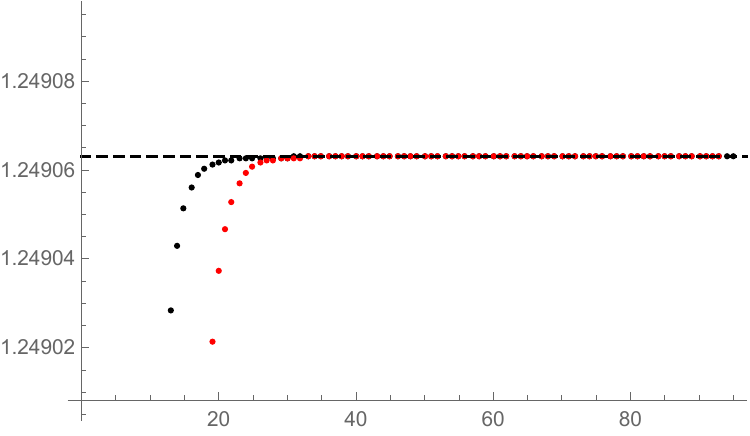} \qquad 
\includegraphics[width=0.45\textwidth]{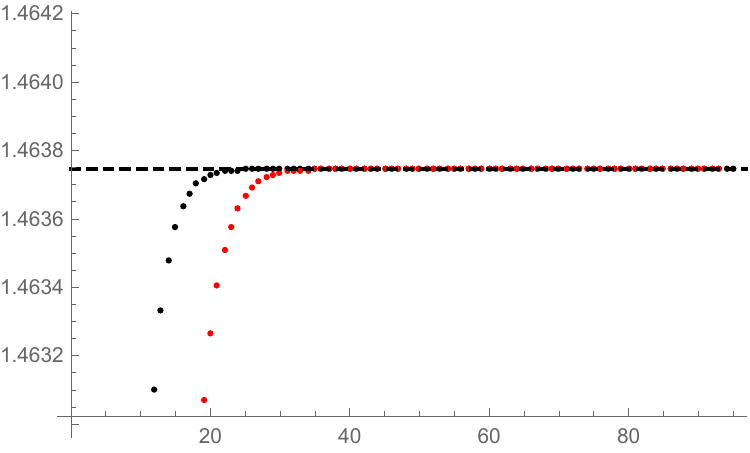}
\caption{Plot of sequence $\tilde{s}_k$ in \eqref{Spcf_p} for the PCF model with $N=4$ (left, black) and $N=5$ (right, black) as well as their respective second Richardson transforms (red). The dashed line is the predicted asymptotic value .
}
\label{fig-asym-beh-PCF}
\end{figure}

\vskip .2cm
(iii) {\it Principal chiral field}. Let us finally test our analytic results for the trans-series in the case of the PCF. 
First of all, we can improve on the results of \cite{mr-ren} and test the Stokes constant associated to the first 
IR singularity at $\zeta=2$. To do this, we consider the sequence 
\begin{equation}
\tilde{s}_k = \frac{2^{2m-1} e_{2m}}{\Gamma(2m)}+\frac{2^{2m} e_{2m+1}}{\Gamma(2m+1)},
\label{Spcf_p}
\end{equation}
which as explained in \cite{mr-ren} removes at leading order the effect of the UV renormalon. By using the value of $\CC_0^\pm$ in (\ref{pcf-cs}), we find the following prediction for the asymptotic value at large $k$,  
\be
\tilde s_k \sim \frac{2}{\re \pi  (1-\Delta ) \Delta }, \qquad k \gg 1. 
\ee
This can be tested by using the perturbative series $\varphi(\alpha)$ and its Richardson transforms. 
A verification for $N=4,5$ is shown in \figref{fig-asym-beh-PCF}. We find 12 digits of agreement for $N=4$, and 10 digits for $N=5$, by using 200 terms of the perturbative series or, equivalently, 99 terms of the sequence $\tilde s_k$.
%\mmm{How many terms were used? How many digits of precision agree?}

\begin{figure}
\center
\includegraphics[width=0.45\textwidth]{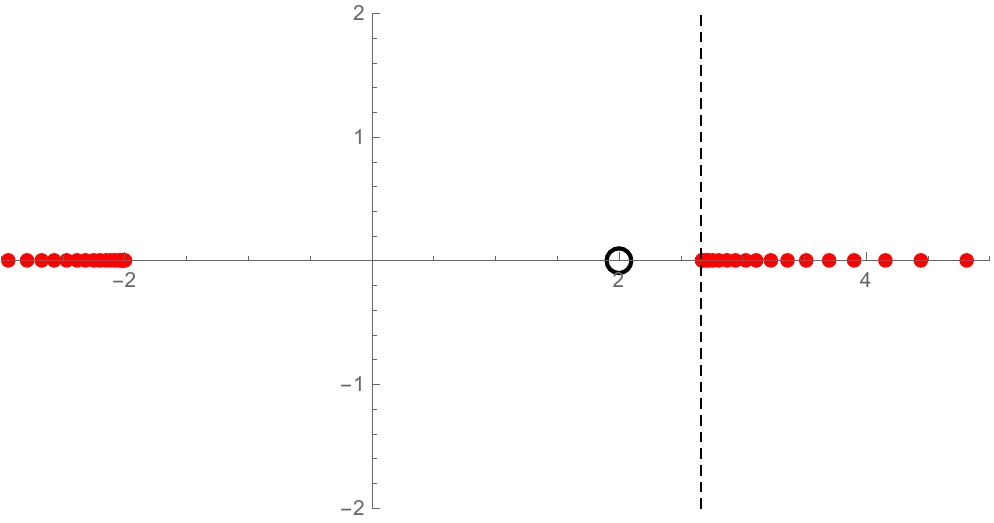} \qquad 
\includegraphics[width=0.45\textwidth]{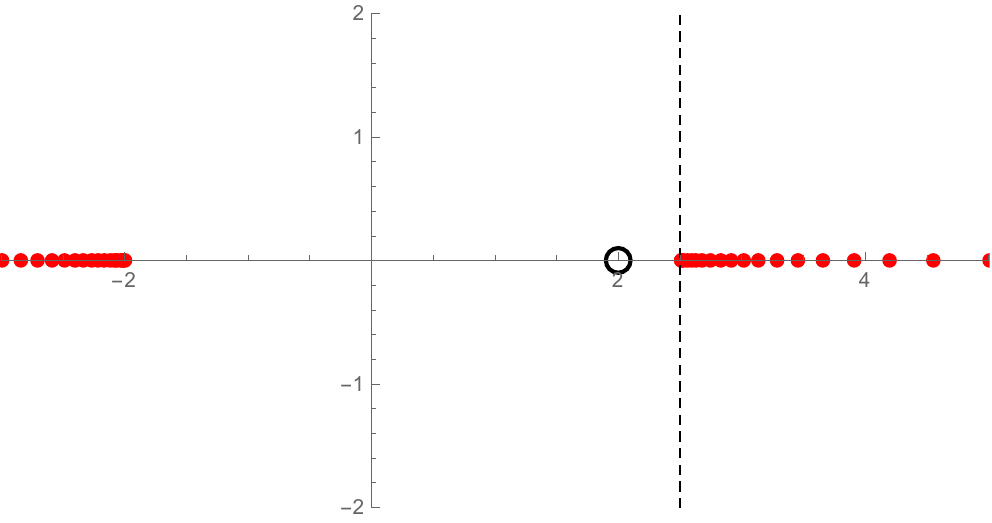}
\caption{The poles of the Borel-Padé approximant of the series $\bar{e}_m$ \eqref{seq_no_IR_bosonic} for the PCF, truncated at 120 terms. The plot on the left corresponds to $N=4$ (left) while the one in the right corresponds to
$N=5$. The dashed line indicates the predicted position of first unconventional renormalon singularity $\zeta = 2/(1-\Delta)$, 
and the black circle indicates the position of the removed IR singularity at $\zeta=2$. %Note that since the residues $\rho_n$ are 0 for $N=5$ there are no poles. The imaginary poles are numerical artifacts and are dislodged by taking more coefficients.
}
\label{fig-borel-poles-PCF}
\end{figure}

\begin{figure}
\center
\includegraphics[width=0.45\textwidth]{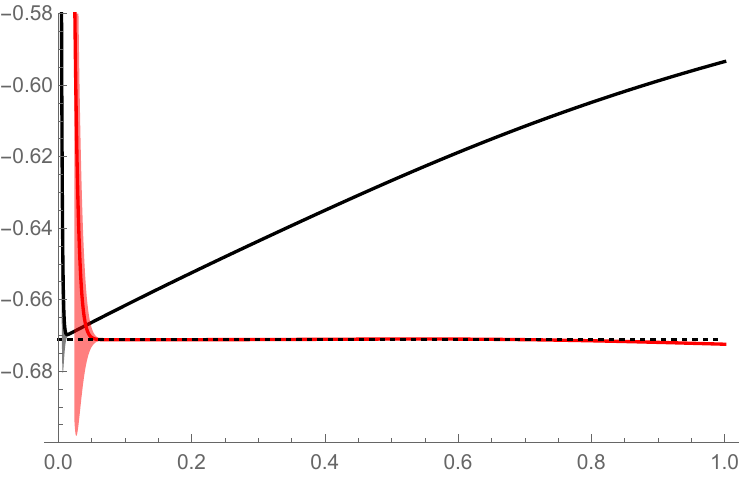} \qquad 
\includegraphics[width=0.45\textwidth]{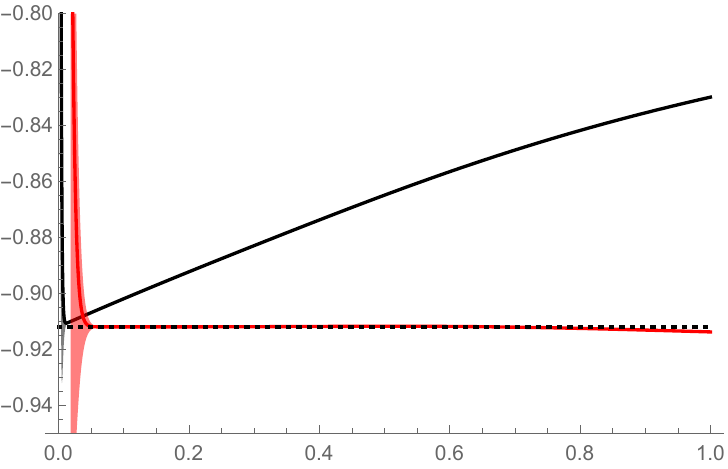}
\caption{Plot of the function $f(\alpha)$ defined in \eqref{f1_bos} for the PCF with $N=4$ 
(left, black) and $N=5$ (right, black), as well as their respective second Richardson 
transforms (red). The dashed line is the predicted asymptotic value 
$(\CC^-_1-\CC^+_1)/(2\pi\ri)$. The shaded areas correspond to error estimates from 
the convergence of the Padé approximant.
}
\label{fig-f1-PCF}
\end{figure}

We can now test the location and the Stokes constant of the first unconventional renormalon at (\ref{nPCF}). 
As in the examples above, the location can be verified from the poles of the Borel--Pad\'e approximant. Two examples are shown in \figref{fig-borel-poles-PCF}, for $N=4,5$.  
To test the Stokes constant of this singularity, we consider again the function (\ref{f1_bos}). In \figref{fig-f1-PCF} we show the numerical extrapolation of this function for $N=4,5$, as compared to the predicted asymptotic 
value $(\CC^-_1-\CC^+_1)/(2 \pi \ri)$. The agreement is again very good, matching the 
coefficient to 6 digits. 

Since $\CC_{0,1}^\pm$ are purely imaginary, we expect the real part of the Borel summation to match $e/\rho^2$, 
up to the effects of the next-to-next-to-leading renormalon sector at $2 \xi_1$ (see (\ref{xiz-pcf})). 
We checked that the difference between the real part of the Borel resummation and the numerical 
calculation of the normalised energy density agrees with $\CC_0^++\CC_0^-$ and 
$\CC_1^++\CC_1^-$ being zero to 6 and 4 digits of precision, respectively. This difference is 
exponentially suppressed beyond order $\re^{-\frac{2\xi_1}{\alpha}}$.
%\mmm{How many digits of precision agree?}

We believe that these tests 
provide very clear evidence for the unconventional renormalon predicted from our analytic formulae.


\sectiono{Trans-series and renormalons in the Gaudin--Yang model}
\label{sec-nr}

The Gaudin--Yang (GY) model \cite{gaudin, yang} describes non-relativistic spin 
$1/2$ fermions interacting through a delta function potential. It can be regarded both 
as an exactly solvable model for a Luttinger liquid, and as toy model for a superconductor. 
We refer to \cite{guan-review} for a review, and to \cite{mr-long} for additional background.
 
We will focus on the basic observable of this model, namely the normalized ground state energy 
density $e(\gamma)$ as a function of the dimensionless coupling constant $\gamma$. This observable can 
be calculated exactly with the Bethe ansatz, and at the same time it has a weak coupling expansion 
as a power series in $\gamma$. The exact answer is obtained from Gaudin's integral equation for the 
density of Bethe roots, which can be written as 
\be
\label{gaudin-int}
{f(x) \over 2} +{1\over 2 \pi} \int_{-B}^B {f(y) \over (x-y)^2+1} \rd y  =1, \qquad -B<x<B. 
\ee
The endpoint of the interval, $B$, is related to the coupling $\gamma$ by
\be
\label{gB}
{1\over \gamma}={1\over \pi} \int_{-B}^B f(x) \rd x, 
\ee
while $e(\gamma)$ is given by
\be
\label{egam-ex}
e(\gamma)= -{\gamma^2\over 4}+ \pi^2 {\int_{-B}^B f(x) x^2 \rd x \over \left(\int_{-B}^B f(x)  \rd x\right)^3}. 
\ee
Note that $e(\gamma)$ depends on $\gamma$ through $B$, and the 
weak coupling limit $\gamma \rightarrow 0$ corresponds to the limit 
of large $B$, as usual in this type of problems. In \cite{mr-ren, mr-ll}, the perturbative 
series $e(\gamma)$ was obtained up to very high order. It was found numerically that 
it is factorially divergent, 
and that its first singularity in the Borel plane is located at 
\be
\zeta= \pi^2. 
\ee
We will now deduce this result analytically, directly from the integral equation (\ref{gaudin-int}), by applying
Wiener--Hopf techniques similar to the ones we have used in the previous sections. 
These techniques were applied to the Gaudin--Yang model in \cite{tw1,tw2}, in order to extract the 
perturbative piece of $e(\gamma)$.  

We first note that we can write Gaudin's integral equation (\ref{gaudin-int}) as in (\ref{ft-ie}), i.e. in the form
\be
\label{int-eq-wh}
{1\over G_+(\omega) G_-(\omega)} \tilde f (\omega) = 
\tilde g(\omega)+ \re^{\ri B \omega} G_+^{-1}(\omega) Q_+ (\omega) + 
 \re^{-\ri B \omega} G_-^{-1}(\omega) Q_- (\omega), 
 \ee
 where $\tilde f(\omega)$ is the Fourier transform of $f(x)$ (which is extended to 
 the zero function outside the interval $[-B,B]$, as we have done in previous examples), and 
 \be
 \tilde g(\omega) = \frac{2\sin(B\omega)}{\omega}. 
 \ee
 In this case, the Wiener--Hopf decomposition of the kernel is given by 
 (see e.g. \cite{tw1,tw2,mr-long})
 \be
G_+(\omega) = \frac{\re^{ \frac{1}{2\pi}\ri\omega [ \log( -\frac{1}{2\pi}\ri\omega ) - 1 ] }}{\sqrt{\pi}}\Gamma\bigg( \frac{1}{2} - \frac{1}{2\pi} \ri\omega \bigg), 
\ee
and due to the parity of the problem $G_-(\omega)= G_+(-\omega)$. 
The quantities $\gamma$, $e(\gamma)$ can be obtained from 
$\tilde f (\omega)$ as follows, 
 \be
 { \pi \over \gamma}= \tilde f(0), \qquad e(\gamma)= -{\gamma^2 \over 4}-\pi^2{\tilde f ''(0) \over (\tilde f(0))^3 }. 
 \ee
 We note that, by using (\ref{int-eq-wh}), $\tilde f''(0)$ can also be 
 obtained as \cite{tw1} 
\begin{equation}
-\tilde f''(0) = \frac{\tilde f(0)}{2} - \tilde g''(0) - 2  \frac{\dd^2}{\dd\omega^2}\left[ \re^{\ri B \omega} \frac{Q_+(\omega)}{G_+(\omega)} \right]_{\omega=0}.
\label{eq_f''hat}
\end{equation}
 The function $Q(\omega)\equiv Q_+(\omega)$ satisfies 
the equation (\ref{Q-eq-np}), where 
\be
\label{poles-GY}
\xi_n = \pi(2n-1), \qquad n \in \mathbb{Z}_{>0},
\end{equation}
and 
\be
\sigma^\pm_n =  {\rm Res}_{\omega = \ri\xi_n \pm 0} \,  \sigma(\omega)= 
\pm \frac{2\pi}{(n-1)!^2}\left( \frac{2n-1}{2\re} \right)^{2n-1}.
%OLD WRONG VERSION %% \mp \frac{2\pi(-1)^{n}}{(n-1)!^2}\left( \frac{2n-1}{2\re} \right)^{2n-1}.
\end{equation}
 The r.h.s. of (\ref{Q-eq-np}), where $g_+(\omega) = \re^{\ri B \omega}\tilde g(\omega)$, can be computed in this case by using the trick explained in \figref{fig-crh}: the term 
 involving the exponential $\re^{2 \ri B \omega}$ is computed by deforming the integration contour into 
a Hankel contour around the imaginary axis, in the complex upper half plane. The remaining term is computed by calculating 
residues in the lower half plane, where $G_-(\omega)$ is analytic. One finds 
\be
%r(\ri \xi)=
\begin{aligned}
{1\over 2 \pi \ri} \int_\IR {G_-(\omega') g_+(\omega' ) \over \omega+ \omega'+ \ri 0}\dd\omega' = \frac{1}{\xi}\left( G_+(\ri\xi) - 1 \right) &+ \sum_{n\ge 1} \frac{\re^{-2B\xi_n}\ri\sigma_n^\pm G_-(\ri\xi_n)}{\xi_n(\xi_n + \xi)}\\
&- \frac{1}{2\pi\ri}\int_{\CC^\pm} \frac{\re^{-2B\xi'}\delta G_-(\ri\xi')}{\xi'(\xi'+\xi)}  \rd\xi'.
\end{aligned}
\label{r-source}
\end{equation}

We want to obtain a trans-series representation of $e(\gamma)$, as we did in the 
relativistic models in previous sections. The perturbative part of the functions appearing in the 
Wiener--Hopf construction has been obtained in \cite{tw1,tw2}. One finds, 
\be
\label{zero-ff}
\ba
\tilde f_{(0)}(0)& = 2B + \frac{\log(B\pi) + 1}{\pi} + \mathcal{O}\big(B^{-1}\big),\\
- \tilde f''_{(0)}(0) &= \frac{2}{3}B^3 + \frac{\log(B\pi) - 1}{\pi} B^2 + \mathcal{O}(B), 
\ea
\ee
where the subscript ${(0)}$ refers to the perturbative part. 
Let us now compute the very first 
exponential correction, at leading order in $1/B$. As in our analysis of the bosonic models, the first ingredient we 
need is $Q_1= Q(\ri \xi_1)$, where $\xi_1= \pi$ is the location of the first singularity. By 
evaluating (\ref{Q-eq-np}) at $\xi=\pi$ we find 
\be
Q_1 = \frac{1}{\pi}\left( G_+(\ri\pi) - 1 \right) - \frac{1}{4\pi^2 B}+\CO\big(B^{-2}\big) + \frac{\ri\sigma_1^\pm \re^{-2B\pi}}{\pi^2} \left( \frac{1}{2} + \frac{1}{8\pi B} +\CO\big(B^{-2}\big)\right).
\label{eq_h0_B^2}
\end{equation}
We now need the first non-perturbative correction to the function $Q(\omega)$, which we will denote by $Q_{(1)}(\omega)$. 
This function satisfies the integral equation (\ref{Q-eq-np}), in which we keep systematically the quantities of order $\re^{-2B\pi}$, and we obtain in this way
\be
\ba
Q_{(1)}(\omega) &=   \frac{\ri\sigma^\pm _1 \re^{-2B\pi}}{\pi(-\ri\omega + \pi)} \left( 1 + \frac{1}{4\pi B} - \frac{3}{32\pi^2 B^2} + \CO\big(B^{-3}\big) \right)\\
&+\frac{1}{2\pi\ri} \int_0^\infty \frac{\re^{-2B\xi'}\delta \sigma (\ri\xi') Q_{(1)}(\ri\xi')}{-\ri\omega + \xi'}\rd\xi'.  
\ea
\label{eq_h(0)}
\end{equation}
 To compute $Q_{(1)}(\omega)$, we iterate the driving term once in the first line of \eqref{eq_h(0)}, and express the resulting integral in terms of the exponential integral function:
\begin{equation}
\int_0^\infty \frac{\re^{-x}}{x+z} x^{a-1} \dd x = \Gamma(a) \re^z E_a(z).
\label{eq_exp_integral}
\end{equation}
It follows from (\ref{int-eq-wh}) that 
\be
\tilde f(0)= 2\big(B+ Q(0)\big), 
\ee
and by using the result for $Q_{(1)}(\omega)$, we obtain
\begin{equation}
\tilde f_{(1)}(0) = \frac{2\ri\sigma^\pm_1 \re^{-2B\pi}}{\pi^2}\left[ 1 + \frac{1}{2\pi B} + \CO\big(B^{-2}\big)- \frac{\ri\sigma_1^\pm \re^{-2B\pi}}{2\pi}\left( 1 + \frac{1}{2\pi B} + \CO\big(B^{-2}\big)\right) \right].
\label{eq_f(0)_result}
\end{equation}
From \eqref{eq_f''hat} we deduce
\begin{equation}
\ba
- \tilde f_{(1)}''(0) &= \frac{2\ri\sigma^\pm_1  \re^{-2B\pi}B^2 }{\pi^2} \Biggl[ 1 + \frac{\log(B\pi) + 3/2}{B\pi} +\CO\big(B^{-2}\big) \\
& \qquad \qquad\qquad \qquad - \frac{\ri\sigma^\pm _1 \re^{-2B\pi}}{2\pi}\left( 1 + \frac{\log(B\pi) + 3/2}{B\pi}  + \CO\big(B^{-2}\big) \right)\Biggr].
\ea
\label{eq_f''(0)_result}
\end{equation}
Combining all these results, we obtain the following expression for the coupling as a function of $B$, including non-perturbative corrections,  
\begin{equation}
\ba
\gamma &= \frac{\pi }{2 B} - \frac{\log (B\pi)+1}{4B^2} + \CO\big(B^{-3}\big) - \frac{\ri\sigma^\pm_1 \re^{-2B\pi}}{2 \pi  B^2} \Big(1+ \CO\big(B^{-1}\big) \Big)  \\
& \qquad - \frac{(\sigma^\pm_1)^2 \re^{-4B\pi}}{4 \pi^2 B^2}\Big( 1+ \CO\big(B^{-1}\big)\Big) + \CO\big(\re^{-6 B \pi}\big).
\ea
\end{equation}
For the energy we find:
\begin{equation}
\ba
e(\gamma) &=  \frac{\pi^2}{12} - \frac{\pi}{4B}  + \CO\big(B^{-2}\big) + \frac{3\ri\sigma^\pm_1 \re^{-2B\pi}}{4 \pi  B^2}  \Big( 1+ \CO\big(B^{-1}\big)\Big) + \frac{5 (\sigma^\pm_1)^2 \re^{-4B\pi}}{8 \pi^2 B^2} \Big( 1+ \CO\big(B^{-1}\big)\Big)\\
& + \CO\big(\re^{-6 B \pi}\big).
\ea
\label{eq_energy_density_exp2}
\end{equation}
Once we express it in terms of $\gamma$, we obtain the result 
\begin{equation}
\label{egy}
e(\gamma) = \frac{\pi^2}{12} - \frac{\gamma}{2} +\CO(\gamma^2) \pm \ri \re^{-\pi^2/\gamma}\gamma \big(1+ \CO(\gamma)\big)+ \frac{\pi^2}{2} \re^{-2\pi^2/\gamma} \big(1+ \CO(\gamma)\big) + \CO\big(\re^{-3\pi^2/\gamma}\big).
\end{equation}
The first exponential correction is precisely what was found in \cite{mr-long,mr-ll}, based on the large order behavior of the perturbative series. We also confirm the conjecture in \cite{mr-long} that the Stokes constant associated to this correction is purely imaginary. As indicated in (\ref{egy}), there are additional exponentially small corrections, corresponding to singularities located at $\zeta= n \pi^2$, $n\in \IZ_{>0}$.  
 
\sectiono{Conclusions and prospects}
\label{sec-conclusions}

In this paper we have developed analytic techniques to find the Borel singularities of the 
free energy in integrable, asymptotically free quantum field theories in two dimensions. 
These techniques are based on the Wiener--Hopf approach to the Bethe 
ansatz integral equations, and they provide a very simple picture of the singularity structure: 
given the Wiener--Hopf factorization of the kernel (\ref{wh-fact}) into two functions $G_\pm (\omega)$, 
the position of IR renormalons is determined by the poles of $G_-(\omega)/ G_+(\omega)$ in the complex upper half plane, while 
the position of UV renormalons is determined by the poles in the lower half plane. 
In addition to these sequences of singularities, there is also 
generically an isolated IR renormalon at the expected position (\ref{slore}) with $\ell=2$. 
We have calculated explicitly the 
very first terms of the trans-series associated to these singularities, and we obtained in 
particular analytic expressions for their Stokes constants. This makes it possible to test in 
detail these predictions with resurgent methods, based on  the large order behavior of 
long perturbative series. 

The first consequence of our analysis is that the location of generic IR renormalon singularities in the free energy 
is not as expected from the pioneering work of Parisi \cite{parisi2,parisi1} and 't Hooft \cite{thooft}. 
According to the standard lore, renormalon singularities are located at (\ref{slore}), and this has been the 
guiding principle in renormalon physics for forty years 
(see e.g. \cite{beneke}). These expectations were based on semi-quantitative analysis, large $N$ estimates, 
and the OPE expansion of \cite{itep}. Our analysis makes it clear that the location of renormalon singularities in QFT is more general than (\ref{slore}). The unconventional renormalons 
uncovered in our study have the property that, at large $N$, they 
become indistinguishable from the conventional ones, therefore they can not be detected with large $N$ techniques. 
In fact, the sequence of large $N$ renormalons appearing in the 
GN model and the PCF, and studied in \cite{fnw1,fnw2,mr-ren,dpmss}, comes from the new, 
unconventional renormalons 
at finite $N$. One of the general lessons of our paper is then that large $N$ estimates, 
based on particular classes of diagrams, are not 
reliable in order to determine the location of Borel singularities, and subleading diagrams in the $1/N$ expansion 
can change this location. A similar phenomenon was recently observed in \cite{dyson-dunne, bdm}, 
where changing the class of diagrams under consideration altered considerably the structure of Borel singularities.

We should note that our calculations have been made with a choice of scheme for the 
coupling constant which arises naturally from the Bethe ansatz 
(this was dubbed the ``TBA scheme" in \cite{dpmss}). 
One could wonder whether our results on the position of the singularities would change if we used, say, the $\overline{\text{MS}}$ scheme\footnote{We would like to thank Marco Serone for a discussion on this point.}. 
The answer to this question depends crucially on the convergence properties (or not) of the beta function in that scheme. 
If the beta function involves itself non-perturbative corrections (as argued in e.g. \cite{dps} in the case of QCD), 
these would change the pattern of singularities in the Borel plane. However, for most of the models considered in this paper, 
the beta function in the $\overline{\text{MS}}$ scheme is known to be a convergent series in the large $N$ limit (see e.g. \cite{gracey-rev}). Therefore, non-perturbative corrections to the beta function, if present, should be suppressed at large $N$, and do 
not have an impact on the unconventional renormalons found in this paper.

Although our results show that the standard expectation (\ref{slore}) is not a universal property of renormalons, it is possible 
that it still holds for obervables which are vevs of products of operators. In this case, it is believed that IR renormalon singularities are associated to the different operators appearing in the OPE \cite{parisi2, itep}. 
Since $\CF(h)$ is not an observable of that type, considerations based on the 
OPE do not apply in principle to the free energy studied in this paper. For this reason, it would be very interesting 
to consider correlation functions in integrable models, where the OPE applies. It is conceivable that information on their renormalon structure could be obtained from their form factor representation. For the PCF at large $N$, a relatively compact 
representation of this type exists for some correlators \cite{orland}. This and similar results might be 
the starting point for a study of the resurgent structure of correlation functions in integrable, asymptotically free theories.

Finding a physical interpretation of the singularities that we have unveiled, and of the associated trans-series, is challenging. 
It is not clear whether they can be interpreted as condensates, since they lead 
to terms of the form (\ref{lam-power}), but where $\ell$ is now rational. Our results are also problematic for semi-classical 
interpretations of renormalons. For example, it has been argued in \cite{cddu} that renormalons in the 
PCF might be related to ``fractons", i.e. fractionalized instantons that appear in twisted compactifications 
(see \cite{dunne-unsal-cpn,du-on} for related work on two-dimensional models). One piece of evidence cited for this 
connection is that the action of fractons matches the location of traditional renormalon singularities. However, 
since the renormalons found in this paper have a different location in the Borel plane, they seem to pose a basic problem for the 
semiclassical interpretation of \cite{cddu}. 

In view of the results in this paper, it would be important to find general principles which 
determine the structure of renormalons in quantum field theory. It has been proposed that 
renormalization group ideas can be used to find the location of Borel singularities (see e.g. 
\cite{parisi-rg,gk} for early work in this direction), and it would be interesting to see how this 
program can be applied to the relatively simple observable studied in this paper.   

In the three bosonic models that we have considered, we have found a subleading sequence 
of singularities whose location is proportional to $N-2$ in the (supersymmetric) sigma model, 
and to $N$ in the principal chiral field. Therefore, they have the correct scaling with $N$ to 
correspond to (unstable) instantons of these theories. It would be very interesting to test whether 
the corresponding trans-series can be reproduced with instanton methods. 

In addition to opening conceptual problems on the interpretation of the new singularities, our work can be 
also much improved on a computational level. The Wiener--Hopf method is very useful to obtain the very 
first coefficients of the trans-series, but it is not practical to compute higher order terms. For that, we would 
need for example an extension of Volin's method which incorporates non-perturbative 
effects. It would be also nice to complete our analysis and study UV renormalons in the bosonic models. 
One should also extend our considerations to the mother of all quantum integrable systems, 
namely the Lieb--Liniger model \cite{ll}. The leading IR singularity of the 
ground state energy of this model was detected numerically in \cite{mr-ll}, but extending our 
tools to this case is not completely straightforward.  

Finally, an important issue is how the findings of this paper relate to the two versions of the 
resurgence program considered in \cite{dpmss}.
%%%%%%%%%%%%%%%%%%%%%%%%%%%%%%%%
According to the weak version of the program, every observable with an asymptotic expansion 
can be written as a Borel-resummed trans-series. Our results fully validate this version 
for the free energy and are backed by numerical evidence. 
Meanwhile,
the strong version of the program requires in addition that all the formal power series appearing in the
trans-series can be eventually produced from a resurgent decoding of the perturbative sector.
In this case, our results are not conclusive.
%%%%%%%%%%%%%%%%%%%%%%%%%%%%%%%%%
As we discussed in section \ref{bos-results}, the $O(3)$ non-linear sigma model might 
be a counter-example for the strong resurgence program, since there is a real, exponentially small 
correction which can not be detected through the large order behavior of the perturbative series. 
However, this might be an exception rather than the norm, and in 
\cite{abbh1, abbh2} substantial evidence was given that the strong resurgence program was valid 
for the $O(4)$ non-linear sigma model. It would be very interesting to know which version of the resurgence program 
is implemented in the different models that we studied, and a more thorough application of our analytic framework 
should be able to answer this question.

\section*{Acknowledgements}
We would like to thank Gerald Dunne, Jie Gu, Marco Serone and Giacomo Sberveglieri for useful 
discussions and comments on the manuscript. This work has been supported by the ERC-SyG project 
``Recursive and Exact New Quantum Theory" (ReNewQuantum), which 
received funding from the European Research Council (ERC) under the European 
Union's Horizon 2020 research and innovation program, 
grant agreement No. 810573.

\appendix

\sectiono{The perturbative expansion of bosonic models}
\label{pert-calculation}

In this Appendix we will present the perturbative calculation of $\CF(h)$ for the bosonic models. 
This calculation was done in \cite{hmn,hn,pcf}, and it leads to remarkable exact results for the 
mass gap of integrable models. 
It has been extensively used in followup papers, like \cite{eh-ssm,eh-scpn}. However, 
as far as we know, the details of the calculation in \cite{hmn,hn,pcf} have not been presented anywhere, 
and we hope that the derivation given here will be useful for 
future studies. In addition, we will use the results in Appendix \ref{airy-kernel} to provide a full analytic calculation, since we will 
be able to solve in closed form the integral equations that were solved numerically in \cite{hmn,hn,pcf}. We should mention 
that Volin's method \cite{volin, volin-thesis} gives a different analytic derivation of the expression for $\CF(h)$.

 \begin{figure}[!ht]
\leavevmode
\begin{center}
\includegraphics[trim= {3.2cm 4.5cm 7.5cm 3.1cm}, clip,height=7cm]{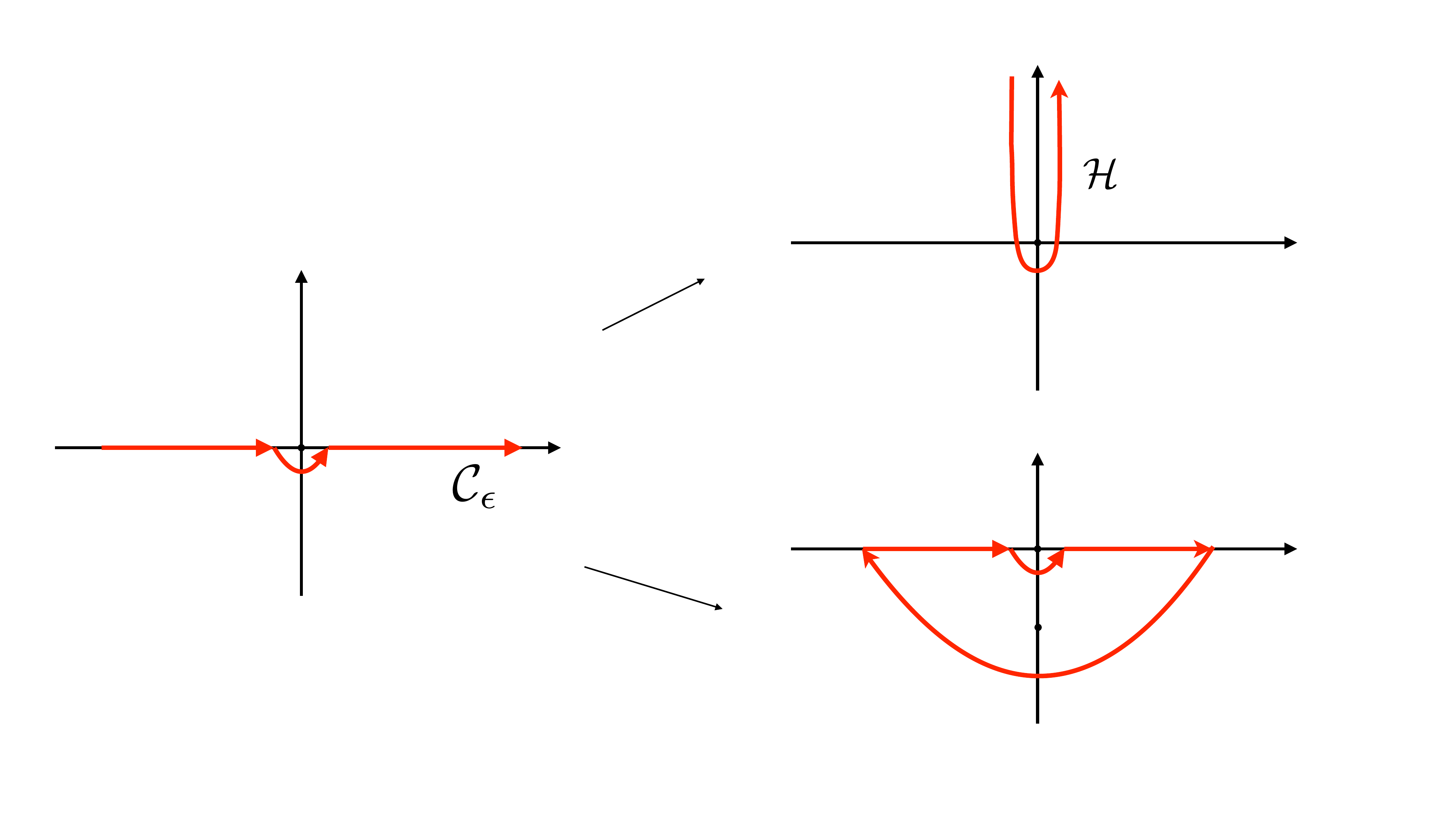}
\end{center}
\caption{The contour of integration in (\ref{eq_r}) is first deformed to $\CC_\epsilon$, which is then further deformed in two different 
ways for the different terms in the integrand.}
\label{fig-crh}
\end{figure} 

The first step in solving \eqref{eq_q} is to compute the driving term $r(x)$, by deriving the perturbative part of (\ref{eq_r}). 
We will do this to next-to-leading order in $1/B$. 
To organize the computation, we split the function $g_+(\omega) $ in two parts:
\begin{align}
g_+^{(h)} (\omega)&= \ri h \frac{ 1 - \re^{2\ri B \omega} }{\omega},\label{eq_g(h)}\\
g_+^{(m)} (\omega)&= {\ri m \re^B \over 2} \left( {\re^{2 \ri B \omega} \over \omega-\ri} -  {1\over \omega + \ri}\right) .\label{eq_g(m)}
\end{align}
To compute the contribution of $g^{(h)}_+(\omega)$ to $r(x)$, we deform the 
contour of integration from $\mathbb{R}$ to ${\cal C}_\epsilon$, in which we add a small 
semicircle around the origin of radius $\epsilon$ in the lower 
half plane, as shown in \figref{fig-crh}. We then split $g^{(h)}_+(\omega)$ into the terms 
$1/\omega$ and $-\re^{2\ri B\omega}/\omega$. For the first term, we deform the contour 
downwards, picking the pole at $\omega' = - \ri x/2B$ in the integrand of \eqref{eq_r}, as shown in the bottom 
right of \figref{fig-crh}. For the second term, we deform the contour upwards, leading to a Hankel contour 
$\mathcal{H}$ around the discontinuity of $G_-(\omega)$, as shown in the top right of \figref{fig-crh}. 
After the change of variable $\omega' = \ri y/2B$ we obtain:
\begin{equation}
r^{(h)}(x) = h G_+(\ri x/2B)\frac{2B}{x} - \frac{h}{2\pi\ri} \int_\mathcal{H}\frac{\re^{-y}}{(y+x)y} (2B) G_-(\ri y/2B) \rd y.
\label{rh1}
\end{equation}
We want now to write the integral over the Hankel contour $\mathcal{H}$ as an integral along 
the discontinuity of $G_-(\ri y/2B)$. Once this is done, we can expand the discontinuity at large $B$ 
and integrate term by term, by using (\ref{Gxi-exp}). However, the first term of this expansion goes like $y^{-3/2}$ and leads to a divergent integral when $y$ approaches 0. So the leading term in the expansion of $G_-(\ri y/2B)$ has to be computed with the following trick:
\be
\frac{kh(2B)^{3/2}}{2\pi} \int_\mathcal{H} \frac{\re^{-y}}{(y+x)y^{3/2}}\rd y = \frac{kh(2B)^{3/2}}{2\pi} \left( \int_\mathcal{H} \frac{\re^{-y}-1}{(y+x)y^{3/2}}\rd y
+ \int_\mathcal{H} \frac{1}{(y+x)y^{3/2}} \rd y\right).
\label{eq_hankel_contour_h_term}
\ee
Here, $k$ is the prefactor appearing in the first line of (\ref{Gxi-exp}). 
The first integral in the r.h.s. is no longer singular at $y=0$ and it can be computed by 
deforming the contour upwards and picking the discontinuity of $y^{-3/2}$. The second integral can be evaluated by 
picking the residue at $y=-x$, and it cancels the leading term of the function $h G_+(\ri x/2B)2B/x$ appearing in \eqref{rh1}. We finally obtain the following expression for 
$r^{(h)}(x)$ at NLO in a $1/B$ expansion:
\begin{equation}
r^{(h)}(x) = -kh (2B)^{1/2}\biggl[2B\frac{\mathsf{K}}{\pi}\frac{\re^x - 1}{x^{3/2}}
+ \left( 1 - \frac{\mathsf{K}}{\pi}\right) \frac{-a\log(2B) + a\log(x) + b}{x^{1/2}} + \mathcal{O}\big(B^{-1/2}\big)\biggr], 
\label{eq_r(h)}
\end{equation}
where $\mK$ is the Airy operator, defined in (\ref{Kop-def}), and $k$, $a$, and $b$ are as in (\ref{Gxi-exp}). 

To compute the contribution from $g^{(m)}_+(\omega)$ to $r$, we deform the contour downwards for the term $-1/(\omega+\ri)$ and upwards for $\re^{2\ri B\omega}/(\omega-\ri)$. This yields
\begin{equation}
r^{(m)}(x) = \frac{m\re^B}{2} \frac{2B}{2B-x}\Big[ G_+(\ri x/2B) - G_+(\ri) \Big]
-\frac{m \re^B}{2} \frac{1}{2\pi\ri} \int_0^\infty \rd y \frac{\re^{-y}(2B)\disc G_-(\ri y/2B)}{(2B-y)(y+x)}.
\end{equation}
Taking into account that $m\re^B$ will give an additional factor $\sqrt{B}$ when 
replaced with $h$ (as we will see in e.g. \eqref{eq_m_h_relation}, in the computation of the boundary condition), the expansion of $r^{(m)}(x)$ at order $B^{1/2}$ is given by
\begin{equation}
r^{(m)}(x) = -\frac{m\re^B G_+(\ri)}{\sqrt{\pi}} \left[ -(2B)^{1/2}\frac{k}{G_+(\ri)}\frac{\sqrt{\pi}}{2} \left(1 + \frac{\mathsf{K}}{\pi}\right)\frac{1}{x^{1/2}} + \frac{\sqrt{\pi}}{2} +\mathcal{O}(B^{-1/2})  \right].
\label{eq_r(m)}
\end{equation}
By combining the two pieces \eqref{eq_r(h)} and \eqref{eq_r(m)}, we can organize $r$ in the following way:
\begin{multline}
r(x) = -kh(2B)^{1/2}\left[ B r_0(x) + \log(2B)r_{2,1}(x) + r_{2,0}(x)\right]\\
- \frac{m\re^B G_+(\ri)}{\sqrt{\pi}}\left[ (2B)^{1/2} r_1(x) + r_2(x) \right] + \mathcal{O}\big(B^0\big),
\end{multline}
where
\begin{align}
r_0(x) &= 2\frac{\mathsf{K}}{\pi}\frac{\re^x - 1}{x^{3/2}},\\
r_1(x) &= -\frac{k}{G_+(\ri)} \frac{\sqrt{\pi}}{2} \left(1 + \frac{\mathsf{K}}{\pi}\right)\frac{1}{x^{1/2}},\\
r_{2,1}(x) &= -a \left( 1 - \frac{\mathsf{K}}{\pi} \right)\frac{1}{x^{1/2}},\\
r_{2,0}(x) &= \left( 1 - \frac{\mathsf{K}}{\pi} \right)\frac{a\log(x) + b}{x^{1/2}},\\
r_2(x) &= \frac{\sqrt{\pi}}{2}.
\end{align}

The next step is to solve the integral equation \eqref{eq_q} order by order in $1/B$. We thus propose the ansatz
\begin{multline}
\label{q-ansatz}
q(x) = -kh(2B)^{1/2}\left[ B q_0(x) + \log(2B)q_{2,1}(x) + q_{2,0}(x)\right]\\
- \frac{m\re^B G_+(\ri)}{\sqrt{\pi}}\left[ (2B)^{1/2} q_1(x) + q_2(x) \right] + \mathcal{O}\big(B^0\big),
\end{multline}
Equating terms of the same order in \eqref{eq_q} yields the following integral equations:
\begin{align}
\left(1+ {\mathsf{K} \over \pi} \right )q_0(x) &=r_0(x),\label{eq_integral_q0}\\
\left(1+ {\mathsf{K} \over \pi} \right )q_1(x) &=r_1(x),\label{eq_integral_q1}\\
\left(1+ {\mathsf{K} \over \pi} \right )q_{2,1}(x) &=r_{2,1}(x) + a {\mathsf{K} \over \pi} (x q_0(x)),\label{eq_integral_q21}\\
\left(1+ {\mathsf{K} \over \pi} \right )q_{2,0}(x) &=r_{2,0}(x) - {\mathsf{K} \over \pi} (a x \log(x) q_0(x) +b x q_0(x)),\label{eq_integral_q20}\\
\left(1 + {\mathsf{K} \over \pi} \right)q_2(x) &= r_2(x). \label{eq_integral_q2}
\end{align}
The function $q_0(x)$ can be solved explicitly, and the solution is written down in (\ref{qx-eq}). 
The integral equation \eqref{eq_integral_q1} is trivially solved by
\begin{equation}
q_1(x) =  -\frac{k}{G_+(\ri)} \frac{\sqrt{\pi}}{2} \frac{1}{x^{1/2}}.
\label{eq_q1_sol}
\end{equation}
As we will see in a moment, we do not need the explicit form of $q_{2,1}(x)$, $q_{2,0}(x)$ and $q_2(x)$, but only their integrals, 
which are calculated in Appendix \ref{airy-kernel}. 

Before considering the computation of $\CF(h)$, we need one last ingredient, which is the 
boundary condition (\ref{bc-eps}). Imposing this condition will yield a relation between $h$, $m$ and $B$. In the following 
we compute $\epsilon_+(\ri\kappa)$ at large $B$. The integral 
\begin{equation}
\frac{1}{2\pi\ri}\int_\mathbb{R}  \frac{G_-(\omega)g_+(\omega)}{\omega - \ri\kappa} \rd\omega
\end{equation}
in \eqref{eps-disc} can be computed with the same methods we used for $r(x)$, and we obtain 
\begin{multline}
\label{eq_int_G-g+}
\frac{1}{2\pi\ri}\int_\mathbb{R} \frac{G_-(\omega)g_+(\omega)}{\omega - \ri\kappa}  \rd\omega
= -\frac{m\re^B}{2(\kappa+1)}G_+(\ri)
+ \frac{kh(2B)^{1/2}}{\pi\kappa} \biggl[ 2\sqrt{\pi} - \frac{\log(2B)}{B} aI_0\\
+ \frac{1}{B}\bigg(aI_1 + bI_0 + \frac{I_0}{\kappa}\bigg)\biggr]
- \frac{km\re^B}{\pi\kappa(2B)^{1/2}}\frac{\sqrt{\pi}}{2} + \mathcal{O}(B^{-1})
\end{multline}
where
\begin{align}
I_0 &= -\frac{1}{2}\int_0^\infty \frac{\re^{-y}}{\sqrt{y}} \rd y = -\frac{\sqrt{\pi}}{2},\\
I_1 &= -\frac{1}{2}\int_0^\infty \frac{\re^{-y}}{\sqrt{y}}\log(y) \rd y = \frac{\sqrt{\pi}}{2}(\gamma_E+\log(4)).
\end{align}
We also need to compute the second piece of \eqref{eps-disc}, which in perturbation theory is given by
\begin{multline}
\frac{1}{\pi} \int_0^\infty \frac{\re^{-x} \gamma(x/2B)}{2B\kappa-x}q(x)\dd x = - \frac{kh(2B)^{1/2}}{2\pi\kappa} \biggl[ \langle q_0 \rangle
+ \frac{1}{B}\biggl( a\log(2B)\langle q_{2,1}^a - xq_0 \rangle
+ a\langle q_{2,0}^a + x\log(x)q_0 \rangle\\
+ b\langle q_{2,0}^b + xq_0 \rangle + \frac{\langle xq_0\rangle}{2\kappa} \biggr)\biggr]
- \frac{m\re^B}{\pi\kappa} \frac{G_+(\ri)}{\sqrt{\pi}} \left[ \frac{\langle q_1 \rangle}{(2B)^{1/2}} + \frac{\langle q_2 \rangle}{2B}  \right] + \mathcal{O}(B^{-1}),
\label{eq_int_gamma_q}
\end{multline}
In this equation we have expressed the integrals in terms of the moments introduced in (\ref{norm}), and we split the functions $q_{2,1}(x)$ and $q_{2,0}(x)$ in 
terms proportional to the constants $a$, $b$ appearing in (\ref{Gxi-exp}):
\begin{align}
q_{2,1}(x) &= a q_{2,1}^a(x),\\
q_{2,0}(x) &= a q_{2,0}^a(x) + b q_{2,0}^b(x),
\end{align}
%according to \eqref{eq_integral_q21}--\eqref{eq_integral_q20}.

Let us now propose the ansatz
\begin{equation}
m\re^B = kh(2B)^{1/2} \frac{\sqrt{\pi}}{G_+(\ri)}\left[ c_0 + \frac{c_1}{\sqrt{B}} + \frac{\log(2B)c_{2,1}}{B} + \frac{c_{2,0}}{B} + \mathcal{O}\big(B^{-3/2}\big) \right].
\label{eq_m_h_relation}
\end{equation}
The boundary condition (\ref{bc-eps}) then yields four equations, obtained by equating contributions of the same order in $B$. The solution to the system is given by
\begin{align}
c_0 &= 1, \qquad c_1 = 0,\\
c_{2,1} &= -\frac{2aI_0}{\pi^{3/2}} - \frac{a}{\pi^{3/2}} \langle q_{2,1}^a - xq_0 \rangle,\\
c_{2,0} &= \frac{2(bI_0+aI_1)}{\pi^{3/2}} - \frac{1}{\pi^{3/2}} \bigg[ a\langle q_{2,0}^a + x\log(x)q_0 \rangle + b\langle q_{2,0}^b +xq_0 \rangle + \langle q_2 \rangle \bigg].
\end{align}
In particular, for the computation of $c_1$ we need the moment $\langle q_1 \rangle$, which can be straightforwardly computed from its expression in \eqref{eq_q1_sol}:
\begin{equation}
\langle q_1 \rangle = - \frac{k}{G_+(\ri)} \frac{\pi}{2}.
\end{equation}
The remaining moments are presented in Appendix \ref{airy-kernel}.
%
%The boundary condition (\ref{bc-eps}) then yields four equations, obtained by equating contributions of the same order in $B$:
%\begin{align}
%-\frac{\sqrt{\pi}}{2} c_0 + \frac{2}{\sqrt{\pi}} - \frac{1}{2\pi} \langle q_0 \rangle &= 0,\label{eq_c0}\\
%-\frac{\sqrt{\pi}}{2} c_1 - \frac{k}{2\sqrt{2}G_+(\ri)} - \frac{1}{2\pi} \langle q_1 \rangle &= 0,\label{eq_c1}\\
%-\frac{\sqrt{\pi}}{2} c_{2,1} - \frac{aI_0}{\pi} - \frac{a}{2\pi} \langle q_{2,1}^a - xq_0 \rangle &= 0,\label{eq_c21}\\
%-\frac{\sqrt{\pi}}{2} c_{2,0} + \frac{bI_0 + aI_1}{\pi} - \frac{1}{2\pi} \bigg[ a\langle q_{2,0}^a + x\log(x)q_0 \rangle + b\langle q_{2,0}^b +xq_0 \rangle + \langle q_{2,0}^c \rangle \bigg] &= 0\label{eq_c20}.
%\end{align}
%The last two equations are only valid assuming that $c_1 = 0$, which we will see is true in a moment. 
%From (\ref{int-result}) and \eqref{eq_c0} we deduce
%%
%\begin{equation}
%c_0=1.
%\end{equation}
%%
%The function $q_1(x)$ is given in \eqref{eq_q1_sol} and 
%%
%\begin{equation}
%\langle q_1 \rangle =  -\frac{k}{G_+(\ri)} \frac{\pi}{\sqrt{2}}.
%\end{equation}
%%
%Thus, \eqref{eq_c1} yields
%%
%\begin{equation}
%c_1=0.
%\end{equation}
%%
%Finally, \eqref{eq_c21} and \eqref{eq_c20} yield
%\begin{align}
%c_{2,1} &= -\frac{2aI_0}{\pi^{3/2}} - \frac{a}{\pi^{3/2}} \langle q_{2,1}^a - xq_0 \rangle,\\
%c_{2,0} &= \frac{2(bI_0+aI_1)}{\pi^{3/2}} - \frac{1}{\pi^{3/2}} \bigg[ a\langle q_{2,0}^a + x\log(x)q_0 \rangle + b\langle q_{2,0}^b +xq_0 \rangle + \langle q_{2,0}^c \rangle \bigg].
%\end{align}
The boundary condition \eqref{eq_m_h_relation} can also be inverted to express $B$ as a function of $\log(h/m)$:
\begin{multline}
B = \log\left( \frac{h}{m} \right) + \frac{1}{2}\log\log\left( \frac{h}{m} \right) + \log\left( \frac{k\sqrt{2\pi}}{G_+(\ri)}\right)\\
+ \frac{\big(c_{2,1}+\tfrac{1}{4}\big)\log\log\big(\tfrac{h}{m}\big)}{\log\big(\tfrac{h}{m}\big)} + \frac{\tfrac{1}{2}\log\big( \tfrac{k\sqrt{2\pi}}{G_+(\ri)}\big) + c_{2,1}\log(2)+c_{2,0}}{\log\big(\tfrac{h}{m}\big)}+ \mathcal{O}\big(\log^{-2}(h/m)\big).
\label{eq_B_hm_relation}
\end{multline}

We now have all the ingredients to compute $\CF(h)$, which is obtained by evaluating \eqref{eq_int_G-g+} and \eqref{eq_int_gamma_q} at $\kappa=1$, and using the boundary condition \eqref{eq_m_h_relation}. We note that the coefficients $c_{2,1}$ and $c_{2,0}$ would a priori contribute to the order we are working, but they cancel in this step. The result is
\begin{multline}
\mathcal{F}(h) = - \frac{k^2 h^2}{4} \biggl\{ B - a \log(2B)\frac{4}{\pi^{3/2}}\big( I_0 + \tfrac{1}{2}\langle q_{2,1}^a - xq_0 \rangle\big)
+ \frac{4}{\pi^{3/2}}\Bigl[ a\big(I_1 - \tfrac{1}{2}\langle q_{2,0}^a + x\log(x)q_0 \rangle \big)\\
+ b\big(I_0 - \tfrac{1}{2}\langle q_{2,0}^b + xq_0 \rangle\big)
+ I_0 - \tfrac{1}{2}\langle q_2 \rangle - \tfrac{1}{4}\langle xq_0\rangle \Bigr] + \mathcal{O}\big(B^{-1/2}\big) \biggr\}.
\label{eq_ground_energy_density_parameters}
\end{multline}
Using the results of Appendix \ref{airy-kernel}, examples \ref{q0ex} and \ref{log-ints}, we can evaluate all the integrals and we find
\begin{equation}
\mathcal{F}(h) = - \frac{k^2 h^2}{4} \biggl\{ B + a \log(2B)
+ a \bigl(\gamma_E  -1 + \log(4) \bigr)\\
- b
-1 + \mathcal{O}\big(B^{-1/2}\big) \biggr\}.
\end{equation}
Finally, we express $B$ in terms of $\log(h/m)$ using \eqref{eq_B_hm_relation}, and we obtain
\begin{multline}
\mathcal{F}(h) = -{k^2 h^2 \over 4} \biggl\{ \log\left( {h \over m} \right)+\left( a+{1\over 2} \right) \log \log\left( {h \over m} \right)+ \log \left( {{k\sqrt{2 \pi}}  \over G_+(\ri)} \right)\\
+ a \big( \gamma_E -1+ \log(8)\big) - b -1 + \mathcal{O}\big(\log^{-1/2}(h/m)\big) \biggr\}.
\end{multline}
This is the result quoted in \cite{pcf}.

\sectiono{The Airy operator}
 
 \label{airy-kernel}
 
 Let us consider the integral operator $\mK$ defined by 
\be
\label{Kop-def}
(\mK f)(x)= \int_0^\infty {\re^{-y} \over x+ y} f(y) \rd y. 
\ee
We will call $\mK$ the Airy operator, since it is closely related to the Airy functions (see e.g. \cite{fendley,oper}). It has a continuous spectrum and its eigenvalues and eigenfunctions are known explicitly \cite{mtw},
\be
\label{kchi}
\int_0^\infty {\re^{-y} \over x+y} \chi_p(y) \rd y= \lambda_p \chi_p(x),
\ee
where 
\be
\lambda_p= \pi {\rm sech} (\pi p), \qquad p \ge 0, 
\ee
and 
  \be
  \label{chiK}
  \chi_p(x) ={  {\sqrt{2 p \sinh(\pi p)}} \over \pi} {\re^{x/2} \over {\sqrt{x}}} K_{\ri p}\left({x \over 2} \right). 
  \ee
Here, $K_\nu(x)$ is the modified Bessel function. These eigenfunctions satisfy the normalization condition 
\be
\int_0^\infty \re^{-x} \chi_p(x) \chi_{p'}(x)\, \rd x =\delta(p-p'). 
\ee
The expression (\ref{chiK}) is more useful than the one appearing in \cite{mtw}. One way to obtain it is to use the observation in \cite{Kostovo2} that $\mK$ commutes with the Schr\"odinger operator
  \be
  \mH= -{\rd^2 \over \rd u^2} + \re^{2u}, 
  \ee
 whose eigenfunctions are well-known to be given by (\ref{chiK}) (see e.g. \cite{cfz}). The following integral identities are useful to perform explicit computations, 
  \be
  \label{chi-ints}
  \ba
 c(\mu;p)&= \int_0^\infty  x^\mu \re^{-x} \chi_p(x)\, \rd x= (2 \lambda_p)^{1/2}  C_{p,0} { \left({1\over 2}+ \ri p\right)_\mu \left({1\over 2}- \ri p\right)_\mu \over \Gamma(\mu+1)}, \\
 c'(\mu;p)&=\int_0^\infty  x^\mu \log(x) \re^{-x} \chi_p(x)\, \rd x\\
 &=c(\mu;p) \left(\psi\left(\frac{1}{2}-\ri p+\mu
\right) + \psi\left(\frac{1}{2}+\ri p+\mu\right)-\psi (\mu+1) \right), 
  \ea
  \ee
 and they hold for ${\rm Re}(\mu) > -1$. In this equation, $\psi(z)$ is the digamma function, and 
  \be
C_{p,0}=(p \tanh(\pi p))^{1/2}. 
\ee

We can use these results to give an explicit solution to the integral equation, 
\be
\label{ie-q}
q(x)+ \frac{1}{\pi} \int_0^\infty {\re^{-y} \over x+ y} q(y) \rd y = r(x), 
\ee
as
\be
\label{q-sol}
q(x)= \int_0^\infty  \rd p \, \left(1 + {\lambda_p \over \pi} \right)^{-1} \langle \chi_p | r\rangle \chi_p(x), 
\ee
where
\be
\langle \chi_p | r\rangle=\int_0^\infty \re^{-x} \chi_p(x) r(x)\, \rd x. 
\ee
In particular, the moments of $q(x)$ can be computed in closed form in terms of integrals over $p$:
\be
\label{n-moment}
\langle q \rangle_n= \int_0^\infty x^n \re^{-x} q(x) \rd x = \int_0^\infty \left(1 + {\lambda_p \over \pi} \right)^{-1} \langle \chi_p | r\rangle c(n;p)\,  \rd p, 
\ee
where $c(n;p)$ is given by (\ref{chi-ints}). We will denote
\begin{equation}
\label{norm}
\langle q \rangle =\langle q \rangle_0= \int_0^\infty  \re^{-x} q(x)\,\rd x. 
\end{equation}

\begin{example} \label{q0ex} Let us solve the integral equation (\ref{eq_integral_q0}). We denote 
\be
s(x)=  {\re^x-1 \over 2 x^{3/2}}. 
\ee
We have, 
\be
\langle r_0|\chi_p \rangle ={4 \over \pi} \lambda_p \langle s(x) | \chi_p \rangle, 
\ee
where 
\be
 \langle s| \chi_p \rangle=\frac{\coth \left(\frac{\pi  p}{2}\right)}{2^{3/2} \left(p^2+1\right) \sqrt{p\,
   \text{csch}(\pi  p)}}. 
   \ee
We now use the expression (\ref{chiK}) and the well-known integral representation of the Bessel function
     \be
     \label{bes-def}
     K_\nu (x) = \int_0^\infty \rd u \, \re^{-x \cosh(u)} \cosh(\nu u) \rd u 
     \ee
to conclude that 
    \be
    \label{qx-eq}
    \ba
    q_0(x)&={2 \over \pi} {\re^{x/2} \over  {\sqrt{x}}} \int_0^\infty \rd u \, \re^{-{x \over 2}  \cosh(u)} \int_0^\infty \rd p { \cos(p u) \over 1+ p^2}= {\re^{x/2} \over  {\sqrt{x}}} \int_0^\infty \rd u \, \re^{-{x \over 2}  \cosh(u)-u}\\
    &= {\re^{x/2} \over \sqrt{x}} \left(K_1\left(\frac{x}{2}\right)-\frac{2 \, \re^{-x/2}}{x}\right).  
    \ea
     \ee
In particular,
    \be
    \label{int-result}
    \langle q_0 \rangle=
     \int_0^\infty \re^{-x} q_0(x) \rd x =\sqrt{\pi}(4-\pi).  
     \ee
     This integral can also be computed by using (\ref{n-moment}). One also finds, 
  \begin{equation}
\langle x q_0 \rangle = \sqrt{\pi} \int_\mathbb{R} \dd p\, \textrm{sech}(\pi p) \frac{p^2+1/4}{p^2+1}= \sqrt{\pi}  \left( \frac{3\pi}{4} - 2\right).
\end{equation}   
\qed
 \end{example}
 
\begin{example} \label{log-ints}
We will now compute the remaining integrals appearing in (\ref{eq_ground_energy_density_parameters}). 
Combining the integral equations \eqref{eq_integral_q0} and \eqref{eq_integral_q21}, we obtain:
\begin{equation}
q_{2,1}^a(x) - x q_0(x) = -\left( 1 + \frac{\mathsf{K}}{\pi} \right)^{-1} \left( 1 - \frac{\mathsf{K}}{\pi} \right)\frac{1}{\sqrt{x}} - \left( 1 + \frac{\mathsf{K}}{\pi} \right)^{-1}\big(xq_0(x)\big).
\label{eq_q21_q0_integral_equation}
\end{equation}
By using again (\ref{n-moment}) we find, 
\begin{equation}
\begin{aligned}
\left\langle \left( 1 + \frac{\mathsf{K}}{\pi} \right)^{-1} \left( 1 - \frac{\mathsf{K}}{\pi} \right)\frac{1}{\sqrt{x}} \right\rangle &= \int_0^\infty \dd p \left( 1 + \frac{\lambda_p}{\pi} \right)^{-1} \left( 1 - \frac{\lambda_p}{\pi} \right) \braket{\chi_p|1/\sqrt{x}} \braket{\chi_p}\\
&= 2\sqrt{\pi} \int_0^\infty \dd p\, \textrm{sech}(p\pi)\tanh^2(p\pi/2) = \sqrt{\pi}\left(\frac{4}{\pi} - 1 \right),
\label{eq_sqrt_moment_21}
\end{aligned}
\end{equation}
as well as
\begin{equation}
\begin{aligned}
\left\langle \left( 1 + \frac{\mathsf{K}}{\pi} \right)^{-1}\big(xq_0(x)\big) \right\rangle &= \int_0^\infty \dd p \left( 1 + \frac{\lambda_p}{\pi} \right)^{-1} \braket{\chi_p|xq_0}\langle \chi_p \rangle\\
&= \int_0^\infty \dd p \frac{2p \tanh(p\pi/2)}{\sqrt{\pi}}\left[ \frac{\pi^2 \textrm{csch}^2(p\pi/2)}{2} - \frac{2\pi\textrm{csch}(p\pi)}{p}\right]\\
&= \sqrt{\pi}\left( \frac{\pi}{2} - \frac{4}{\pi} \right).
\end{aligned}
\label{eq_q0_moment_21}
\end{equation}
The inner product $\braket{\chi_p|xq_0}$ can be computed by using the explicit expressions (\ref{chiK}) and (\ref{qx-eq}). We conclude that 
\begin{equation}
\langle q_{2,1}^a - x q_0 \rangle =  -\sqrt{\pi}\left(\frac{\pi}{2} - 1\right). 
\end{equation}

To compute $\langle q_{2,0}^b + xq_0 \rangle$, we realize that this combination of $q$ functions satisfies the same equation as in \eqref{eq_q21_q0_integral_equation}, but with a change of sign in the r.h.s.:
\begin{equation}
q_{2,0}^b(x) + x q_0(x) = \left( 1 + \frac{\mathsf{K}}{\pi} \right)^{-1} \left( 1 - \frac{\mathsf{K}}{\pi} \right)\frac{1}{\sqrt{x}} + \left( 1 + \frac{\mathsf{K}}{\pi} \right)^{-1}\big(xq_0(x)\big).
\end{equation}
Therefore, we have
\begin{equation}
\langle q_{2,0}^b + x q_0 \rangle = \sqrt{\pi}\left( \frac{\pi}{2} - 1 \right).
\end{equation}
We also have
\begin{equation}
q_2(x) = \frac{\sqrt{\pi}}{2} \left( 1 + \frac{\mathsf{K}}{\pi} \right)^{-1} 1,
\end{equation}
and we obtain
\begin{equation}
\langle q_2 \rangle = \frac{\sqrt{\pi}}{2}\int_0^\infty \dd p \left( 1 + \frac{\lambda_p}{\pi} \right)^{-1} \langle \chi_p \rangle^2 = \frac{\pi^{3/2}}{8}.
\end{equation}

The most difficult calculation involves $q_{2,0}^a (x) + x \log(x) q_0(x)$. From (\ref{eq_integral_q0}) and (\ref{eq_integral_q20}) we obtain
\be
q_{2,0}^a (x) + x \log(x) q_0(x) =\left( 1+ {\mK \over \pi} \right)^{-1} \left(  \left(1-{\mK \over \pi} \right) {\log (x) \over {\sqrt{x}}} + x \log(x) q_0(x) \right). 
\ee
We are ultimately interested in computing $\langle q_{2,0}^a+ x \log(x) q_0\rangle$. By using (\ref{chi-ints}) and other integral formulae, we find
\be
\ba
\langle \chi_p| x \log(x) q_0 \rangle&= { \sqrt{p \sinh (\pi  p)}  \over  \sqrt{2} } \biggl\{ \pi  \text{csch}^2\left(\frac{\pi  p}{2}\right)
   \left(\psi\left(-\frac{\ri p}{2}\right)+\psi \left(\frac{\ri
   p}{2}\right)+\gamma_E +\log (4)\right) \\
   & \qquad  \qquad -{4 \text{csch}(\pi  p) \over p} \left(\psi(-\ri
   p)+\psi (\ri p)+\gamma_E +\log (4)\right)\biggr\}, 
   \ea
   \ee
   and one eventually obtains 
  \be
\langle q_{2,0}^a + x \log(x) q_0 \rangle= \sqrt{\pi}  \big( \left(\gamma_E + \log(4) \right)  I_e+ \pi I_s- I_c \big), 
  \ee
  where
  \be
  I_e=\int_\IR \rd p \left(  {\pi p \over \sinh(\pi p)}- {1\over \cosh(p)} \right)={\pi \over 2}-1
  \ee
  can be calculated easily, and
  \be
  \label{g-ints}
  \ba
    I_c&=\int_\IR \rd p {1 \over \cosh(\pi p)} \left(\psi\left(-\ri p\right)+\psi \left(\ri
   p\right) \right),\\
  I_s&= \int_\IR \rd p {p \over \sinh(\pi p)} \left(\psi\left(-\frac{\ri p}{2}\right)+\psi \left(\frac{\ri
   p}{2}\right) \right).
    \ea
   \ee
\begin{figure}[!ht]
\leavevmode
\begin{center}
\includegraphics[height=4.5cm]{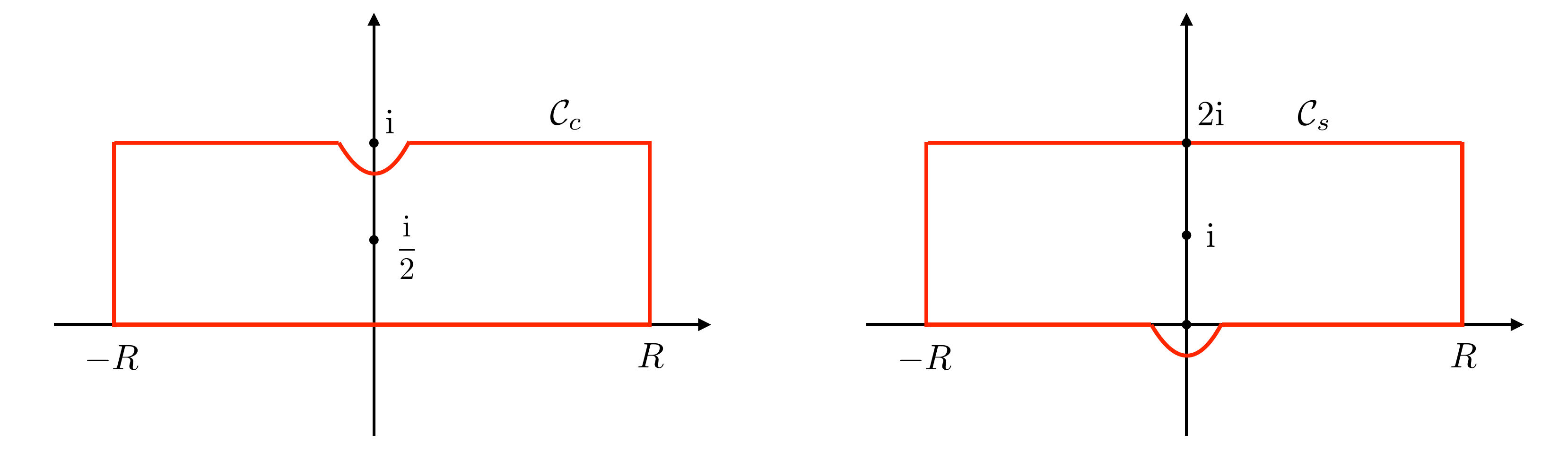}
\end{center}
\caption{Integration contours for the calculation of the integrals (\ref{cg-ints}). They are oriented counterclockwise.}
\label{ccs-fig}
\end{figure}%
To calculate these integrals, we consider the contours $\CC_c$ and $\CC_s$, shown respectively in the left 
and right drawings in \figref{ccs-fig}, and the corresponding integrals 
   \be
   \label{cg-ints}
   \ba
   \CI_c &= \int_{\CC_c} {\rd z \over \cosh(\pi z)} \left(\psi\left(-\ri z\right)+\psi \left(\ri
   z\right) \right),\\
   \CI_s&=  \int_{\CC_s} \rd z{ (z-2 \ri)^2 \over \sinh(\pi z)} \left(\psi\left(-{\ri z \over 2} \right)+\psi \left({\ri
   z \over 2} \right) \right).
   \ea
   \ee
 By using the well-known property of the digamma function, 
 \be
 \psi(z+1)= \psi(z) +{1\over z}
 \ee
 one can reduce the calculation of $\CI_{c,s}$ to elementary integrals and residues. One finds in the end, 
 \be
 I_c= -2 \gamma_E + \pi -4 \log(2), \qquad 
 I_s= -\gamma_E -2 \log(2) +{3\over 2}. 
 \ee
 We conclude that
 \be
\langle q_{2,0}^a + x \log(x) q_0 \rangle = \frac{\sqrt{\pi}}{2}  \bigl(-\gamma_E  (\pi -2)+\pi -2\pi  \log (2)+4 \log (2)\bigr). 
 \ee
\qed
\end{example}

\bibliographystyle{JHEP}

\linespread{0.6}
\bibliography{antrans-bib}

\end{document}